\documentclass[12pt]{article}
\pdfoutput=1

\usepackage{draft} 
\usepackage{hyperref}
\usepackage{graphicx,color,subfig}
\usepackage{cite}
\usepackage{mciteplus}
\usepackage{skak}
\usepackage{empheq}
\usepackage{tikz}
\usepackage{booktabs}
\usepackage{siunitx}
\usepackage{bbm}
\usepackage{makecell}
\usepackage{float}

\usepackage{bm}
\usepackage{braket}
\usepackage{slashed}

\DeclareFontFamily{OT1}{pzc}{}
\DeclareFontShape{OT1}{pzc}{m}{it}{<-> s * [1.10] pzcmi7t}{}
\DeclareMathAlphabet{\mathpzc}{OT1}{pzc}{m}{it}

\def\be#1\ee{\begin{align}#1\end{align}}

\makeatletter
\newcommand{\bdryno}{\mathpalette\bdry@no\relax}
\newcommand{\bdry@no}[2]{%
  \mspace{1mu}%
  \vbox{%
    \hbox{$\m@th#1\scriptstyle{\ast}$}
    \nointerlineskip
    \kern.25ex
    \hbox{$\m@th#1\scriptstyle{\ast}$}
    \kern-.06ex
  }%
  \mspace{1mu}%
}
\makeatother
    
\usetikzlibrary{snakes}
\usetikzlibrary{arrows}
\usetikzlibrary{decorations.pathmorphing}
\usetikzlibrary{decorations.markings}
\tikzset{snake it/.style={decorate, decoration=snake}}
\usetikzlibrary{shapes.misc}

\tikzset{cross/.style={cross out, draw=black, minimum size=2*(#1-\pgflinewidth), inner sep=0pt, outer sep=0pt},
cross/.default={1pt}}

\tikzset{crossbl/.style={cross out, draw=black, minimum size=3*(#1-\pgflinewidth), inner sep=0pt, outer sep=0pt},
cross/.default={1pt}}

\tikzset{gcross/.style={cross out, draw=magenta, minimum size=3*(#1-\pgflinewidth), inner sep=0pt, outer sep=0pt},
cross/.default={1pt}}

\tikzset{bcross/.style={cross out, draw=cyan, minimum size=3*(#1-\pgflinewidth), inner sep=0pt, outer sep=0pt},
cross/.default={1pt}}

\usetikzlibrary{shapes.geometric}

\tikzset{
  pics/diske/.style n args={1}{
    code = { %
        \filldraw[color=black, fill=black!30, thick] circle (0.3);
        \node[color=black] at (0,0.5) {\scriptsize ${#1}$};
    }
  }
}
\tikzset{
  pics/disker/.style n args={1}{
    code = { %
        \filldraw[color=red, fill=black!30, thick] circle (0.3);
        \node[color=black] at (0,0.5) {\scriptsize ${#1}$};
    }
  }
}
\tikzset{
  pics/diskeb/.style n args={1}{
    code = { %
        \filldraw[color=blue, fill=black!30, thick] circle (0.3);
        \node[color=black] at (0,0.5) {\scriptsize ${#1}$};
    }
  }
}   
\tikzset{
  pics/disk1/.style n args={1}{
    code = { %
        \filldraw[color=black, fill=black!30, thick] circle (0.3);
        \node[color=black] at (0,0.5) {\scriptsize ${#1}$};
        \draw node[cross=3pt, very thick] {};
    }
  }
}   
\tikzset{
  pics/disk1r/.style n args={1}{
    code = { %
        \filldraw[color=red, fill=black!30, thick] circle (0.3);
        \node[color=black] at (0,0.5) {\scriptsize ${#1}$};
        \draw node[cross=3pt, very thick] {};
    }
  }
}
\tikzset{
  pics/disk1b/.style n args={1}{
    code = { %
        \filldraw[color=blue, fill=black!30, thick] circle (0.3);
        \node[color=black] at (0,0.5) {\scriptsize ${#1}$};
        \draw node[cross=3pt, very thick] {};
    }
  }
}
\tikzset{
  pics/disk2/.style n args={1}{
    code = { %
        \filldraw[color=black, fill=black!30, thick] circle (0.3);
        \node[color=black] at (0,0.5) {\scriptsize ${#1}$};
        \node[cross=3pt, very thick] at (0.125,0) {};
        \node[cross=3pt, very thick] at (-0.125,0) {};
    }
  }
}
\tikzset{
  pics/disk2r/.style n args={1}{
    code = { %
        \filldraw[color=red, fill=black!30, thick] circle (0.3);
        \node[color=black] at (0,0.5) {\scriptsize ${#1}$};
        \node[cross=3pt, very thick] at (0.125,0) {};
        \node[cross=3pt, very thick] at (-0.125,0) {};
    }
  }
}
\tikzset{
  pics/disk2b/.style n args={1}{
    code = { %
        \filldraw[color=blue, fill=black!30, thick] circle (0.3);
        \node[color=black] at (0,0.5) {\scriptsize ${#1}$};
        \node[cross=3pt, very thick] at (0.125,0) {};
        \node[cross=3pt, very thick] at (-0.125,0) {};
    }
  }
}   

\tikzset{
  pics/cylnn0/.style n args={2}{
    code = { %
        \filldraw[color=black,fill=black!30, thick] circle (0.35);
        \filldraw[color=black, fill=white, thick]  circle (0.17);
        \node[color=black] at (0,0.51) {\scriptsize ${#1}$};
        \node[color=black] at (0,0) {\scriptsize ${#2}$};

    }
  }
}
\tikzset{
  pics/cylnnrb0/.style n args={2}{
    code = { %
        \filldraw[color=red,fill=black!30, thick] circle (0.35);
        \filldraw[color=blue, fill=white, thick]  circle (0.17);
        \node[color=black] at (0,0.51) {\scriptsize ${#1}$};
        \node[color=black] at (0,0) {\scriptsize ${#2}$};

    }
  }
}

\tikzset{
  pics/cylnnrb/.style n args={2}{
    code = { %
        \filldraw[color=red, densely dashed,fill=black!30, thick] circle (0.35);
        \filldraw[color=blue, densely dashed, fill=white, thick]  circle (0.17);
        \node[color=black] at (0,0.51) {\scriptsize ${#1}$};
        \node[color=black] at (0,0) {\scriptsize ${#2}$};

    }
  }
}
\tikzset{
  pics/cylnnrr0/.style n args={2}{
    code = { %
        \filldraw[color=red,fill=black!30, thick] circle (0.35);
        \filldraw[color=red, fill=white, thick]  circle (0.17);
        \node[color=black] at (0,0.51) {\scriptsize ${#1}$};
        \node[color=black] at (0,0) {\scriptsize ${#2}$};

    }
  }
}
\tikzset{
  pics/cylnnrr/.style n args={2}{
    code = { %
        \filldraw[color=red, densely dashed,fill=black!30, thick] circle (0.35);
        \filldraw[color=red, densely dashed, fill=white, thick]  circle (0.17);
        \node[color=black] at (0,0.51) {\scriptsize ${#1}$};
        \node[color=black] at (0,0) {\scriptsize ${#2}$};

    }
  }
}
\tikzset{
  pics/cylnnbb0/.style n args={2}{
    code = { %
        \filldraw[color=blue,fill=black!30, thick] circle (0.35);
        \filldraw[color=blue, fill=white, thick]  circle (0.17);
        \node[color=black] at (0,0.51) {\scriptsize ${#1}$};
        \node[color=black] at (0,0) {\scriptsize ${#2}$};

    }
  }
}
\tikzset{
  pics/cylnnbb/.style n args={2}{
    code = { %
        \filldraw[color=blue, densely dashed,fill=black!30, thick] circle (0.35);
        \filldraw[color=blue, densely dashed, fill=white, thick]  circle (0.17);
        \node[color=black] at (0,0.51) {\scriptsize ${#1}$};
        \node[color=black] at (0,0) {\scriptsize ${#2}$};

    }
  }
}
\tikzset{
  pics/cylnn1/.style n args={2}{
    code = { %
        \filldraw[color=black,fill=black!30, thick] circle (0.35);
        \filldraw[color=black, fill=white, thick] (0.075,0) circle (0.15);
        \node[color=black] at (0,0.51) {\scriptsize ${#1}$};
        \node[color=black] at (0,0) {\scriptsize ${#2}$};
        \node[cross=3pt, very thick] at (-0.225,0) {};
    }
  }
}
\tikzset{
  pics/cylnn1rr/.style n args={2}{
    code = { %
        \filldraw[color=red,fill=black!30, thick] circle (0.35);
        \filldraw[color=red, fill=white, thick] (0.075,0) circle (0.15);
        \node[color=black] at (0,0.51) {\scriptsize ${#1}$};
        \node[color=black] at (0,0) {\scriptsize ${#2}$};
        \node[cross=3pt, very thick] at (-0.225,0) {};
    }
  }
}
\tikzset{
  pics/cylnn1bb/.style n args={2}{
    code = { %
        \filldraw[color=blue,fill=black!30, thick] circle (0.35);
        \filldraw[color=blue, fill=white, thick] (0.075,0) circle (0.15);
        \node[color=black] at (0,0.51) {\scriptsize ${#1}$};
        \node[color=black] at (0,0) {\scriptsize ${#2}$};
        \node[cross=3pt, very thick] at (-0.225,0) {};
    }
  }
}

\tikzset{
  pics/disc2holes/.style n args={2}{
    code = { %
        \filldraw[color=black,fill=black!30, thick] circle (0.35);
        \filldraw[color=black, fill=white, thick] (-0.15,0) circle (0.1);
        \filldraw[color=black, fill=white, thick] (0.15,0) circle (0.1);
        \node[color=black] at (-0.15,0.1) {\scriptsize ${#1}$};
        \node[color=black] at (0.15,0.1) {\scriptsize ${#2}$};
    }
  }
}

\tikzset{
  pics/torus1hole/.style n args={2}{
    code = { %
\draw [thick, fill=black!30] (-0.6,0) to [out=85,in=95] (0.4,0) to [out=-95,in=-85] (-0.6,0);
\filldraw [color=white, fill=white] (-0.15,0.01) to [out=40,in=130] (0.15,0.01) to [out=-160,in=-20] (-0.15,0.01);
\draw [thick] (-0.2,0.05) to [out=-40,in=-130] (0.2,0.05);
\draw [thick] (-0.15,0.01) to [out=40,in=130] (0.15,0.01);
\filldraw[color=black, fill=white, thick] (-0.375,0) circle (0.125);
\node[color=black] at (-0.15,0.1) {\scriptsize ${#1}$};
\node[color=black] at (0.15,0.1) {\scriptsize ${#2}$};
    }
  }
}

\begin{document}

\unitlength = .8mm

\begin{titlepage}

\begin{center}

\hfill \\
\hfill \\
\vskip 1cm

\title{D-instanton Effects in Type IIB String Theory}

\author{Nathan B. Agmon$^\spadesuit$, Bruno Balthazar$^\diamondsuit$, Minjae Cho$^{\clubsuit}$, Victor A. Rodriguez$^{\heartsuit}$, Xi Yin$^\spadesuit$}

\address{
$^\spadesuit$Jefferson Physical Laboratory, Harvard University, \\
Cambridge, MA 02138 USA
\\
$^\diamondsuit$Enrico Fermi Institute \& Kadanoff Center for Theoretical Physics,\\
University of Chicago, Chicago, IL 60637, USA
\\
$^\clubsuit$Princeton Center for Theoretical Science, Princeton University, \\ Princeton, NJ 08544,
USA
\\
$^\heartsuit$Joseph Henry Laboratories, Princeton University, \\ Princeton, NJ 08544, USA \\
}

\email{nagmon@g.harvard.edu, brunobalthazar@uchicago.edu, minjae@princeton.edu, vrodriguez@princeton.edu, xiyin@fas.harvard.edu}

\end{center}

\abstract{We study D-instanton contributions to supergraviton scattering amplitudes in the ten-dimensional type IIB superstring theory beyond the leading non-perturbative order. In particular, we determine the one-D-instanton contribution to maximal R-symmetry violating (MRV) amplitudes with arbitrary momenta at the first subleading order in string coupling, as well as the effects of a D-/anti-D-instanton pair at leading nontrivial order in the momentum expansion. These results confirm a number of predictions of S-duality, and unveil some previously unknown pieces of type IIB string amplitudes. 

Our computation is based on the Neveu-Schwarz-Ramond formalism with picture changing operators and vertical integration. The naive on-shell prescription for D-instanton mediated amplitudes, based on integration over the moduli space of worldsheet geometries as well as the moduli space of D-instanton boundary conditions, suffers from potential open string divergences and regularization ambiguities that are in principle resolved in the framework of open+closed superstring field theory. In this paper, the ``on-shell ambiguities" of one-D-instanton MRV amplitudes are resolved by arguments involving supersymmetry and soft limits, part of which is verified by a string field theoretic computation in a highly nontrivial manner.
}

\vfill

\end{titlepage}

\eject

\begingroup
\hypersetup{linkcolor=black}

\tableofcontents

\endgroup

\vskip 2cm

\section{Introduction} 

Critical string theories are defined at the level of the perturbative spacetime S-matrix by a two-dimensional worldsheet conformal field theory with BRST-exact stress-energy tensor and a prescription for integration over the moduli space of punctured Riemann surfaces \cite{Friedan:1985ge, Green:1987sp, Polchinski:1998rq}. The worldsheet formalism is also known to capture certain non-perturbative effects, known as D-instantons, by including worldsheets with boundaries subject to suitable Dirichlet-type boundary conditions \cite{Polchinski:1994fq}. While D-instanton effects are fundamental to the dynamics of string theory \cite{Becker:1995kb, Kachru:2003aw}, and have been determined in some cases by a combination of on-shell worldsheet methods and string dualities \cite{Green:1997tv, Green:1997di, Kiritsis:1997em, Pioline:1997pu, Alexandrov:2010ca, Balthazar:2019rnh, Balthazar:2019ypi, Balthazar:2022atu, Balthazar:2022apu}, a systematic framework for computing them has been lacking until recent work of Sen \cite{Sen:2019qqg, Sen:2020cef, Sen:2020eck, Sen:2021qdk, Sen:2021tpp, Sen:2021jbr} based on open+closed string field theory \cite{FarooghMoosavian:2019yke}.

The goal of this paper is to apply the framework of D-instanton perturbation theory to compute D-instanton contributions to the S-matrix of type IIB superstring theory in ten-dimensional Minkowskian spacetime, beyond the leading order results of \cite{Green:1997tv, Sen:2021tpp}. Before outlining our strategy and results, a few comments are in order. There are two known types of non-perturbative effects in string theory: D-instanton effects of order $e^{-1/g_s}$, and gravitational (including NS brane) instanton effects of order $e^{-1/g_s^2}$. The latter are analogous to instantons in quantum field theory in the sense that they correspond to saddle points of the Euclidean functional integral based on an (effective) action of spacetime fields. The D-instantons, on the other hand, cannot be understood as saddle points of an action of closed string fields. Rather, they must be included as extra contributions to a scattering amplitude through worldsheets with boundaries, or ``holes", that are attached to the D-instanton. 

As with any non-perturbative corrections, one can meaningfully speak of the D-instanton effects only if there is a way to distinguish them from the perturbative results, or if there is a prescription for summing up the perturbative expansion. In the 2D $c=1$ string theory analyzed in \cite{Balthazar:2019rnh, Balthazar:2019ypi}, the perturbative series for closed string scattering amplitudes are Borel-summable (assuming the conjectured matrix model dual), and the D-instanton contributions were understood to be corrections on top of the Borel-resummed perturbative results. In the present paper, we will focus on D-instanton effects in type IIB string theory that are either distinguished from the perturbative contributions in that the former violate certain perturbative global symmetries, or correct a perturbative expansion that is known to terminate at a finite order.

\subsection{The general structure of D-instanton perturbation theory}

Our working hypothesis is that with a suitable prescription for summing up the perturbative series, there is a well-defined contribution from each D-instanton sector, i.e. a family of BRST-invariant boundary conditions of Dirichlet type on the worldsheet.

The boundary conformal field theory of the D-instanton gives rise to the space of open string fields ${\cal H}_o$ in the sense of Batalin-Vilkovisky (BV) formalism \cite{FarooghMoosavian:2019yke}. For type IIB string theory, states of ${\cal H}_o$ are subject to GSO invariance and the usual restriction on picture number, namely $(-1)$-picture in the Neveu-Schwarz (NS) sector and $(-{1\over 2})$- and $(-{3\over2})$-picture in the Ramond (R) sector.

Let ${\cal H}_c$ be the space of closed string fields, as defined in \cite{deLacroix:2017lif}. We will denote by $\Psi_c\in {\cal H}_c$ a closed string field and $\Psi_o\in {\cal H}_o$ an open string field on the D-instanton. The space of string fields is equipped with a Grassmann-odd symplectic structure defined through the BV anti-bracket. In performing the functional integral over open string fields, one should restrict to a Lagrangian subspace $L$ (i.e. a BV gauge condition).\footnote{For instance, one may split $\Psi_o$ into the ``regular" field $\Phi_o$ and BV anti-field $\Phi_o^\ddagger$ based on ghost number grading, and define the subspace $L$ by the constraint $\Phi_o^\ddagger = {\delta V\over \delta \Phi_o}$ for some choice of functional $V[\Phi_o]$. Further details on the BV gauge condition for D-instanton perturbation theory are discussed in Section~\ref{sec:bvgauge}.} The contribution from the D-instanton to the closed string field Euclidean 1PI effective action $\Gamma[\Psi_c]$ takes the schematic form
\ie\label{ocpathint}
\left. e^{-\Gamma[\Psi_c]} \right|_{\rm D-inst} = e^{-{C\over g_s}} \int \left.D\Phi_o\right|_L \exp\left( - S_{oc}[\Psi_o, \Psi_c]\right).
\fe
Here we have separated the D-instanton action $C/g_s$ from the rest of the open+closed string field action $S_{oc}[\Psi_o,\Psi_c]$, defined ``perturbatively" by integration over string vertices \cite{FarooghMoosavian:2019yke} that are 1PI with respect to the closed string fields and Wilsonian with respect to the open string fields.

To perform the functional integral of (\ref{ocpathint}), one separates the open string modes into two types: the ``massive" modes with non-degenerate kinetic terms, and the ``massless" modes with degenerate kinetic terms. The massive modes can be integrated out perturbatively, giving rise to Feynman diagrams with open string loops. Note that the worldsheet configuration corresponding to a Feynman diagram need not be connected, but rather should be ``connected modulo boundary" (e.g. in (\ref{eq:diagramexpansion})). 

The integration over massless open string modes, on the other hand, cannot be treated perturbatively. This includes the collective modes $\Phi_o^m$, whose integration is analogous to integrating over the D-instanton moduli space, and a mode that corresponds to the Faddeev-Popov ghost associated with fixing the $U(1)$ gauge symmetry on the D-instanton. A consistent treatment of the integration over massless open string fields, as explained in \cite{Sen:2020cef}, will be discussed in Section~\ref{sftsec} for the type IIB string amplitudes of interest.

The open+closed string field theory provides a framework for computing D-instanton effects on the closed string amplitude, at the level of perturbation theory around a D-instanton configuration. Such a perturbation theory may break down when open string modes on the D-instantons become tachyonic. The latter occurs for the ZZ-instantons in $c=1$ string theory \cite{Balthazar:2019rnh, Balthazar:2019ypi, Sen:2020eck}, where a Wick rotation prescription for the integration contour in the open string tachyon field has been proposed, as is anticipated from the general prescription for summing over saddle point contributions based on steepest descent contours. In type IIB superstring theory, a pair consisting of a D-instanton and an anti-D-instanton that are sufficiently nearby one another gives rise to tachyonic open string modes. It is unclear whether the string field theory formalism provides an unambiguous result in this situation, as there may be intrinsic ambiguities in such D-instanton contributions that are tied to the resummation prescription for string perturbation theory.

\subsection{The (somewhat naive) on-shell prescription}

Working explicitly with all components of the open string field on the D-instanton can be exceedingly tedious. It is often possible to take a shortcut, in which one extends the rules of on-shell string perturbation theory, based on integrating correlators of BRST-closed string vertex operators over the moduli space of Riemann surfaces, to include worldsheets with boundaries ending on the D-instanton. As was pointed out in \cite{Polchinski:1994fq, Green:1997tv}, such an on-shell prescription is subject to ambiguities due to divergences where the worldsheet degenerates. In this paper, we view the on-shell prescription as nothing more than a computational shortcut for obtaining partial results that could be in principle recovered from string field theory. The remaining ``regularization ambiguity" may either be determined by consideration of symmetries that are not manifest in the on-shell prescription, such as spacetime supersymmetry (as will be the case for the results presented in this paper), or a genuine string field theoretic computation (part of which is performed in Section~\ref{sftsec}).

For type IIB superstring theory in ten-dimensional Minkowskian spacetime, the known D-instantons include both the ``BPS" D-instanton that carries unit charge with respect to the Ramond-Ramond axion (0-form potential), and the corresponding anti-D-instanton that carries the opposite RR charge. One expects any closed string amplitude to receive contributions from arbitrary configurations of $n$ D-instantons and $m$ anti-D-instantons,\footnote{Here we focus only on D$(-1)$ branes in Minkowski spacetime. Both the string field theory and on-shell approaches generalize to D-instantons appearing in other backgrounds, such as Euclidean D$p$-branes wrapping non-contractible cycles in the type II string theories (see e.g. \cite{Alexandrov:2021shf, Alexandrov:2021dyl}).} given schematically by
\ie
{\cal A}_{(n,m)} = \int_{\widetilde{\cal M}_{n,m}} d\widetilde{\mu} & ~ \exp\Big( \tikz[baseline={([yshift=-1.5ex]current bounding box.center)}]{\pic{diske={~}}} \,+\, \tikz[baseline={([yshift=-1.5ex]current bounding box.center)}]{\pic{cylnn0={}{}}} \,+\, \tikz[baseline={([yshift=-0.8ex]current bounding box.center)}]{\pic{torus1hole={}{}}} \,+\,
\tikz[baseline={([yshift=-0.8ex]current bounding box.center)}]{\pic{disc2holes={}{}}} \,+\, \cdots \Big) \\
&\times \left[ \, \tikz[baseline={([yshift=-1.5ex]current bounding box.center)}]{\pic{disk1={~}}} \cdots 
\tikz[baseline={([yshift=-1.5ex]current bounding box.center)}]{\pic{disk1={~}}} \,+\, \tikz[baseline={([yshift=-1.5ex]current bounding box.center)}]{\pic{disk2={~}}} \cdot \tikz[baseline={([yshift=-1.5ex]current bounding box.center)}]{\pic{disk1={~}}} \cdots \tikz[baseline={([yshift=-1.5ex]current bounding box.center)}]{\pic{disk1={~}}} \,+\, \tikz[baseline={([yshift=-1.5ex]current bounding box.center)}]{\pic{cylnn1={}{}}} \cdot \tikz[baseline={([yshift=-1.5ex]current bounding box.center)}]{\pic{disk1={~}}} \cdots \tikz[baseline={([yshift=-1.5ex]current bounding box.center)}]{\pic{disk1={~}}} \,+\, \cdots \right]
\label{eq:diagramexpansion}
\fe
in the on-shell prescription. Here, $\widetilde{\cal M}_{n,m}$ is the super-moduli space of the $(n,m)$ D-instanton boundary conditions, whose bosonic and fermionic collective coordinates are in correspondence with the massless open string BRST cohomology in the Neveu-Schwarz and Ramond sectors, respectively. 
The measure $d\widetilde\mu$, without including any of the empty disconnected diagrams, is the natural one determined by the Zamolodchikov metric of boundary deformations (flat in the present example), up to a constant normalization to be specified later. The integrand meanwhile takes the form of a sum over the topologies of (generally disconnected) Riemann surfaces $\Sigma$, whose connected components all share the same D-instanton boundary condition, with a given number of vertex operator insertions that correspond to closed string asymptotic states. Each diagram comes with the string coupling dependence $g_s^{-\chi(\Sigma)}$, where $\chi$ is the Euler characteristic (including $-1$ from each puncture/vertex operator). Of note is the sum over empty discs, each of which evaluates to minus the D-instanton action $-(n+m){C\over g_s}$, that exponentiates to give the prefactor $e^{-(n+m){C\over g_s}}$ in \eqref{ocpathint}. The sum over remaining topologies is then interpreted as a perturbative series in the $(n,m)$ type D-instanton background.

A general deformation of the D-instanton along its super-moduli space is formally characterized by a boundary deformation of the worldsheet action, which takes the form
\ie
\Delta S_{\text{WS}} =  \int_{\partial\Sigma} dx \: U^{(0)}_{\text{NS},m}(x) + \int_{\partial\Sigma}dx\:\mathbf{P}_{1\over 2}U^{(-\frac12)}_{\text{R}}(x) .
\label{eq:naivewsdef}
\fe
Here $U^{(0)}_{\text{NS},m}$ is a pictured-raised unfixed vertex operator in the NS sector, constructed purely from the matter sector of the worldsheet CFT, that is BRST invariant modulo total derivatives. $U^{(-\frac12)}_{{\rm R}}$ meanwhile is an unfixed Ramond sector vertex operator in the $(-{1\over 2})$-picture, which necessarily involves the ghost fields. To ensure that the deformation has picture number 0, one introduces a formal ``${1\over 2}$-picture rasing" operator ${\bf P}_{1\over 2}$ \cite{Berenstein:1999jq,Berenstein:1999ip}, defined in such a way that a pair of ${\bf P}_{1\over 2}$ insertions amounts to that of a single picture-changing operator ${\cal X}$ (PCO). For instance, one may propose to replace \eqref{eq:naivewsdef} with the insertion of \cite{Cho:2018nfn}\footnote{Note that \eqref{eq:betterwsdef} amounts to expanding the exponential R deformation in \eqref{eq:naivewsdef}, and keeping only even powers of the R sector deformation, with half of them in picture $-\frac{1}{2}$ and the other half in picture $+\frac{1}{2}$ through ${\bf P}_{1\over 2}^2 = {\cal X}$. This new insertion is not a local deformation of the worldsheet CFT, but can be justified by consistent factorization of string amplitudes. }
\ie
e^{-\Delta S_{\text{WS}}} \equiv \exp \left(-\int_{\partial\Sigma}dx\: U^{(0)}_{\text{NS},m}(x) \right) \sum_{j=0}^\infty {1 \over (2j)!} \left[ \int_{\partial\Sigma}dx\: U^{(+\frac12)}_{\text{R}}(x) \right]^j \left[\int_{\partial\Sigma}dx\: U^{(-\frac12)}_{{\rm R}}(x) \right]^j ,
\label{eq:betterwsdef}
\fe 
into worldsheet correlators, where $U^{(+\frac12)}_{{\rm R}}$ is the picture-raised version of $U^{(-\frac12)}_{{\rm R}}$, in the $(+{1\over 2})$-picture. Note that such a deformation leads to a non-local boundary CFT.

The deformation operators appearing on the RHS of (\ref{eq:naivewsdef}) can be expanded in a basis of local boundary operators, with Grassmann even coefficients $x$ in the NS sector and Grassmann odd coefficients $\theta$ in the R sector. Together, $(x,\theta)$ comprise the collective coordinates on a patch of the super-moduli space of the D-instanton, whose origin corresponds to the undeformed boundary CFT. In the case of $(n,m)$ D-instanton in flat spacetime, the supermoduli space $\widetilde {\cal M}_{(n,m)}$ is parameterized by $10(n+m)$ bosonic and $16(n+m)$ fermionic collective coordinates, at least when the D-instantons and anti-D-instantons are sufficiently far separated so that all modes of the open strings stretched between them are ``massive," or off-shell.

Starting from the string field theory for the D-instantons, one may integrate out the massive open string modes perturbatively, leaving only integration over the massless open string modes. One might expect the physical massless open string degrees of freedom to be in correspondence with the moduli of the D-instanton, and that the quantum effective action for massless open string modes to be roughly equivalent to the diagrammatic expansion appearing on the RHS of (\ref{eq:diagramexpansion}). However, this expectation fails to hold in two important ways.

First, the diagrammatic expansion \eqref{eq:diagramexpansion} suffers from logarithmic divergences due to the propagation of massless open string modes\footnote{This should be contrasted with open strings on a D$p$-brane for $p\geq 0$ that generically carry nonzero momenta along the worldvolume of the brane.} along a thin strip that pinches near the boundary of the moduli space of the worldsheet geometry. One may regularize such divergences by cutting off near the boundary of the moduli space; indeed, the leading divergences cancel between different diagrams that contribute to on-shell closed string amplitudes mediated by the D-instanton \cite{Polchinski:1994fq}. Such a prescription leaves a finite regulator-dependent ambiguity whose resolution generally requires string field theory. Nonetheless, the on-shell prescription can still capture a meaningful part of the D-instanton amplitude, as will be explained later in this paper. 

Second, not all massless, or on-shell, modes of the open string field correspond to moduli of the D-instanton. This occurs for the multi-D-instanton, say of type $(n,0)$ where $n\geq 2$, where the open string fields carry $u(n)$ Chan-Paton factors, whereas the moduli space is ${\rm Sym}^n(\mathbb{R}^{10|16})$. In this case, the integrand appearing on the RHS (\ref{eq:diagramexpansion}) is singular along the loci where D-instantons collide, and the naive on-shell open string perturbation theory breaks down. This is already apparent in the low energy limit, where for suitable observables, the open string field theory reduces to the IKKT matrix model \cite{Ishibashi:1996xs, Green:1997tn, Sen:2021jbr}, which does not admit a perturbative expansion \cite{Yi:1997eg, Sethi:1997pa, Moore:1998et}.

\subsection{Supergraviton amplitudes in type IIB string theory}

A fundamental observable of type IIB string theory is the S-matrix in asymptotically ten-dimensional Minkowskian spacetime. At the non-perturbative level, the S-matrix elements are expected to be well-defined for asymptotic states spanned by the Fock space of supergravitons. The simplest nontrivial S-matrix element is that of $2\to 2$ supergraviton scattering, well known to be constrained by supersymmetry to be of the form \cite{Elvang:2013cua}
\ie\label{susyconstrfour}
{\cal A}_{2\to 2} = {\cal A}_{2\to 2}^{SUGRA} M(s,t;\tau,\bar\tau),
\fe
where ${\cal A}_{2\to 2}^{SUGRA}$ is the corresponding amplitude in tree-level (two-derivative) type IIB supergravity, and $M(s,t;\tau,\bar\tau)$ is a single function that approaches 1 in the low energy limit $s,t\to 0$. Here $s$ and $t$ are Mandelstam variables, and $\tau = \tau_1+i\tau_2$ ($\tau_2=1/g_s$) is the axion-dilaton expectation value parameterizing type IIB vacua, on which the low-energy accidental $SL(2,\mathbb{R})$ symmetry acts by M\"obius transformation.

Note that the full function $M(s,t;\tau,\bar\tau)$ is exceedingly complicated, as it encapsulates all possible resonances that are produced by scattering a pair of gravitons, including black hole states. It can be organized in two different expansions: in energy/momentum, or in string coupling. The momentum expansion takes the form
\ie
M(s,t;\tau,\bar\tau) &= \underline{s}\underline{t}\underline{u}\bigg[ {1\over \underline{s}\underline{t}\underline{u}}+ f_0(\tau,\bar\tau) + H_2(\underline{s},\underline{t}) + f_4(\tau,\bar\tau)(\underline{s}^2+\underline{t}^2+\underline{u}^2) + f_6(\tau,\bar\tau)(\underline{s}^3+\underline{t}^3+\underline{u}^3)
\\
&~~~ + f_8(\tau,\bar\tau) (\underline{s}^4+\underline{t}^4+\underline{u}^4) + f_0(\tau,\bar\tau)H_8(\underline{s},\underline{t}) + \cdots  \bigg]
\label{eq:Mstuexp}
\fe
where the underlined notation $\underline{s}, \underline{t},\underline{u}$ stands for the Mandelstam variables in units of the ten-dimensional Planck mass. The momentum-independent coefficients $f_0, f_4, f_6, f_8,\cdots$ are commonly referred to as the coefficients of $R^4, D^4R^4, D^6R^4, D^8R^4, \cdots$ terms in the quantum effective action of type IIB string theory. The function $H_2(\underline{s},\underline{t})$, which is independent of the moduli $\tau$, scales like two powers of momenta with logarithmic branch cuts extending to zero momentum, and is determined by the supergravity 1-loop amplitude. The function $H_8(\underline{s},\underline{t})$ similarly scales like eight powers of momenta with non-analyticity at zero momentum, and whose discontinuity factorizes into a supergravity tree amplitude and an $R^4$ vertex.

The coefficients $f_0, f_4, f_6$ are known to be constrained by supersymmetry \cite{Green:1998by, Wang:2015jna} to satisfied second-order differential equations on the moduli space of IIB vacua. For instance, $f_0(\tau,\bar\tau)$ obeys the equation
\ie
\left( \tau_2^2 \partial_\tau \partial_{\bar\tau} - {3\over 16} \right) f_0(\tau,\bar\tau) = 0.
\fe
Such equations dictate that the perturbative string contributions to $f_0, f_4, f_6$ truncates at a finite loop order, and the combination of perturbative results together with the assumption of S-duality invariance fixes these functions completely. $f_8$, on the other hand, is not known to be constrained by supersymmetry, and is not even known in perturbation theory starting at 3-loop order.

The string coupling expansion of $M(s,t;\tau,\bar\tau)$, on the other hand, is expected to take the form
\ie\label{mdinstexp}
\tau_2^{-2} M(s,t;\tau,\bar\tau) &= \sum_{h=0}^\infty \tau_2^{-2-2h} M_h(\A's,\A't) + \sum_{n,m} e^{2\pi i (n\tau-m\bar\tau)} M^{(n,m)}(\A' s,\A' t; \tau_2) + \cdots.
\fe
Here $M_h$ stands for the genus $h$ perturbative string amplitude, $M^{(n,m)}$ stands for the contribution from $n$ D-instantons and $m$ anti-D-instantons, and $\cdots$ stands for possible gravitational instanton effects. Note that the relation between the string tension and Planck mass is such that $\A' s =\tau_2^{1\over 2} \underline{s}$. As already alluded to, the instanton corrections are unambiguously defined only if there is a prescription for summing up the perturbative series, or if the perturbative contributions of certain momentum and/or coupling dependence are absent.

In the naive on-shell prescription, each D-instanton sector contribution $M^{(n,m)}$ is given by a sum over worldsheet diagrams with D-instanton boundary conditions, integrated over the moduli space of the D- and anti-D-instantons, with the structure
\ie\label{mnmnaive}
M^{(n,m)}(\A' s,\A' t; \tau_2) = \tau_2^{-{7\over 2}(n+m)} \sum_{L=0}^\infty \tau_2^{-L} M_L^{(n,m)}(\A's, \A' t),
\fe
where the ``open string loop order" $L$ is minus the Euler characteristic of the worldsheet diagram (with closed string insertions as punctures). The overall factor $\tau_2^{-{7\over 2}(n+m)}$ comes from the normalization of the measure on the D-instanton moduli space. (\ref{mnmnaive}) is expected to hold when the D-instanton moduli integration is non-singular, as will be the case for the contributions explicitly computed in this paper, including $M^{(1,0)}$ and certain terms in $M^{(1,1)}$. On the other hand, it is known to fail when there are singularities in the D-instanton moduli space, which occurs in $M^{(n,0)}$ for $n\geq 2$. Such singularities are due to the appearance of new ``massless" open string modes, and may be resolved in the open+closed string field theory approach \cite{Sen:2021jbr}.

\subsection{Summary of results}

The leading one-D-instanton contribution to the four-graviton amplitude, namely $M_0^{(1,0)}$ in the notation of (\ref{mdinstexp}), (\ref{mnmnaive}) (and similarly $M_0^{(0,1)}$ for the anti-D-instanton), was studied in \cite{Green:1997tv} and shown to be $stu$ times a constant. This constant is determined to be 
\ie\label{leadmst}
{1\over \alpha'^3 stu} M_0^{(1,0)}(\alpha's,\alpha't)={\pi\over 16}
\fe
by S-duality and consideration of supersymmetry \cite{Green:1998by, Wang:2015jna}, and was reproduced from a first principles string field theoretic computation recently in \cite{Sen:2021tpp}.

In Section~\ref{sec:1D4pt}, we present the first main result of this paper: the next-to-leading order single D-instanton contribution,
\ie\label{nextleadmst}
{1\over \alpha'^3 stu} M_1^{(1,0)}(\alpha's,\alpha't) &= C_1 + \sum_{p=2}^\infty {\zeta(p)\over 2^{2p+3}} \alpha'^p(s^p+t^p+u^p) ,~~~~~{\rm where}~~ C_1 ={3\over 256}.
\fe
The momentum dependent terms on the RHS of (\ref{nextleadmst}) come from the worldsheet diagram consisting of three discs with boundary on the D-instanton, where one of the discs contains two closed string vertex operators, while the other two each contain one closed string insertion. The constant term $C_1$ appearing on the RHS of (\ref{nextleadmst}) cannot be computed directly in the on-shell approach due to ambiguities in the regularization scheme. In fact, we are not aware of any simple regularization scheme based on cutting off the worldsheet moduli integral that produces the correct answer. On the other hand, $C_1$ has been argued in \cite{Wang:2015jna} to be fixed by consideration of supersymmetry Ward identities for 6-point amplitudes and a soft relation between the 6- and 4-point amplitudes, giving the result of (\ref{nextleadmst}). 

The open+closed string field theory approach developed in \cite{Sen:2019qqg, Sen:2020cef, Sen:2020eck, Sen:2021qdk, Sen:2021tpp, Sen:2021jbr} is free of the afore mentioned regularization ambiguity. In Section~\ref{sftsec}, we will analyze various pieces of the string field theoretic computation of (\ref{nextleadmst}) and argue that the SFT result can at most differ from the naive on-shell computation by the constant term $C_1$. Whether SFT in fact produces the correct value of $C_1$ amounts to the dynamical question of whether the super-Poincar\'e symmetry is preserved by the Minkowskian vacua of type IIB string theory at the non-perturbative level. While the latter is certainly expected, it is not manifest in the SFT formulation of D-instanton perturbation theory, where the closed string field vacuum is determined by extremizing the quantum effective action $\Gamma[\Psi_c]$ with all open string fields integrated out. 

It is illuminating to consider a generalization of $C_1$, namely the coefficient of the $N$-point ``maximal R-symmetry violating" (MRV) coupling of the schematic form $(\delta\tau)^{N-4} R^4$ at order $e^{2\pi i \tau} \tau_2^{-1}$, which is fixed by consideration of supersymmetry and soft limits in Section~\ref{sec:cnnlo} to be (see also \cite{Green:2019rhz})
\ie\label{conerest}
C_1^{(N)}  \equiv {N(N-1)\over 2} A_0 + N A_1 + A_2 = {3\over 256} - {(N-4)(N-5)\over 64}.
\fe
From the SFT perspective, $A_0$ comes entirely from the worldsheet diagram that involves a disc with two closed string insertions, whereas $A_1$ includes contributions from an annulus with one closed string insertion, and the Jacobian factor due to the change of integration variables from open string collective modes to the D-instanton moduli. The constant $A_2$, having the most complicated origin, includes contributions from worldsheet diagrams of the topology of a 3-holed sphere or a 1-holed torus.

In Section~\ref{sftsec}, we will perform the explicit string field theoretic computation of $A_0$, with the result $A_0 = -{1\over 32}$ in perfect agreement with the RHS of (\ref{conerest}). The SFT analysis of $A_1$ and $A_2$ will be left for the future.

In Section~\ref{sec:higherptmrv}, we apply our results on certain connected worldsheet diagrams (with boundary) appearing in D-instanton perturbation theory to analyze the single D-instanton contribution to $N$-point MRV amplitudes. In particular, we obtain the leading-order contribution and the full momentum dependence of the next-to-leading order contribution.This includes the 6-point 14-derivative order MRV amplitude considered in \cite{Green:2019rhz}, pinning down a previously unknown coefficient.\footnote{This coefficient is $c_1$ in the notation of \cite{Green:2019rhz}, section 6.}

In Section~\ref{sec:dantid}, we analyze the contribution from a D-instanton/anti-D-instanton pair, at leading order in the open string expansion, namely $M_0^{(1,1)}$ in (\ref{mdinstexp}). The main new idea here is the nontrivial measure on the instanton moduli space, as determined from the annulus diagram with two boundary components on the D- and anti-D-instanton respectively. We will see that the moduli space integrand is singular at finite distance from the origin, and that the on-shell computation of $M_0^{(1,1)}$ is ill-defined. This is perhaps unsurprising given that the precise definition of $M_0^{(1,1)}$ requires a (as of yet unknown) prescription for summing up the perturbative closed string amplitudes. Nonetheless, we will argue that the leading term in the momentum expansion of $M_0^{(1,1)}$ is unambiguously determined by the integration over the instanton moduli space at asymptotically large separation between the D- and anti-D-instanton, giving the result 
\ie
{1\over \A'^3 stu} M_0^{(1,1)}(\A' s,\A' t) = - 2^{-11}\A'^3(s^3+t^3+u^3) + {\cal O}(\A'^4).
\label{eq:M11}
\fe
This confirms, in a highly nontrivial manner,  a prediction of S-duality for the $D^6R^4$ effective coupling \cite{Green:2005ba}.

Furthermore, in Section~\ref{sec:DDbarunitarity}, we perform a check of non-perturbative unitarity for D-instanton scattering amplitudes. In particular, we demonstrate that certain terms in the low energy expansion of $M_0^{(1,1)}$ have non-analytic dependence on the momenta,  and are related to amplitudes mediated by a single D-instanton or anti-D-instanton, namely $M_0^{(1,0)}$ or $M_0^{(0,1)}$, through unitarity cuts.

\section{Effects of a D-instanton: an on-shell computation}
\label{sec:1D4pt}

Before we begin, let us outline our conventions for superamplitudes in type IIB string theory following the spinor helicity formalism of \cite{Boels:2012ie}. Given a massless external particle $i$, its momentum $p_i$ can be expressed in terms of auxiliary spinor helicity variables $\lambda^\alpha_{ia}$ according to
\ie{}
p_i^\mu (\gamma_\mu)^{\alpha\beta} = \lambda_{ia}^\alpha \lambda_i^{\beta a},
\fe
where $\alpha$ denotes an $SO(1,9)$ 16-dimensional chiral spinor index, and  $a$ denotes an $SO(8)$ little group index. Using these variables, we can also construct the supermomenta
\ie{}
q_{i+}^\alpha = \lambda_{ia}^\alpha \eta^a_i, \quad q_{i-}^\alpha = \lambda_i^{\alpha a}{\partial \over \partial \eta_i^a},
\fe
where $\eta^a_i$ is a Grassmann odd object satisfying $\{\eta^a_i,{\partial \over \partial \eta_i^b}\} = \delta^a_b$. From this, it follows that the supercharges
\ie{}
Q_+^\alpha = \sum_i q_{i+}^\alpha , \quad Q_-^\alpha = \sum_i q_{i-}^\alpha,
\fe
satisfy the ${\cal N}=(2,0)$ super-Poincar\'e algebra
\ie{}
\{Q_+^\alpha, Q_-^\beta\} = -(\gamma_\mu)^{\alpha\beta} P^\mu,
\label{eq:abstractalg}
\fe
where $P^\mu = \sum_i p_i^\mu$ is the total momentum. 

Using this construction, the $2^8=256$ one-particle states of the supergraviton multiplet can be embedded into a super-state
\ie{}
\Phi_i= \Phi_i^{(0)} + \eta^a_i \Phi^{(1)}_{ia} + {1 \over 2!} \eta^a_i \eta^b_i \Phi^{(2)}_{iab} + \cdots + {1 \over 8!} \eta_i^{a_1} \cdots \eta_i^{a_8} \Phi^{(8)}_{ia_1 \cdots a_8}.
\fe
We can assign each of the components definite weights $q_R$ under the $U(1)_R$ that acts as an outer-automorphism of the supersymmetry algebra in \eqref{eq:abstractalg}, which also appears as an accidental R-symmetry in the low energy limit. We shall take $\Phi_i$ to have weight $q_R = -1$ and $\eta$ to have weight $q_R = -{1 \over 4}$, which then fixes the weights of the rest of the components. Of particular interest are the axion-dilaton $\delta\tau$ and its complex-conjugate $\delta\bar\tau$, which appear at the ends of the multiplet via
\ie{}
\Phi^{(0)}_i = \ket{\delta\tau,p_i}, \quad \Phi^{(8)}_{ia_1 \cdots a_8} = \epsilon_{a_1 \cdots a_8} \ket{\delta\bar\tau,p_i},
\fe
where $\ket{\delta\tau,p_i}$ and $\ket{\delta\bar\tau,p_i}$ are the associated 1-particle states. From this it follows that $\delta\tau$ is assigned weight $q_R = -1$ and $\delta\bar\tau$ weight $q_R = +1$.

Given a set of external super-states, it is meaningful to talk about the superamplitude ${\cal A}(\Phi_i)$, which generically takes the form
\ie{}
{\cal A}(\Phi_i) = i(2\pi)^{10}\delta^{10}(P)Q_+^{16} {\cal F}(\lambda_i, \eta_i),
\fe
where $Q_\pm^{16}$ is defined as
\ie{}
Q_\pm^{16}  = {1 \over 16!} \epsilon_{\alpha_1 \cdots \alpha_{16}} Q_\pm^{\alpha_1} \cdots Q_\pm^{\alpha_{16}}.
\fe
We will often work with the reduced amplitude $\widetilde{{\cal A}}$, defined by
\ie{}
{\cal A}(\Phi_i) = i(2\pi)^{10}\delta^{10}(P) \widetilde{{\cal A}}(\Phi_i) .
\fe
The superamplitude is constrained by supersymmetry to obey various Ward identities, which can be conveniently packaged into the expression
\ie{}
\delta^{10}(P)Q_+^{16} Q^\alpha_- {\cal F}(\lambda_i, \eta_i) = 0.
\fe

In this section we shall focus on the case of $2\to 2$ supergraviton scattering, whose corresponding superamplitude takes the form
\ie{}
{\cal A}_{2\to 2}(\Phi_i) =  i(2\pi)^{10}\delta^{10}(P)Q_+^{16}F(s,t),
\label{eq:F2to2}
\fe
where $F(s,t)$ is a single function of the Mandelstam invariants
\ie{}
s = -(p_1+p_2)^2, \quad t = -(p_1+p_3)^2, \quad u = -(p_1+p_4)^2.
\fe
For instance, in our conventions the $2 \to 2$ type IIB supergravity amplitude is given by
\ie{}
F^{\text{SUGRA}}(s,t) = {\kappa^2 \over stu}.
\fe
To determine $F(s,t)$ in general, it suffices to restrict the asymptotic states to the scalar components. We shall therefore focus on the scattering process $\delta\bar\tau(p_1) + \delta\tau(p_2) \to \delta\bar\tau(p_3) + \delta\tau(p_4)$ corresponding to the $\eta_1^8 \eta_3^8$ component of the amplitude, which appears in
\ie{}
Q_+^{16} = \eta_1^8 \eta_2^8 s^4 + \eta_1^8 \eta_3^8 t^4 +  \eta_1^8 \eta_4^8 u^4 +   \cdots,
\label{eq:superchageexp}
\fe
where~$\cdots$ captures contributions from other states in the multiplet, and $\eta_i^8 \equiv \epsilon_{a_1 \cdots a_8}\eta_i^{a_1} \cdots \eta_i^{a_8}$.  As a convenient shorthand, we introduce the notation
\ie{}
{\cal A}_{\delta\bar\tau\delta\tau\delta\bar\tau\delta\tau} \equiv {\cal A}_{2\to 2}\big|_{\eta_1^8 \eta_3^8}
\fe
with the understanding that this corresponds to the specific assignment of momenta as discussed above.

\subsection{Diagrammatic expansion}
\label{sec:diagrammatics}

In the on-shell approach, we define the contribution of a single D-instanton to the 2$\to$2 axion-dilaton scattering amplitude as
\ie
{\cal A}^{(1,0)}_{\delta\bar\tau\delta\tau\delta\bar\tau\delta\tau}= {\cal N}_{D} e^{2\pi i \tau} \int d^{10}x d^{16}\theta \sum_{L=0}^\infty \tau_2^{-L} A^{(L)}_{\delta\bar\tau\delta\tau\delta\bar\tau\delta\tau}(x,\theta) ,
\label{eq:(1,0)amplitude}
\fe
where the superscript $(n,m) = (1,0)$ labels the number of D-instantons $n$ and number of anti-D-instantons $m$. The super-moduli space $\widetilde{{\cal M}}_{1,0} = \mathbb{R}^{10|16}$ is parametrized by ten bosonic collective coordinates $x^\mu$ and sixteen fermionic collective coordinates $\theta_\alpha$. $\tau$ is the vacuum expectation value of the axion-dilaton field, defined in such a way that it transforms in the M\"obius form with respect to the low energy $SL(2,\mathbb{R})$ symmetry. The overall normalization of the measure on super-moduli space, namely the factor ${\cal N}_D$, a priori depends on $\tau$ and is not fixed by consideration of unitarity alone. Note that it can be determined either in the on-shell formalism with the assumption of S-duality \cite{Green:1997tv}, or from first principles in the string field theory formalism \cite{Sen:2021qdk, Sen:2021tpp}.

The integrand on the RHS of (\ref{eq:(1,0)amplitude}) takes the form of a sum over worldsheet diagrams. The $e^{2\pi i \tau}$ factor comes from the exponentiated empty disc diagram.\footnote{The worldsheet computation is performed  for $\tau_1=0$, whose result generalizes straightforwardly to general $\tau$. }The sum over empty annuli exponentiates to some constant of order $\tau_2^0$, which has been absorbed into ${\cal N}_D$.
$A^{(L)}_{\delta\bar\tau\delta\tau\delta\bar\tau\delta\tau}(x,\theta)$ captures the contribution from the remaining components of the worldsheet diagrams with Euler characteristic $-L$ (taking into account $-1$ from each puncture), including empty connected components with negative Euler characteristic.

The leading term comes from the diagram consisting of four disconnected 1-punctured discs,
\ie
A^{(0)}_{\delta\bar\tau\delta\tau\delta\bar\tau\delta\tau}(x,\theta) = \left\langle \! \left\langle e^{-\Delta S_{\text{WS,R}}(\theta)} V_{\delta\bar\tau(p_1)} V_{\delta\tau(p_2)}  V_{\delta\bar\tau(p_3)} V_{\delta\tau(p_4)} \right\rangle \! \right\rangle^{D^2_1 \sqcup D^2_1 \sqcup D^2_1 \sqcup D^2_1}_{x} .
\label{eq:leadingamp}
\fe
Here $V_{\delta\tau/\delta\bar\tau(p)} \equiv {1\over \sqrt{2}} (V_{\delta\tau_1(p)} \pm iV_{\delta\tau_2(p)})$, where $V_{\delta \tau_1(p)}$ and $V_{\delta\tau_2(p)}$ are the fixed vertex operators for the RR axion and NSNS dilaton respectively (see Appendix~\ref{app:sugramultiplet} for a review of the supergraviton vertex operators).  The worldsheet diagrams have a D-instanton boundary condition in which the bosonic worldsheet matter fields satisfy $X^\mu|_{\partial\Sigma} = x^\mu$. The dependence on the fermionic collective coordinates of the D-instanton, on the other hand, is introduced through the insertion of the R sector boundary deformation 
\ie 
e^{-\Delta S_{\text{WS,R}}(\theta)} &\equiv \sum_{j=0}^\infty {1 \over (2j)!} \left[ \int_{\partial\Sigma}dx\: \theta_\alpha (\gamma^\mu)^{\alpha\beta} e^{+\frac12\phi} S_\beta i \partial X_\mu \right]^j \left[\int_{\partial\Sigma}dx\: \theta_\alpha e^{-\frac12\phi} S^\alpha(x) \right]^j .
\label{(1,0)def}
\fe
The double bracket in \eqref{eq:leadingamp} should be understood as an ordered amplitude of vertex operators distributed over the disconnected surfaces, with insertions of $b$,$c$ ghosts associated to worldsheet moduli and conformal Killing vectors kept implicit. That is, $V_{\delta\bar\tau(p_1)}$ is to be inserted on the first disc, $V_{\delta\tau(p_2)}$ on the second, etc., such that\footnote{Each of the vertex operators carries a factor of $g_c \propto \tau_2^{-1}$, which cancels against the factor of $\tau_2$ provided by each disc, ensuring that the $A_0(x,\theta)$ has no $\tau_2$ dependence.}
\ie
A^{(0)}_{\delta\bar\tau\delta\tau\delta\bar\tau\delta\tau}(x,\theta) &= \langle e^{-\Delta S_{\text{WS,R}}(\theta)} c_0V_{\delta\bar\tau(p_1)}\rangle^{D^2}_x \langle c_0 V_{\delta\tau(p_2)}\rangle^{D^2}_x  \langle c_0V_{\delta\bar\tau(p_3)}\rangle^{D^2}_x \langle c_0V_{\delta\tau(p_4)}\rangle^{D^2}_x \\
&~~+  \langle c_0V_{\delta\bar\tau(p_1)}\rangle^{D^2}_x \langle  e^{-\Delta S_{\text{WS,R}}(\theta)}c_0 V_{\delta\tau(p_2)}\rangle^{D^2}_x  \langle c_0V_{\delta\bar\tau(p_3)}\rangle^{D^2}_x \langle c_0V_{\delta\tau(p_4)}\rangle^{D^2}_x \\
&~~+ \langle c_0V_{\delta\bar\tau(p_1)}\rangle^{D^2}_x \langle  c_0 V_{\delta\tau(p_2)}\rangle^{D^2}_x  \langle e^{-\Delta S_{\text{WS,R}}(\theta)}c_0V_{\delta\bar\tau(p_3)}\rangle^{D^2}_x \langle c_0V_{\delta\tau(p_4)}\rangle^{D^2}_x \\
&~~+ \langle c_0V_{\delta\bar\tau(p_1)}\rangle^{D^2}_x \langle  c_0 V_{\delta\tau(p_2)}\rangle^{D^2}_x  \langle c_0V_{\delta\bar\tau(p_3)}\rangle^{D^2}_x \langle e^{-\Delta S_{\text{WS,R}}(\theta)} c_0V_{\delta\tau(p_4)}\rangle^{D^2}_x,
\label{eq:leadingD4pt}
\fe
where the contour associated to the R sector deformation decomposes into a sum over contours taken along the boundaries of each of the four discs. Here, $\langle \cdot \rangle_{x}^{D^2}$ is the standard worldsheet correlator on the disc with the associated boundary conditions. Each vertex operator is accompanied by an additional $c_0$ insertion that fixes the residual $U(1)\subset PSL(2,\mathbb{R})$ conformal symmetry of the 1-punctured disc $D_1^2$, which ensures that the 1-point amplitude is both BRST-invariant and independent of the operation position. 

At next-to-leading order in $\tau_2^{-1}$, the D-instanton contributes through the 2-punctured disc $D_2^2$ as well as the 1-punctured annulus $A_1^2$
\ie
\tau_2^{-1}A^{(1)}_{\delta\bar\tau\delta\tau\delta\bar\tau\delta\tau}(x,\theta) &=  \left\langle \! \left\langle e^{-\Delta S_{\text{WS,R}}(\theta)} V_{\delta\bar\tau(p_1)} V_{\delta\tau(p_2)}  V_{\delta\bar\tau(p_3)} V_{\delta\tau(p_4)} \right\rangle \! \right\rangle^{D^2_2 \sqcup D_1^2 \sqcup D_1^2}_x \\ &~~~+  \left\langle \! \left\langle e^{-\Delta S_{\text{WS,R}}(\theta)} V_{\delta\bar\tau(p_1)} V_{\delta\tau(p_2)}  V_{\delta\bar\tau(p_3)} V_{\delta\tau(p_4)} \right\rangle \! \right\rangle^{A^2_1 \sqcup D_1^2 \sqcup D_1^2}_x \\&~~~+ \text{permutations},
\label{eq:NLOdiagrams}
\fe
where the above expression includes distinct permutations of the four closed string vertex operators. The disc 2-point amplitude takes the schematic form
\ie
\int dy \left\langle {\cal B}_y V_{\delta\tau/\delta\bar\tau(p_i)}(z_1,\bar{z}_1) V_{\delta\tau/\delta\bar\tau(p_j)}(z_2,\bar{z}_2)\right\rangle^{D^2}_{x},
\fe
where $y$ is the single modulus of the 2-punctured disc, and ${\cal B}_y$ is a contour integral associated $b$-ghost insertion. Similarly, the annulus 1-point amplitude takes the schematic form
\ie
\int dt \int du \left\langle {\cal B}_t {\cal B}_u V_{\delta\tau/\delta\bar\tau(p_i)}(z,\bar{z}) \right\rangle^{A^2}_{x},
\fe
where $t$ is the annulus modulus and $u$ is the modulus associated with the position of the puncture relative to the boundary. In principle, there are additional contributions from the product of four discs, each with a closed string insertion, and any empty diagram that goes like $\tau_2^{-1}$, namely the 3-holed sphere and the 1-holed torus. Contributions from such empty diagrams can be interpreted as order $\tau_2^{-1}$ corrections to the D-instanton action $-2 \pi i \tau$. However, spacetime supersymmetry implies that the action is not renormalized , and so we expect all higher-order empty diagrams to vanish.

\subsection{Integration over the D-instanton moduli}
\label{sec:fermmod}

For the case of the single D-instanton, it is possible to handle integration over the moduli before computing any worldsheet diagrams. Let us first turn tackle the effects of the bosonic moduli $x^\mu$. Each of the closed string vertex operators depends on the zero mode of $X^\mu$ as $e^{ip_i \cdot x}$, and so $A^{(L)}_{\delta\bar\tau\delta\tau\delta\bar\tau\delta\tau}(x,\theta)$ necessarily takes the form $ e^{iP \cdot x}f(\theta)$, where $f(\theta)$ is independent of $x^\mu$. Integrating over $x^\mu$ thus gives $f(\theta)$ multiplied by a momentum-conserving delta function $i(2\pi)^{10}\delta^{10}(P)$, where the factor of $i$ arises from Wick rotation to Lorentzian signature. In this way, integration over the bosonic moduli restores the target space translation symmetry.

Next let us turn our attention to the fermionic moduli. Performing the Berezin integral over $\theta_\alpha$ gives
\ie{}
\int d^{16}\theta \: e^{-\Delta S_{\text{WS,R}}(\theta)} = \pi^{16}\widehat{Q}_-^{16},
\label{eq:Rinsertion}
\fe
where the operator $\widehat{Q}_-^{16}$ is given by a formal product over the broken supercharges
\ie{}
\widehat{Q}_\pm^{16} \equiv {1 \over 16!} \epsilon_{\alpha_1 \cdots \alpha_{16}} \widehat{Q}^{\alpha_1}_{(+\frac12),\pm}\widehat{Q}^{\alpha_2}_{(-\frac12),\pm} \cdots \widehat{Q}^{\alpha_{15}}_{(+\frac12),\pm}\widehat{Q}^{\alpha_{16}}_{(-\frac12),\pm},
\label{eq:deltaQ}
\fe
with $\widehat{Q}^\alpha_{(\pm\frac12),\pm}$ taking the form of an integrated picture-$(\pm\frac12)$ supercurrent along the boundary $\partial \Sigma$, as defined in Appendix~\ref{app:wstheory}. (For brevity we shall drop picture subscript whenever both choices apply.) The RHS of  \eqref{eq:deltaQ} is ill-defined due to divergences arising from operator collisions on the boundary. It can be rendered well-defined by separately deforming each of the contours into the bulk, while keeping them away from one another as well as any closed string vertex operators.\footnote{To be precise, in deforming the contour of $\widehat{Q}^\alpha_{-}$ from the boundary into the bulk, the operator becomes a generic linear combination of the form $\widehat{Q}^\alpha_{-} + a \widehat{Q}^\alpha_{+}$ with $a \in \mathbb{C}$. This is related to the fact that the $\widehat{Q}^\alpha_{+}$ uniquely correspond to the supercharges preserved (annihilated) by the D-instanton boundary condition, whereas the ``broken supercharges'' are a priori ambiguous. We shall find it convenient to set $a = 0$, referring to $\widehat{Q}^\alpha_{-}$ as the broken supercharges.} The expression in \eqref{eq:deltaQ} should thus be interpreted as an ordered product of contours in the bulk, with the contour of a given operator surrounding those of its neighbors on the right, and in turn being surrounded by those of its neighbors on the left. Note that while $\widehat{Q}_\pm^{16}$ requires a choice of ordering to be well-defined, the individual supercharges anti-commute, and so the operator is ultimately free of possible ordering ambiguities. 

\subsubsection*{Broken supercharges and the axion-dilaton}

Similar to the bosonic moduli, integration over the fermionic moduli is generally expected to restore spacetime supersymmetry. This can be observed in practice by shrinking the contours of $\widehat{Q}_-^{16}$ on the vertex operators and determining the supersymmetry transformation of the 1-particle states. Its action on the axion-dilaton vertex operators can be determined from that of a single supercharge, which takes the form
\ie{}
&\widehat{Q}^\alpha_{(-\frac12),\mp} V_{\delta \tau_1(p)} \pm i \widehat{Q}^\alpha_{(+\frac12),\mp} V_{\delta \tau_2(p)} = Q_B\Lambda
\fe
for some vertex operator $\Lambda$. In other words, the preserved supercharges $\widehat{Q}^\alpha_+$ annihilate $V_{\delta\bar\tau(p)}$ and the broken supercharges $\widehat{Q}^\alpha_-$ annihilate $V_{\delta\tau(p)}$, up to the addition of BRST-exact terms. (Such terms do not contribute to the unambiguous part of this amplitude, and so we shall ignore them in the following discussion.) Consequently, the non-vanishing configurations are given by eight of the broken supercharges acting on $V_{\delta\bar\tau(p_1)}$ and the other eight on $V_{\delta\bar\tau(p_3)}$, which converts each of them to $V_{\delta\tau(p_i)}$ for $p_i=1,3$ together with an overall momentum-dependent coefficient proportional to $(p_1 \cdot p_3)^4$. 

In order to determine its value, it is simplest to work in the center-of-mass (COM) frame where the momenta of the closed strings are
\ie
p_1^\mu &= E(1, 0, \ldots, 0, 1)^\mu, \\
p_2^\mu &= E(1, 0, \ldots, 0, -1)^\mu, \\
p_3^\mu &= E(-1, 0, \ldots,\sin \theta,  \cos \theta)^\mu, \\
p_4^\mu &= E(-1, 0, \ldots, -\sin \theta, -\cos \theta)^\mu,
\label{eq:momframe}
\fe
where $2E$ is the COM energy and $\theta$ is the scattering angle. In this frame, the action of the super-Poincar\'e algebra for $\delta\bar\tau(p_1)$ reduces to
\ie
\{ \widehat{Q}^\alpha_{(\pm\frac12),-}, \widehat{Q}^\beta_{(\mp\frac12),-} \}V_{\delta\bar\tau(p_1)} = 2 E\begin{pmatrix}0_{8\times8} & 0_{8\times8} \\ 0_{8\times8} & 1_{8\times8} \end{pmatrix}^{\alpha\beta}V_{\delta\bar\tau(p_1)}.
\label{eq:fermionicRelations1}
\fe
From this, we observe that $\widehat{Q}^{9,\cdots,16}_{-}$ act as lowering operators on the supergraviton multiplet associated to $\delta\bar\tau(p_1)$, whereas $\widehat{Q}^{1,\ldots,8}_{-}$ annihilate the entire multiplet. This implies that there is a unique nontrivial configuration of supercharges in the COM frame, with $\widehat{Q}^{9,\cdots,16}_{-}$ acting on $V_{\delta\bar\tau(p_1)}$ and $\widehat{Q}^{1,\ldots,8}_{-}$ on $V_{\delta\bar\tau(p_3)}$, both of which are proportional to $V_{\delta\tau(p_1)}$. Using \eqref{eq:fermionicRelations1}, we find 
\ie{}
\widehat{Q}^{9}_{(+\frac12),-}\widehat{Q}^{10}_{(-\frac12),-} \cdots \widehat{Q}^{15}_{(+\frac12),-}\widehat{Q}^{16}_{(-\frac12),-}V_{\delta\bar\tau(p_1)} = 16 E^4 \: V_{\delta\tau(p_1)}.
\fe

In order to determine the action of the broken supercharges on $V_{\delta\bar\tau(p_3)}$, we perform a $\theta$ clockwise rotation in the $89$ plane combined with a time reversal such that $p_3^\mu \to p_3^{\mu'} = p_1^\mu$. As chiral spinors, the supercharges transform under this rotation as
\ie{} 
\begin{pmatrix} \widehat{Q}^\alpha_{(\pm\frac12),-} \\ \widehat{Q}^{\alpha+8}_{(\pm\frac12),-}\end{pmatrix} \to \begin{pmatrix} \widehat{Q}^{\alpha'}_{(\pm\frac12),-} \\ \widehat{Q}^{\alpha'+8}_{(\pm\frac12),-}\end{pmatrix} = \begin{pmatrix} \cos (\tfrac{\theta}{2}) & -\sin (\tfrac{\theta}{2}) \\ \noalign{\vspace{1ex}} \sin (\tfrac{\theta}{2}) & \cos (\tfrac{\theta}{2}) \end{pmatrix} \begin{pmatrix} \widehat{Q}^\alpha_{(\pm\frac12),-} \\ \widehat{Q}^{\alpha+8}_{(\pm\frac12),-}\end{pmatrix}  , \quad \alpha=1,\ldots 8.
\label{eq:rotatedSUSY}
\fe
The supercharges in the rotated frame obey the same commutation relations as in \eqref{eq:fermionicRelations1} up to an irrelevant factor of $i$, with $\delta\bar\tau(p_1)$ replaced by $\delta\bar\tau(p_3)$. This implies that the rotated supercharges $\widehat{Q}^{\alpha'=9,\ldots,16}_{-}$ act as lowering operators on $\delta\bar\tau(p_3)$, while $\widehat{Q}^{\alpha'=1,\ldots,8}_{-}$ annihilate it. The original supercharges are given by linear combinations of the rotated ones, and so using (the inverse of) \eqref{eq:rotatedSUSY} we have
\ie{}
\widehat{Q}^{1}_{(+\frac12),-}\widehat{Q}^{2}_{(-\frac12),-} \cdots \widehat{Q}^{7}_{(+\frac12),-}\widehat{Q}^{8}_{(-\frac12),-}V_{\delta\bar\tau(p_3)} = 16E^4 \cos^8(\tfrac{\theta}{2}) \: V_{\delta\tau(p_3)}.
\fe

Using \eqref{eq:fermionicRelations1} and \eqref{eq:rotatedSUSY}, we find the desired result for the action of $\widehat{Q}^{16}_-$ on the vertex operators, namely
\ie{}
\widehat{Q}^{16}_- \big(V_{\delta\bar\tau(p_1)} V_{\delta\tau(p_2)}V_{\delta\bar\tau(p_3)}V_{\delta\tau(p_4)} \big) = t^4 V_{\delta\tau(p_1)} V_{\delta\tau(p_2)}V_{\delta\tau(p_3)}V_{\delta\tau(p_4)},
\label{eq:susy1D}
\fe
where we replaced $256 E^8\cos({\theta \over 2})^8$ with the manifestly Lorentz-invariant quantity $t^4 = (p_1+p_3)^8$. In arriving at this expression, we have exploited the fact that our argument does not rely on the ordering of the supercharges, which contributes an overall factor of $16!$ that cancels a similar factor in the numerator of \eqref{eq:deltaQ}.  

As promised, integration over the fermionic moduli has restored spacetime supersymmetry, at least up to the ambiguities present in the on-shell formalism. We can identify the $t^4$ factor in \eqref{eq:susy1D} as precisely the $\eta_1^8\eta_3^8$ component of the superamplitude contained in the supersymmetry factor $Q_+^{16}$, see e.g. \eqref{eq:superchageexp}.\footnote{We emphasize that although this result was derived for a specific component of the 4-point superamplitude, the appearance of $\widehat{Q}_-^{16}$ indicates that the same type of argument should hold for any choice of asymptotic states. That is, all of the vertex operators are converted to $V_{\delta\tau(p_i)}$, with the supersymmetry factor $t^4$ corresponding to the $\eta_1^8\eta_3^8$ component replaced by the analogous quantity in $Q_+^{16}$ \eqref{eq:superchageexp}.} 

\subsubsection*{Simplified diagrammatics}

Combining our results for integration over the bosonic moduli and fermionic moduli in \eqref{eq:susy1D}, we find that the D-instanton contribution to the amplitude in \eqref{eq:(1,0)amplitude} simplifies to
\ie
\widetilde{{\cal A}}^{(1,0)}_{\delta\bar\tau\delta\tau\delta\bar\tau\delta\tau} = \pi^{16} t^4 {\cal N}_{D} e^{2\pi i \tau} \sum_{L=0}^\infty \tau_2^{-L} A^{(L)}_{\delta\tau\delta\tau\delta\tau\delta\tau} ,
\label{eq:(1,0)amplitudeNew}
\fe
where $A^{(L)}_{\delta\tau\delta\tau\delta\tau\delta\tau} $ captures contributions from worldsheet diagrams with Euler characteristic $-L$ (with the empty disc and annulus excluded as before). As there are no moduli remaining, it is computed using all $\delta\tau$ vertex operators $V_{\delta\tau(p_i)}$ and with fixed boundary condition for $X^\mu$ such that $x^\mu = 0$. 

\subsection{Leading order contribution}

We now compute the leading order contribution from a single D-instanton to the $2 \to 2$ scattering amplitude. Following \eqref{eq:(1,0)amplitudeNew} with $L=0$, it consists of the product of four discs, each with a $\delta\tau$ insertion. 

\subsubsection{Disc 1-point diagram}
\label{sec:disconepoint}

We are interested in the disc 1-point amplitude of $\delta\tau$ with boundary lying on the D-instanton. The $PSL(2,\bR)$ gauge redundancy can be used to fix the closed string puncture to $z = i$. Without any additional closed/open string insertions, there is a residual $U(1)$ that rotates $D^2$ around its origin, i.e. leaves the closed string puncture at $z = i$ invariant. We account for this remaining gauge redundancy by dividing by the volume of the gauge group $\text{Vol}(U(1)) = 2\pi$, whose normalization is unambiguous with respect to the open-closed disc 2-point amplitude, i.e. with one bulk puncture and one boundary puncture. We must also insert $c_0$ to soak up the remaining zero mode of $c(z)$ in the $bc$ path integral. 

\paragraph{The dilaton:} First consider the 1-point disc diagram for a single dilaton. The picture anomaly on the disc requires total picture $-2$, and so we work with the NSNS vertex operator in picture $(-1,-1)$ as provided in \eqref{NSNS(-1,-1)}. The 1-point disc diagram is given by the correlator
\ie{}
A_{\delta\tau_2(p)}^{D^2} = {g_c \over 2\pi} e_{\mu\nu}(p) \left\langle c_0 c \widetilde{c} e^{-\phi}\psi^\mu e^{-\widetilde{\phi}}\widetilde{\psi}^\nu e^{ip \cdot X}(i,-i) \right\rangle^{D^2}_{x^\mu = 0}.
\fe
Using the doubling trick, we can replace all of the antiholomorphic operators in the upper half plane with their holomorphic counterparts at the reflected point in the lower half plane. In doing so, we must also include any phase factors present in the boundary conditions relating the holomorphic and antiholomorphic operators. For the NSNS vertex operator, the doubling trick gives an overall factor of $-1$ from the replacement $\widetilde{\psi}^\mu(-i) \to - \psi^\mu(-i)$. The 1-point disc diagram thus reads
\ie
A_{\delta\tau_2(p)}^{D^2} &= -\frac{g_c C_{D^2}}{2\pi}  e_{\mu\nu}(p) \left\langle  \partial c ce^{-\phi} \psi^\mu e^{ip \cdot X_L}(i)  ce^{-\phi} \psi^\nu e^{-ip \cdot X_L}(-i) \right\rangle^{S^2}_{\mathbf{chiral}},
\fe
where $C_{D^2}$ is a constant multiplying all disc correlators that cannot be fixed purely from CFT considerations. Here, $\langle \cdot \rangle_{\mathbf{chiral}}$ stands for correlator evaluated in the chiral sector of the CFT, i.e. only for holomorphic operators.  It is normalized such that $\langle c(z_1)c(z_2)c(z_3)e^{-2\phi}(z)\rangle^{S^2}_{\mathbf{chiral}} = z_{12}z_{13}z_{23}$. Note that the holomorphic operator $X_L^\mu$ is defined with the zero mode $x^\mu$ removed. The correlator is easily evaluated using Wick contractions, which gives
\ie  
A_{\delta\tau_2(p)}^{D^2} &=  {g_c C_{D^2} \over 2\pi}\eta^{\mu\nu}e_{\mu\nu}(p) .
\fe
Using the dilaton polarization tensor in \eqref{dilatonPolarization} we have $\eta^{\mu\nu}e_{\mu\nu}(p) = 2\sqrt{2}$, and so the 1-point diagram reduces to
\ie  
A_{\delta\tau_2(p)}^{D^2} &=  {\sqrt{2}g_c C_{D^2} \over \pi}.
\label{eq:dilaton1pt}
\fe

\paragraph{The axion:} Next we tackle the axion 1-point diagram. Strictly speaking, it is ill-defined since the RR vertex operator appears in the $(-{1 \over 2},-{1 \over 2})$ picture, and yet there are various ways to assign it a definitive value. For instance, in certain circumstances one can define an ``inverse'' PCO that carries picture $-1$. We shall take a more natural approach from the on-shell perspective and analyze the open-closed disc 2-point diagram with a single bosonic collective coordinate $\delta x^\mu$ with vertex operator $ce^{-\phi}\psi^\mu$ in the $(-1)$-picture. This diagram reduces to the disc 1-point diagram multiplied by an overall factor proportional to the momentum $p^\mu$ of the closed string insertion. 

In order to determine this proportionality constant, we first consider the open-closed disc amplitude with a dilaton, taking the NSNS vertex operator to be in the $(0,-1)$ picture \eqref{NSNS(0,-1)}. The residual $PSL(2,\bR)$ gauge redundancy can be used to fix the closed string vertex operator to $z = i$ and the open string vertex operator to $z = 0$. The doubling trick in this case gives the same factor of $-1$ from the antiholomorphic fermion. It follows that the diagram reads
\ie{}
A_{\delta x^\mu\delta\tau_2(p)}^{D^2} &= \sqrt{2} g_c C_{D^2} e_{\sigma\rho}(p) \\
&~~~\times \left\langle ce^{-\phi}\psi^\mu(0) c\left(i \partial X_L^\sigma + \frac12 p \cdot \psi \psi^\sigma \right)e^{ip \cdot X_L}(i)ce^{-\phi}\psi^\rho e^{-ip \cdot X_L}(-i) \right\rangle^{S^2}_{\mathbf{chiral}}.
\fe
To evaluate the correlator, we make use of the fact that the transversality constraint $p^\mu e_{\mu\nu}(p) = 0$ and mass-shell constraint $p^2 = 0$ imply that the $X^\mu$ CFT does not contribute. What remains is a correlator of four fermions together with the $c$ ghosts, which evaluates to
\ie
A_{\delta x^\mu\delta\tau_2(p)}^{D^2}   = 2g_c C_{D^2} p^\mu.
\fe
Comparing this with dilaton disc 1-point amplitude \eqref{eq:dilaton1pt}, we immediately determine $\sqrt{2}\pi p^\mu$ as the coefficient of proportionality. 

With the coefficient in hand, let us return to the case of the axion. The setup of the open-closed disc amplitude is the same as for the dilaton, with the NSNS vertex operator replaced by the RR vertex operator in picture $(-\frac12,-\frac12)$, for which the amplitude reads
\ie{}
A_{\delta x^\mu\delta\tau_1(p)}^{D^2}  = g_c\left\langle ce^{-\phi}\psi^\mu(0) f_{\alpha\beta}(p) c \widetilde{c}e^{-\frac12\phi}S^\alpha e^{-\frac12\widetilde{\phi}}\widetilde{S}^\beta e^{ip \cdot X}(i,-i)\right\rangle_{x^\mu=0}^{D^2}.
\fe
The doubling trick gives a factor of $-i$ from the replacement $\widetilde{S}^\beta(i) \to -i S^\beta(-i)$, and so we have
\ie{}
A_{\delta x^\mu\delta\tau_1(p)}^{D^2} &=  i g_c C_{D^2} \left\langle ce^{-\phi}\psi^\mu(0) ce^{-\frac12\phi}S^\alpha e^{ip \cdot X_L}(i) ce^{-\frac12\phi}S^\beta e^{-ip \cdot X_L}(-i)\right\rangle^{S^2}_{\mathbf{chiral}} .
\fe
Using the $\psi^\mu S^\alpha S^\beta$ correlator in \eqref{eq:psiSS}, we find
\ie{}
A_{\delta x^\mu\delta\tau_1(p)}^{D^2} &= -{i g_c C_{D^2} \over \sqrt{2}} \tr(f(p)\gamma^\mu) \\
&= 2i g_c C_{D^2} p^\mu,
\fe
where we have used the axion polarization tensor, which leads to the factor $\tr(\slashed{p}\gamma^\mu) = 16p^\mu$. From this, we can divide by $\sqrt{2}\pi p^\mu$ to extract the desired disc 1-point diagram 
\ie 
A_{\delta\tau_1(p)}^{D^2} =i {\sqrt{2} g_c C_{D^2} \over \pi}.
\label{eq:axion1pt}
\fe

\paragraph{The axion-dilaton:} The disc 1-point diagram for the axion-dilaton follows from $V_{\delta\tau} = {1 \over \sqrt{2}}(V_{\delta\tau_1} + i V_{\delta\tau_2})$. The disc constant $C_{D^2}$, which is proportional to the tension of the D-instanton, can be fixed by unitarity of the perturbative string S-matrix. In our conventions, it reads \cite{Polchinski:1998rq,Polchinski:1998rr}
\ie 
C_{D^2} = 4\pi^3 \tau_2 .
\fe
Similarly, the closed string coupling $g_c$ is related to the dilaton expectation value as
\ie{}
g_c = {4\pi^{\frac52}\alpha'^2 \over \tau_2},
\fe
where we have restored the value of $\alpha'$. Using these together with the diagrams computed in \eqref{eq:dilaton1pt} and \eqref{eq:axion1pt}, it follows that the axion-dilaton disc 1-point diagram is
\ie{}
A_{\delta\tau(p)}^{D^2} &= 32\pi^{\frac92} \alpha'^2 i .
\label{eq:disc1pt}
\fe 
In order to work with quantities where the $SL(2,\bZ)$ symmetry is manifest, we substitute out $\alpha'$ using the relation\cite{Polchinski:1998rr}
\ie{}
\kappa^2 = \frac12(2 \pi)^7 \tau_2^{-2} \alpha'^{4},
\label{eq:newtonconst}
\fe
where $\kappa$ is Newton's constant. It follows that \eqref{eq:disc1pt} can be rewritten as
\ie{}
A_{\delta\tau(p)}^{D^2} &= 4\pi i \kappa \tau_2. 
\label{eq:disc1ptSLZ}
\fe 

\subsubsection{The 4-point amplitude}

The leading contribution of a single D-instanton to the 4-point amplitude follows from a straightforward application of \eqref{eq:(1,0)amplitudeNew} together with four copies of the disc 1-point diagram \eqref{eq:disc1pt}, which gives
\ie
\widetilde{{\cal A}}^{(1,0)}_{\delta\bar\tau\delta\tau\delta\bar\tau\delta\tau}\big|_{\text{LO}} &= \pi^{16} t^4 {\cal N}_D e^{2\pi i \tau} \left( A_{\delta\tau(p)}^{D^2} \right)^4 \\
  &= 256 \pi^{20} t^4 {\cal N}_D \kappa^4 \tau_2^4 e^{2\pi i \tau}.
\label{eq:LOamp}
\fe

\subsection{Next-to-leading order contribution}

The next-to-leading order contribution from a single D-instanton to the $2 \to 2$ scattering amplitude consists of the following worldsheet diagrams,
\ie
{}&\scalebox{2}{\tikz[baseline={([yshift=-1.3ex]current bounding box.center)}]{\pic{disk2={~}}}} \cdot \scalebox{2}{\tikz[baseline={([yshift=-1.3ex]current bounding box.center)}]{\pic{disk1={~}}}} \cdot \scalebox{2}{\tikz[baseline={([yshift=-1.3ex]current bounding box.center)}]{\pic{disk1={~}}}} \,+\ \scalebox{2}{\tikz[baseline={([yshift=-1.3ex]current bounding box.center)}]{\pic{cylnn1={}{}}}} \cdot \scalebox{2}{\tikz[baseline={([yshift=-1.3ex]current bounding box.center)}]{\pic{disk1={~}}}} \cdot \scalebox{2}{\tikz[baseline={([yshift=-1.3ex]current bounding box.center)}]{\pic{disk1={~}}}} \cdot \scalebox{2}{\tikz[baseline={([yshift=-1.3ex]current bounding box.center)}]{\pic{disk1={~}}}} \\
&\,+\ \left(\scalebox{2}{\tikz[baseline={([yshift=-0.4ex]current bounding box.center)}]{\pic{torus1hole={}{}}}} \,+\,
\scalebox{2}{\tikz[baseline={([yshift=-0.4ex]current bounding box.center)}]{\pic{disc2holes={}{}}}} \right) \cdot \scalebox{2}{\tikz[baseline={([yshift=-1.3ex]current bounding box.center)}]{\pic{disk1={~}}}} \cdot \scalebox{2}{\tikz[baseline={([yshift=-1.3ex]current bounding box.center)}]{\pic{disk1={~}}}} \cdot \scalebox{2}{\tikz[baseline={([yshift=-1.3ex]current bounding box.center)}]{\pic{disk1={~}}}} \cdot \scalebox{2}{\tikz[baseline={([yshift=-1.3ex]current bounding box.center)}]{\pic{disk1={~}}}}~~,
\label{eq:wsdiagramsNLO}
\fe 
which includes a sum over distinct permutations of the on-shell closed string vertex operator insertions $V_{\delta\tau_i}$ with $i=1,\ldots,4$. 
In the following subsections we compute the connected components of \eqref{eq:wsdiagramsNLO} individually, save for the 1-holed torus and 3-holed sphere, which are not expected to contribute to the amplitude. The final result is presented in equation \eqref{eq:NLOamp}.

\subsubsection{Disc 2-point diagram}

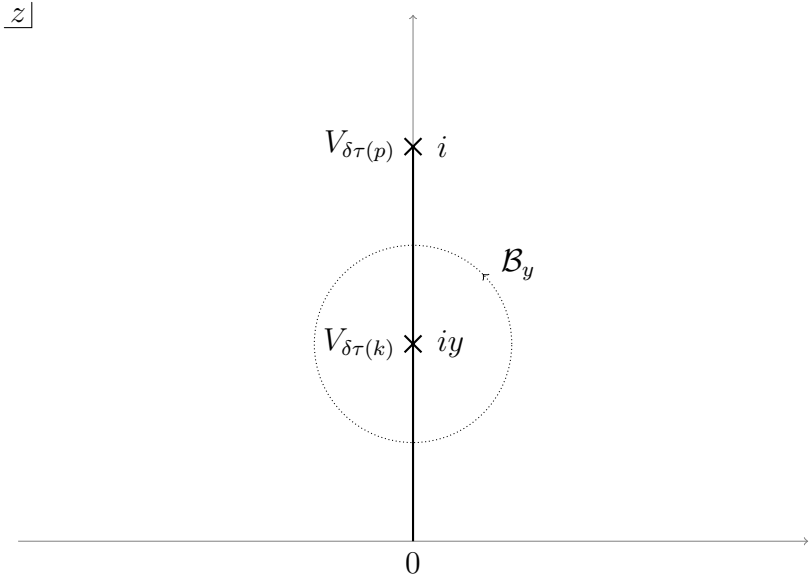
\begin{figure}[t!]
\centering
\begin{tikzpicture}
[scale=1.75, decoration={markings,mark=at position 0.125 with {\arrow{>}}}]
  \draw [help lines,->] (-3,0) -- (3,0) {};
  \draw [help lines,->] (0,0) -- (0,4) {};
  \path (0,3) node[cross=4pt, thick] {} ;
  \path (0,1.5) node[cross=4pt, thick] {} ;
  \draw [densely dotted, postaction={decorate}] (0,1.5) circle (0.75);
  \draw [thick] (0,0) -- (0,3) {} ;
  \draw (-0.05,3) node[left] {$V_{\delta\tau(p)}$} ;
  \draw (0.1,3) node[right] {$i$} ;
  \draw (-0.05,1.5) node[left] {$V_{\delta\tau(k)}$} ;
  \draw (0.1,1.5) node[right] {$iy$} ;
  \node [below] {$0$} ;
  \node at (0.8,2.1) {${\cal B}_y$} ;
  \node(n)[inner sep=2pt] at (-3,4) {$z$} ;
  \draw [line cap=round] (n.south west) -- (n.south east) -- (n.north east) ;
\end{tikzpicture}
\caption{The two-punctured disc represented as the upper half plane with global coordinate $z$. One vertex operator $V_{\delta\tau(p)}$ is fixed at position $z=i$ while the second vertex operator $V_{\delta\tau(k)}$ is fixed at position $z=iy$, integrating over the modulus $y\in [0,1]$. The $b$ ghost contour ${\cal B}_y$ surrounds the integrated vertex operator as drawn.}
\label{fig:disk2ptmodspace}
\end{figure}

Our first goal is compute the disc amplitude for two $\delta\tau$ insertions. The residual $PSL(2,\bR)$ gauge redundancy can be used to fix one closed string puncture to $z = i$ and the other to $z = iy$ with $0 \leq y \leq 1$, as shown in Figure \ref{fig:disk2ptmodspace}. For this choice, the disc diagram is given by
\ie{}
A_{\delta\tau(p)\delta\tau(k)}^{D^2}  = \int_0^1 dy \left\langle {\cal B}_y V_{\delta\tau(p)}(i,-i) V_{\delta\tau(k)}(iy,-iy) \right\rangle^{D^2}_{x^\mu=0}.
\fe
The $b$ ghost insertion ${\cal B}_y$ accompanying the modulus $y$ takes the form
\ie{}
{\cal B}_y = {1 \over 2\pi}\oint_{C_{iy}} \left(dz b(z) + d\bar{z} \widetilde{b}(\bar{z})\right),
\fe
where $C_{iy}$ is a small circular contour taken counterclockwise around the point $z = v$. Each vertex operator takes the form $V_{\delta\tau(p)} = c \widetilde{c} U_{\delta\tau(p)}$ plus terms with nonzero $\eta,\xi$ charge that do not contribute. It follows that the contribution of the $b$,$c$ ghost system is
\ie{}
\left\langle {\cal B}_y c \widetilde{c}(i,-i) c\widetilde{c}(iy,-iy)\right\rangle^{D^2}_0 = -4(1-y^2),
\label{eq:bcdisc2pt}
\fe
and so the amplitude reduces to 
\ie{}
A_{\delta\tau(p)\delta\tau(k)}^{D^2}  =-4\int_0^1 dy (1-y^2) \left\langle c_{-1}c_0 c_1 U_{\delta\tau}(i,-i) U_{\delta\tau}(iy,-iy) \right\rangle^{D^2}_{x^\mu=0},
\label{eq:bcout}
\fe
where the inclusion of $c_{-1}c_0 c_1$ ensures that the correlator evaluated in the full matter+ghost CFT is nonzero.

\paragraph{Two axions:} We first evaluate the amplitude for two axions with both RR vertex operators in the $(-1,-1)$-picture. In this case, the doubling trick gives an overall factor of $+1$ from the two antiholomorphic spin fields. From \eqref{eq:bcout} it follows that the amplitude is given by
\ie{}
A_{\delta\tau_1(p)\delta\tau_1(k)}^{D^2}  &= 4g_c^2 f_{\alpha\beta}(p)f_{\gamma\delta}(k) \int_0^1 dy (1-y^2)\\ &\times \bigg\langle c_{-1} c_0 c_{1} e^{-\frac12\phi}S^{\alpha}e^{ip \cdot X_L}(i) e^{-\frac12\phi}S^{\beta}e^{-ip \cdot X_L}(-i) \\ &~~~~~~e^{-\frac12\phi}S^{\gamma}e^{ik \cdot X_L}(iy)e^{-\frac12\phi}S^{\delta}e^{-ik \cdot X_L}(-iy) \bigg\rangle^{S^2}_{\mathbf{chiral}}.
\fe
Using the correlator for four holomorphic spin fields \eqref{eq:SSSS} together with Wick contractions for the free fields, we find
\ie{}
A_{\delta\tau_1(p)\delta\tau_1(k)}^{D^2} &= -\frac12 g_c^2 C_{D^2} \int_0^1 dy \left({1-y \over 1+y}\right)^{p \cdot k} \bigg[{\tr\{f(p)\gamma^\mu\}\:\tr\{f(k)\gamma_\mu\} \over y}  \\&~~~+ {2\:\tr\left\{f(p)\gamma^\mu [f(k)+f(k)^T]\gamma_\mu\right\} \over y(1-y^2)} + {2\:\tr\left\{f(p)\gamma^\mu [f(k)-f(k)^T]\gamma_\mu\right\}\over (1-y^2)} \bigg].
\fe
Specializing to the axion polarization tensor \eqref{eq:axionpolarization} then gives
\ie{}
A_{\delta\tau_1(p)\delta\tau_1(k)}^{D^2} 
&=-{1 \over 64} g_c^2 C_{D^2} \int_0^1 dy \left({1-y \over 1+y}\right)^{p \cdot k} \bigg({4\:\tr(\slashed{p}\gamma^\mu \slashed{k}\gamma_\mu)\over y(1-y^2)}  + {\tr(\slashed{p}\gamma^\mu)\:\tr(\slashed{k}\gamma_\mu)  \over y}\bigg) \\
&= 4 g_c^2 C_{D^2} I(p\cdot k),
\label{eq:RR2pt}
\fe
where we have introduced the worldsheet integral
\ie{}
I(z) \equiv z \int_0^1 dy  { (1+y^2)(1-y)^{z-1} \over y(1+y)^{z+1}}.
\label{eq:integral}
\fe

\paragraph{One dilaton, one axion:} Next consider the contribution from one dilaton and one axion. We take the NSNS vertex operator to be in the $(0,-1)$ picture and the axion vertex operator to be in the $(-\frac12,-\frac12)$ picture. In this case the doubling trick gives a factor of $+i$ from the antiholomorphic fermion and spin field,  see \eqref{eq:BCspinfields}. From \eqref{eq:bcout}, the disc 2-point amplitude is
\ie{}
A^{D^2}_{\delta\tau_2(p)\delta\tau_1(k)} &= 4\sqrt{2}g_c^2 C_{D^2} e_{\mu\nu}(p)f_{\alpha\beta}(k) \int_0^1 dy (1-y)^2 \bigg\langle c_{-1} c_0 c_{1}\left(i \partial X_L^\mu + \frac12 p \cdot \psi \psi^\mu\right)e^{ip \cdot X_L}(i) \\&~~~\times e^{-\phi}\psi^\nu e^{-ip \cdot X_L}(-i) e^{-\frac12\phi} S^\alpha e^{ik \cdot X_L}(iy) e^{-\frac12\phi} S^\beta e^{-ik \cdot X_L}(-iy)\bigg\rangle^{S^2}_{\mathbf{chiral}}.
\fe
This correlator can be evaluated by applying the Ward identities for the translation current $\partial X_L^\mu$ and the Lorentz current $\psi^\mu\psi^\nu$, which yields
\ie{}
A_{\delta\tau_2(p)\delta\tau_1(k)}^{D^2}  &=  -g_c^2 C_{D^2} e_{\mu\nu}(p)\slashed{k}_{\alpha\beta} \int_0^1 dy (1-y)^2  \left(M^{\mu\nu}_\sigma(y) \delta^\alpha_\gamma \delta^\beta_\delta + \delta^\nu_\sigma N^{\mu;\alpha\beta}_{\gamma\delta}(y)\right) \\&\times\bigg\langle c_{-1} c_0 c_{1} e^{ip \cdot X_L}(i)e^{-\phi} \psi^\sigma e^{-ip \cdot X_L}(-i)e^{-\frac12\phi}S^\gamma e^{ik \cdot X_L}(iy)e^{-\frac12\phi}S^\delta e^{-ik \cdot X}(-iy)\bigg\rangle^{S^2}_{\mathbf{chiral}},
\label{eq:intcorr}
\fe
where $M^{\mu\nu}_\sigma(y)$ and $N^{\mu;\alpha\beta}_{\gamma\delta}(y)$ are $c$-numbers given by
\ie{}
M^{\mu\nu}_\sigma(y) &= \frac12 \left({ k^\mu \delta^\nu_\sigma \over i-iy} + {-k^\mu \delta^\nu_\sigma \over i+iy}  +  {\eta^{\mu\nu}p_{\sigma} - p^\nu \delta_\sigma^\mu \over 2i} \right),\\
N^{\mu;\alpha\beta}_{\gamma\delta}(y) &= -\frac14 \left({p_{\sigma}(\gamma^{\sigma\mu})^\alpha_{\:\:\:\gamma}\delta^\beta_\delta \over i - iy}+{p_{\sigma}(\gamma^{\sigma\mu})^\beta_{\:\:\:\delta}\delta^\alpha_\gamma \over i + iy} \right).
\fe 
The expression in \eqref{eq:intcorr} is readily evaluated using the $\psi^\mu S^\alpha S^\beta$ correlator \eqref{eq:psiSS}, from which we find
\ie{}
A_{\delta\tau_2(p)\delta\tau_1(k)}^{D^2} &=
 \frac12 g_c^2 C_{D^2} e_{\mu\nu}(p)f_{\alpha\beta}(k)  \int_0^1 dy \left( {1-y \over 1+y} \right)^{p \cdot k} \left( \slashed{p}^{\alpha\beta} \eta^{\mu\nu} {1+y^2 \over 1-y^2} - 4k^\mu(\gamma^\nu)^{\alpha\beta}  {1 \over 1-y^2} \right).
\fe
Specializing to the case of the axion and dilaton polarization tensors,  we are left with
\ie
A_{\delta\tau_2(p)\delta\tau_1(k)}^{D^2} &= 4g_c^2 C_{D^2} \left(I(p\cdot k) - {e_{\mu\nu}(p) k^\mu k^\nu \over \sqrt{2} p \cdot k}\right). 
\label{eq:RNS2pt}
\fe

\paragraph{Two dilatons:} We now consider the case of two dilatons, taking both NSNS vertex operators to be in the $(0,-1)$ picture. In this case, the doubling trick contributes an overall factor of $(-1)^2 = +1$ from the two antiholomorphic fermions. The disc 2-point amplitude reads
\ie\label{NSNS2ptint}
A_{\delta\tau_2(p)\delta\tau_2(k)}^{D^2} &= -8g_c^2 C_{D^2} e_{\mu\nu}^1e_{\sigma\rho}^2 \int_0^1 dy (1-y)^2 \\
&\times \bigg\langle c_{-1} c_0 c_{1} \left(i \partial X_L^\mu + \frac12 p \cdot \psi \psi^\mu\right)e^{ip \cdot X_L}(i) e^{-\phi}\psi^\nu e^{-ip \cdot X_L}(-i)\\
&~~~~\left(i \partial X_L^\rho + \frac12 k \cdot \psi \psi^\rho\right)e^{ik \cdot X_L}(iy) e^{-\phi}\psi^\sigma e^{-ik \cdot X_L}(-iy)\bigg\rangle^{S^2}_{\mathbf{chiral}} .
\fe
The correlator is readily evaluated using Wick contractions, giving 
\ie
A_{\delta\tau_2(p)\delta\tau_2(k)}^{D^2}   &= -\frac12 g_c^2 C_{D^2} e_\mu^m\mu(p) e_\nu^\nu(k) (p \cdot k) \int_0^1 dy {1 \over y} \left( {1-y \over 1+y}\right)^{p \cdot k - 1}   \\
&~~~+ 4\sqrt{2} g_c^2  C_{D^2} \left( e_{\mu\nu}(p) k^\mu k^\nu + e_{\sigma\rho}(k) p^\sigma p^\rho \right) \int_0^1 dy {(1-y)^{p \cdot k-1} \over (1+y)^{p \cdot k+1}}\\
&~~~+ 4 g_c^2 C_{D^2} e_{\mu\nu}(p)e^{\mu\nu}(k)\int_0^1 dy {\partial \over \partial y} \left( {y (1+y)^{p \cdot k-2} \over (1-y)^{p \cdot k}} \right).
\fe
The form of the integral in the first line can be massaged to give $I(p\cdot k)+1$, while the other integrals can be evaluated directly, with the one in the third line vanishing altogether. Specializing to the case of the dilaton polarization tensor,  we subsequently find
\ie
A_{\delta\tau_2(p)\delta\tau_2(k)}^{D^2}  &= -4g_c^2 C_{D^2} \left(I(p\cdot k) + 1 - { e_{\sigma\rho}(p) k^\sigma k^\rho + e_{\sigma\rho}(k) p^\sigma p^\rho \over \sqrt{2} p \cdot k}  \right).
\label{eq:NSNS2pt}
\fe

\paragraph{Two axion-dilatons:} We can combine our results for the axion and dilaton disc 1-point amplitudes to find the disc 2-point amplitude for two $\delta\tau$ insertions. Using \eqref{eq:RR2pt}, \eqref{eq:NSNS2pt}, and \eqref{eq:RNS2pt}, it follows that
\ie{}
A_{\delta\tau(p)\delta\tau(k)}^{D^2}  = 2 g_c^2 C_{D^2} \left(4I(p\cdot k)+1 - 2{ e_{\sigma\rho}(p) k^\sigma k^\rho + e_{\sigma\rho}(k) p^\sigma p^\rho \over \sqrt{2}p \cdot k} \right).
\fe
The integral $I(p\cdot k)$ in \eqref{eq:integral} diverges logarithmically near the boundary of moduli space $y=0$ as
\ie{}
p \cdot k \int_0 {dy\over y},
\fe
which occurs when the disc with two bulk punctures degenerates into two separate discs, each with a single bulk puncture, connected via an infinitely long strip. As mentioned in the Introduction, this limit corresponds to intermediate ``massless'' open string states with $L_0 = 0$, which formally contribute $\infty$ to the amplitude. In the SFT approach, such massless states are forbidden from propagating, and so the amplitude is manifestly finite. In the naive on-shell prescription, we can tame such divergences through the introduction of a cutoff $\epsilon > 0$. The regularized integral evaluates to
\ie{}
I(p\cdot k) &\to p \cdot k \int_\epsilon^1 dy  { (1+y^2)(1-y)^{p \cdot k-1} \over y(1+y)^{p \cdot k+1}} \\
&= -p \cdot k \left[\psi\left(1+{p \cdot k \over 2}\right) +  \gamma + \log(4\epsilon)\right]+ 1,
\fe
where $\gamma=0.577 \ldots$ is the Euler--Mascheroni constant and $\psi(z) = \partial_z\log(\Gamma(z))$ is the digamma function. We therefore find that the disc 2-point amplitude is
\ie{}
A_{\delta\tau(p)\delta\tau(k)}^{D^2}  &= {128\pi^8 \alpha'^4 \over \tau_2} \bigg[5 - 2{ e_{\sigma\rho}(p) k^\sigma k^\rho + e_{\sigma\rho}(k) p^\sigma p^\rho \over \sqrt{2}p \cdot k}   \\&~~~~~~~~~~~~~~ - 4\alpha'(p \cdot k) \left(\psi\left(1+{\alpha' p \cdot k \over 2}\right)+ \gamma + \log(4\epsilon) \right) \bigg],
\label{eq:disc2pt}
\fe
where we have used $g_c^2 C_{D^2} = 64\pi^8\alpha'^4/\tau_2$ \cite{Polchinski:1998rq,Polchinski:1998rr} and restored the correct factor of $\alpha'$. Substituting it (partially) out using \eqref{eq:newtonconst} gives
\ie{}
A_{\delta\tau(p)\delta\tau(k)}^{D^2}  &= {2\pi \kappa^2\tau_2} \bigg[5 - \sqrt{2}{ e_{\sigma\rho}(p) k^\sigma k^\rho + e_{\sigma\rho}(k) p^\sigma p^\rho \over p \cdot k}   \\&~~~~~~~~~~~~~~ - 4\alpha'(p \cdot k) \left(\psi\left(1+{\alpha' p \cdot k \over 2}\right)+ \gamma + \log(4\epsilon) \right) \bigg].
\label{eq:disc2ptSLZ}
\fe

\subsubsection{Annulus 1-point diagram}

In this section we compute the annulus amplitude with a single $\delta\tau$ insertion. Based on supersymmetry arguments, this diagram is also expected to vanish in the on-shell approach, which we demonstrate explicitly.

The annulus $A^2$ with modulus $t \in (0,\infty)$ can be parametrized by the strip with coordinate $w$ satisfying $0 \leq \text{Re}(w) \leq \pi$ with the identification $w \simeq w + 2\pi it$. Alternatively, it can be described in terms of the torus $T^2$ with modulus $it$ under the identification $w \simeq -\bar{w}$. We take both boundaries to lie on the same D-instanton. The residual conformal symmetry $S^1 \times \bZ_2$ acts by periodic translations $\text{Im}(w) \mapsto \text{Im}(w) + v$ for $v \in (0,2\pi)$  and reflections $\text{Re}(w) \to \pi - \text{Re}(w)$ that exchange the two boundaries. The annulus admits four spin structures corresponding to the choice of boundary conditions for the fermionic fields as well as their periodicity under $w \mapsto w + 2\pi i t$. It is convenient to label the spin structures in terms of those on the torus, with $\nu=1$ denoting the odd spin structure and $\nu = 2,3,4$ the even spin structures.

\begin{figure}[t!]
\centering
\begin{tikzpicture}
[scale=1.75, decoration={markings,mark=at position 0.125 with {\arrow{>}}}]
  %axes
  \draw [help lines,->] (0,0) -- (0,3.5) {};
  \draw [help lines,->] (0,0) -- (4.5,0) {};
  %annulus
  \draw [thin] (0,0) -- (4,0) -- (4,3) -- (0,3) -- (0,0) {};
  \draw [thin] (2-0.05,0-0.05) -- (2+0.05,0+0.05) {};
  \draw [thin] (2-0.05,3-0.05) -- (2+0.05,3+0.05) {};
  %vertex operator
  \draw (2,1.75) node[left] {$V_{\delta\tau(p)}$} ;  
  \path (2,1.5) node[cross=4pt, thick] {} ;
  \draw (2,1.3) node {$u$} ;
  % b contours
  \node at (1.,2.65) {${\cal B}_t$} ;
  \draw[densely dotted, postaction={decorate}] (0,2.5) -- (4,2.5);
  \node at (2.75,2.1) {${\cal B}_u$} ;
  \draw [densely dotted, postaction={decorate}] (2,1.5) circle (0.75);
  % moduli integration contour
  \draw [thick] (0,1.5) -- (4,1.5) {} ;
  %labels
  \node [below] {$0$} ;
  \draw (-0.25,3) node {$2\pi i t$} ;
  \draw (4,-0.1) node {$\pi$} ;
  \node(n)[inner sep=2pt] at (-0.5,3.5) {$w$} ;
  \draw [line cap=round] (n.south west) -- (n.south east) -- (n.north east) ;
\end{tikzpicture}\caption{The annulus $A^2(t)$ represented by a rectangle $w \in [0,\pi] \times [0,2\pi i t]$ with opposite sides $w \simeq w + 2\pi it$ identified. There is a single closed string puncture at $w=u$ for $u \in \mathbb{R}$, which has been drawn off the real axis for clarity. The $b$ ghost contour ${\cal B}_u$ surrounds the integrated vertex operator $V_{\delta\tau(p)}$, while ${\cal B}_{t}$ runs along a horizontal line segment.}
\label{fig:Annulus1ptModspace}
\end{figure}
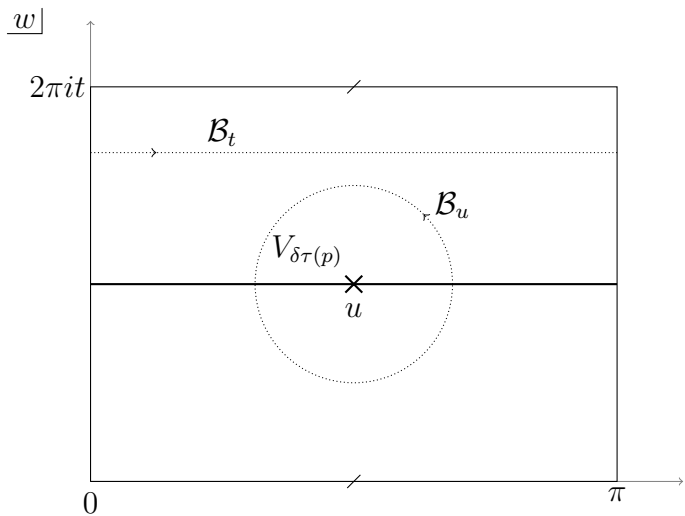

Now consider the annulus with a single closed string puncture. We use the residual conformal symmetry to fix its location to $w = u$ with $0 \leq u \leq \pi$. The $\mathbb{Z}_2$ reflection symmetry can be accounted for multiplying the amplitude by a factor of $\frac12$. We implement the type IIB GSO projection by inserting $\frac12 (-1)^\nu$ into the annulus correlator and summing over spin structures $\nu$. The amplitude requires a single PCO for an RR insertion and two PCOs for an NSNS insertion, which we take to be coincident with the vertex operators., such that they are in the $(+\frac12,-\frac12)$- and $(0,0)$-pictures, respectively. With these preliminaries in mind, the amplitude takes the form
\ie
A_{\delta\tau(p)}^{A^2}  =  \frac14\sum_{\nu=1}^4 (-1)^\nu \int_0^\infty dt \int_0^{\pi} du \: \left\langle {\cal B}_t {\cal B}_u V_{\delta\tau(p)}(u) \right\rangle^{A^2(t),\nu}_{x^\mu=0} ,
\fe
where as usual the choice of picture is kept implicit. Here, the $b$ contours associated to the moduli $u,t$ are given by
\ie{}
{\cal B}_u &=  {1 \over 2\pi i}\oint_{C_u} \left( b(w)dw - \widetilde{b}(\bar{w})d\bar{w}\right), \\
{\cal B}_t &= \int_S \left( b(w)dw + \widetilde{b}(\bar{w})d\bar{w} \right),
\fe
where $C_u$ is a counterclockwise contour surrounding $V_{\delta\tau(p)}(u)$, and $S$ is a line segment at some fixed vertical position that runs horizontally from $\text{Re}(w) =0$ to $\text{Re}(w)= \pi$. 

Both the NSNS and RR vertex operators take the form $c \widetilde{c} e^{q \phi}e^{-q \widetilde{\phi}}{\cal O}_{\delta\tau_{1/2}(p)}$ modulo extra operators which have vanishing correlator, where ${\cal O}_{\delta\tau_{1/2}(p)}$ is a conformal primary in the RR/NSNS sector of the matter SCFT corresponding to the axion/dilaton. It follows that the contribution of the $b,c$ ghosts is
\ie
\big\langle {\cal B}_t {\cal B}_u c\widetilde{c}(u)\big\rangle_{bc}^{A^2(t),\nu} = 2\pi i \eta(it)^2.
\fe
Similarly, the contribution of the $\phi$,$\eta$,$\xi$ system is
\ie
\left\langle e^{q\phi}e^{-q\widetilde{\phi}}(u)\right\rangle^{A^2(t),\nu}_{\phi\eta\xi} = {\eta(it) \over \vartheta_\nu(2q u|it)} \left( { \vartheta_1(2u|it) \over \vartheta'_1(it)} \right)^{q^2},
\fe
where $\vartheta_\nu(w|\tau)$ are the Jacobi theta functions with characteristics, with $\vartheta_1(w|\tau)$ being the unique odd function in $w$, and $\vartheta_\nu(\tau) \equiv \vartheta_\nu(0|\tau)$. Up to an overall phase, the amplitude thus reduces to an integrated correlator in the matter CFT given by
\ie
A_{\delta\tau_{1/2}(p)}^{A^2}  = {\pi \over 2} \sum_{\nu=1}^4 (-1)^{\nu+1} \int_0^\infty dt {\eta(it)^3 \over \vartheta'_1(it)^{q^2}} \int_0^{\pi} du {\vartheta_1(2u|it)^{q^2} \over \vartheta_\nu(2q u|it)} \left\langle {\cal O}_{\delta\tau_{1/2}(p)}(u) \right\rangle^{A^2(t),\nu}_{X^\mu\psi^\nu,\:x^\mu=0} .
\label{eq:annulusghosts}
\fe

\paragraph{One dilaton:} First consider the annulus 1-point amplitude for the dilaton. We strip off the matter part of the vertex operator $V_{\delta\tau_2(p)}$ in picture $(0,0)$,  see \eqref{NSNS(0,0)}, which contributes to the amplitude as
\ie
{\cal O}_{\delta\tau_2(p)}(u) = e_{\mu\nu}(p) \left( i \partial X^\mu + \frac12 p \cdot \psi \psi^\mu \right)\left( i {\bar \partial} X^\nu + \frac12 p \cdot \widetilde{\psi} \widetilde{\psi}^\nu \right) e^{ip \cdot X}(u).
\fe
Using the doubling trick on the annulus, we can trade all of antiholomorphic operators at $\bar{w} = u$ with their holomorphic counterparts at the reflected point $-\bar{w} = -u$ on the torus $T^2$ with complex modulus $it$. By Lorentz invariance, the four-fermion correlator can only involve kinematic structures of the form $p^2 e_{\mu\nu}(p)\eta^{\mu\nu}$ or $e_{\mu\nu}(p)p^\mu p^\nu$, which vanish due to the mass-shell and transversality constraints, respectively. The mixed correlators involving both $\psi^\mu$ and $\partial X^\mu$ vanish since the fermion fields have nothing to contract with. The only remaining term involves both copies of $\partial X^\mu$, contributing 
\ie
\bigg\langle e_{\mu\nu} \partial X^\mu e^{i p \cdot X}(u) \partial X^\nu e^{-i p \cdot X}(-u) \bigg\rangle^{T^2(it),\nu}_{\mathbf{chiral},\: X^\mu\psi^\mu} \propto {\vartheta_\nu(it)^5 \over  \eta(it)^{15} } {\partial^2 \over \partial u^2} \log \vartheta_1 (2u|it) .
\label{eq:annulusdilatonmatter}
\fe
In the above expression, we have evaluated the torus correlator in the matter CFT, which in particular receives a contribution $[\vartheta_\nu(it) / \eta(it)]^5$ from the path integral over $\psi^\mu$. 

\paragraph{One axion:} Next consider the annulus 1-point amplitude for the axion. In the $(+{1 \over 2},-{1\over 2})$ picture, the matter field content of ${\cal O}_{\delta\tau_1(p)}$ consists of $(\slashed{p}\gamma^\mu)^{\alpha}_{\:\:\:\beta} S_\alpha \widetilde{S}^\beta \partial X_\mu e^{ip \cdot X}$. It follows that the correlator appearing in the moduli integrand for each spin structure is proportional to $p^2$, which vanishes by the mass-shell constraint, and so the amplitude is zero.

\paragraph{Axion-dilaton 1-point amplitude:} Since the axion correlator vanishes identically, the annulus amplitude reduces to the contribution of the NSNS matter fields \eqref{eq:annulusdilatonmatter}. Using \eqref{eq:annulusghosts}, we find
\ie
A_{\delta\tau(p)}^{A^2}  \propto \sum_{\nu=1}^4 (-1)^\nu \int_0^\infty dt {\vartheta_\nu(it)^4 \over \eta(it)^{12}} \int_0^{\pi} du \: {\partial^2 \over \partial u^2} \vartheta_1(2u|it)   = 0,
\label{eq:annulus1pt}
\fe
which vanishes by the quartic Jacobi identity $\vartheta_2^4 - \vartheta_3^4 + \vartheta_4^4 = 0$.

\subsubsection{The 4-point amplitude}

We now have all the ingredients necessary to assemble the next-to-leading order D-instanton contribution to the 4-point amplitude, computed by the worldsheet diagrams in (\ref{eq:wsdiagramsNLO}). To find the contribution of the disc 2-point diagram, we are instructed to sum over distinct pairs of vertex operators $V_{\delta\tau(p_i)}V_{\delta\tau(p_j)}$ with  $1 \leq i < j \leq 4$ and multiply the result by two copies of the disc 1-point diagram \eqref{eq:disc1ptSLZ}. This takes the form
\ie{}
32 \pi^3 \kappa^4 \tau_2^3 e^{2\pi i \tau}&\sum_{1 \leq i < j \leq 4}\bigg[-5 + \sqrt{2}{ e_{\sigma\rho}(p_i) p_j^\sigma p_j^\rho + e_{\sigma\rho}(p_j) p_i^\sigma p_i^\rho \over p_i \cdot p_j}   \\&~~~~~~~~~~~~~~ + 4\alpha'(p \cdot k) \left(\psi\left(1+{\alpha' p \cdot k \over 2}\right)+ \gamma + \log(4\epsilon) \right) \bigg].
\fe
Following from momentum-conservation, the above expression can be simplified using
\ie{}
&\sum_{1 \leq i < j \leq 4} p_i \cdot p_j = 0,\\
&\sum_{1 \leq i < j \leq 4}{ e_{\mu\nu}(p_i)p_j^\mu p_j^\nu + e_{\mu\nu}(p_j)p_i^\mu p_i^\nu \over p_i \cdot p_j} &=  \sqrt{8}.
\fe
Due to the first identity, the $\ell_i^\mu$ parametrizing the dilaton polarization tensors in \eqref{dilatonPolarization} drop out of the calculation, as expected. 

The annulus 1-point diagram vanishes, as do the contributions from the 1-holed torus and 3-holed sphere. Consequently, the 4-point amplitude is given by
\ie 
\widetilde{{\cal A}}_{\delta\bar\tau\delta\tau\delta\bar\tau\delta\tau}^{(1,0)}\big|_{\text{NLO}} &= \pi^{16} t^4 {\cal N}_D e^{2\pi i \tau} \left(A_{\delta\tau(p)}\right)^2\sum_{1 \leq i<j \leq 4}  A_{\delta\tau(p_i)\delta\tau(p_j)}^{D^2}  \\
&= -128\pi^{19}t^4 {\cal N}_D \kappa^4 \tau_2^3 e^{2\pi i\tau} \\
&~~~~\times\bigg[\alpha' s\psi\left(1-{\alpha' s \over 4}\right)+\alpha' t\psi\left(1-{\alpha' t \over 4}\right)+\alpha' u\psi\left(1-{\alpha' u \over 4}\right) \bigg],
\label{eq:NLOamp}
\fe
where have dropped the constant terms which contribute at leading order in the momentum expansion. As we will discuss in the next section, such terms are ambiguous in the on-shell approach.

\subsection{An ambiguity and its tension with supersymmetry}
\label{sec:susy}

In general, D-instanton amplitudes as computed from the on-shell prescription suffer from ambiguities related to the choice of regularization scheme. This is to be contrasted with perturbative string amplitudes for on-shell external states which, barring any spurious singularities, are independent of such data. For our choice of regularization scheme, these ambiguities partially manifest themselves in the choice of PCO locations. Normally, amplitudes with different arrangement of PCOs can differ by at most a boundary term in the worldsheet moduli space. However, it is precisely these types of boundary terms which are divergent in the naive formulation of D-instanton perturbation theory, and so can lead to regulator-dependent discrepancies. 

For the specific amplitudes under consideration, this ambiguity manifests itself as an unknown momentum-independent constant at next-to-leading order in $\tau_2^{-1}$, which we denote by $C_1$.  As we shall soon see, this ambiguity is in tension with spacetime supersymmetry. One approach to mend this issue is to abandon the on-shell prescription altogether, as string field theory is free of such ambiguities. Instead of computing any off-shell quantities directly, we shall take a slightly more modest approach and simply demand that the physical amplitude respects supersymmetry. This turns out to be sufficient to pin down the exact value of $C_1$.  

The relevant object of study here is the coefficient $f_0(\tau,\bar\tau)$ multiplying the $R^4$ vertex.\footnote{Recall that $f_i(\tau,\bar\tau)$ enter into the expansion of $M(s,t;\tau,\bar\tau)$, defined as the ratio of $F(s,t)$ to the supergravity contribution $F^{\text{SUGRA}}(s,t) = 2^{-\frac92} \pi^{-{21 \over 4}} \kappa^{\frac72} (\underline{stu})^{-1}$.} As mentioned in the Introduction, this coefficient, as well as those of $D^4 R^4$ and $D^6R^4$, is constrained by supersymmetry to obey a second order differential equation on the moduli space of vacuua. A particularly elegant method of deriving these constraints can be found in \cite{Wang:2015jna}, whose logic we briefly summarize as follows. The basic idea is to analyze the constraints imposed by supersymmetry and unitarity on the factorization of supergraviton scattering amplitudes. For the $R^4$ coefficient, the relevant object of study is the 6-point amplitude. Dimensional analysis implies that it can only factorize through a single $R^4$ vertex and a pair of cubic supergravity vertices. The $\delta\tau\delta\bar\tau R^4$ coupling can then be extracted by taking the soft limit where the momenta of $\delta\tau$ and $\delta\bar\tau$ are taken to zero. From this it follows that the coupling is necessarily proportional to $\tau_2^2\partial_{\tau}\partial_{\bar\tau} f_0$, where the factor of $\tau_2^2$ arises from the normalization of the axion-dilaton kinetic term. Supersymmetry dictates that there is no independent $\delta\tau\delta\bar\tau R^4$ coupling, and so it must be proportional to $f_0$ itself. The constant of proportionality can then be fixed by comparing with the $R^4$ coupling in any supersymmetric theory, such as type IIB string theory, which results in the differential equation \cite{Wang:2015jna}
\ie\label{eq:f0diffeq}
\left( \tau_2^2 \partial_\tau \partial_{\bar\tau} - {3\over 16} \right) f_0(\tau,\bar\tau) = 0.
\fe

With the differential equation in hand, we now set out to explore its implications for our D-instanton results. Using \eqref{eq:LOamp} together with \eqref{Dinstnorm}, the leading-order D-instanton amplitude takes the form
\ie{}
F^{(1,0)}(s,t)\big|_{\text{LO}} &= 2^{-{17 \over 2}} \pi^{-{17 \over 4}} \kappa^{\frac72} e^{2\pi i\tau}.
\label{eq:LOampF}
\fe
Similarly, from \eqref{eq:NLOamp} we have that the next-to-leading order contribution is given by
\ie{}
&F^{(1,0)}(s,t)\big|_{\text{NLO}} = 2^{-{9 \over 2}} \pi^{-{21 \over 4}}\kappa^{{7 \over 2}} e^{2\pi i \tau}\tau_2^{-1} \\ &~~~~~~~~~~~~~~~~~~~~~\times\left[C_1 - {\alpha' s \over 32}\psi\left(1-{\alpha' s \over 4}\right) -{\alpha' t \over 32} \psi\left(1-{\alpha' t \over 4}\right)- {\alpha' u\over 32}\psi\left(1-{\alpha' u \over 4}\right) \right],
\label{eq:dinstNLO}
\fe
where we have included the ambiguous constant $C_1$. In writing down the above expression, we have used the known value for the moduli space measure, i.e.\footnote{The expression for ${\cal N}_D$ in \cite{Sen:2021tpp} takes the form $i 2^{-18} \pi^{-26}\tau_2^{-\frac72}\alpha'^{-1}$,  which agrees with our expressoin upon substituting out $\alpha'$. We have chosen to include the factor of $i$ in the momentum-conserving delta function.}
\ie
{\cal N}_D =   2^{-{33 \over 2}}\pi^{-{97 \over 4}}\kappa^{-\frac12}\tau_2^{-4},
\label{Dinstnorm}
\fe
which, while inaccessible to the on-shell approach, can be derived from first principles in the string field theory formalism. (This was first done in \cite{Sen:2021tpp} and was shown to be consistent with the expectation from S-duality.) Altogether, these amplitudes contribute to the $R^4$ vertex via the momentum-independent terms in \eqref{eq:LOampF} and \eqref{eq:dinstNLO} as
\ie{}
F^{(1,0)}(s,t)\big|_{R^4} &= 2^{-{9 \over 2}} \pi^{-{21 \over 4}}\kappa^{{7 \over 2}} e^{2\pi i \tau}\left[ {\pi \over 16} + C_1 \tau_2^{-1} + {\cal O}(\tau_2^{-2}) \right].
\label{eq:R4dinst}
\fe
Per our discussion above, this amplitude is constrained by supersymmetry to obey the differential equation in \eqref{eq:f0diffeq}. Upon substituting \eqref{eq:R4dinst} into this equation, we find
\ie{}
e^{2\pi i \tau}\left(C_1 - {3\over 256} + {\cal O}(\tau_2^{-1}) \right) = 0,
\fe
from which it immediately follows that
\ie{}
C_1 = {3 \over 256}.
\fe

\subsubsection*{Comparison against S-duality} 

The next-to-leading D-instanton amplitude in \eqref{eq:dinstNLO} admits an expansion in powers of momenta given by
\ie{}
F^{(1,0)}(s,t)\big|_{\text{NLO}} &= 2^{-{9 \over 2}} \pi^{-{21 \over 4}}\kappa^{{7 \over 2}} e^{2\pi i \tau} \left[{3 \over 256}\tau_2^{-1} + \sum_{p=2}^\infty {\zeta(p) \over 2^{3+2p}} \tau_2^{{p-2 \over 2}} (\underline{s}^p+\underline{t}^p+\underline{u}^p)\right],
\label{eq:Dinstamp}
\fe
where we have used the value of $C_1 = {3 \over 256}$, as determined by supersymmetry in the previous subsection. Recall that the underlined variables are the Mandelstam invariants as measured in ten-dimensional Planck units, as determined by the relation
\ie
\alpha' s = \underline{s}\tau_2^{\frac12},
\label{eq:planckstrings}
\fe
and similarly for $t,u$. Unlike the constant term, the momentum-dependent terms in the sum are unambiguous and represent the leading-order single D-instanton contribution to higher-order vertices of the form $D^{2p} R^4$, thus serving as a nontrivial test of S-duality. We can carry out such a test explicitly for the supersymmetry-protected terms, namely $D^4R^4$ and $D^6 R^4$, whose coefficients admit weak-coupling expansions of the form
\ie{}
2^{11}f_4(\tau,\bar\tau) &= 2\zeta(5)\tau_2^{\frac32} + {4 \pi^4 \over 135}\tau_2^{-\frac32} + (e^{2\pi i \tau}+e^{-2\pi i \tau}) \left(16\zeta(2) + {\cal O}(\tau_2^{-1}) \right) + {\cal O}(e^{-4\pi\tau_2}), \\
2^{12}f_6(\tau,\bar\tau) &= {2 \zeta(3)^2 \over 3} \tau_2^3 + {4\zeta(2) \zeta(3) \over 3}\tau_2 + {8\zeta(2)^2 \over 5} \tau_2^{-1} + {4 \zeta(6) \over 28}\tau_2^{-3} \\
&~~~+ (e^{2\pi i \tau}+e^{-2\pi i \tau})\left(8\zeta(3)\tau_2^{\frac12} + {\cal O}(1) \right) - e^{-4\pi \tau_2} \left(2 \tau_2^{-2} + {\cal O}(\tau_2^{-3}) \right) \\
&~~~+ {\cal O}(e^{-6\pi\tau_2}) .
\label{eq:D4D6}
\fe
The leading order D-instanton contribution to these vertices, which arises at next-to-leading order in the open string loop expansion, can be identified as the leading terms multiplying $e^{2\pi i\tau}$, i.e.
\ie{}
f^{(1,0)}_4(\tau,\bar\tau)\big|_{\text{NLO}} = {\zeta(2) \over 128}e^{2\pi i \tau}, \quad
f^{(1,0)}_6(\tau,\bar\tau)\big|_{\text{NLO}} = {\zeta(3) \over 512}e^{2\pi i \tau}\tau_2^{\frac12}.
\fe
Upon accounting for the supergravity contribution  $2^{-\frac92} \pi^{-{21 \over 4}} \kappa^{\frac72}$, we find that these match precisely with our results in \eqref{eq:Dinstamp} for $p=2,3$!

\section{Higher-point MRV amplitudes}
\label{sec:higherptmrv}

In this section, we shall generalize our results for the $2 \to 2$ scattering amplitude in Section~\ref{sec:1D4pt} to higher-point supergraviton amplitudes. Recall that each component of the multiplet has a definite weight $q_R$ under the $U(1)_R$ outer-automorphism group of the supersymmetry algebra, which serves as an accidental global symmetry in the low energy limit. Even though this symmetry is broken explicitly by superstring amplitudes, the degree to which it is violated is controlled by supersymmetric Ward identities. In particular, a given $N$-point amplitude is non-vanishing only if its net charge lies in the range $|q_R| \leq |N-4|$ \cite{Boels:2012zr}. We shall restrict our attention to the so-called ``maximal R-symmetry violating'' (MRV) amplitudes that saturate this bound with $q_R = 4-N$, since these share the most similarities with the 4-point amplitude. In particular, they are constrained by supersymmetry to take the form 
\ie{}
{\cal R}_N(\Phi_i) = i (2\pi)^{10} \delta^{10}(P)Q_+^{16} R_{N}(s_{ij}),
\label{eq:A2toN}
\fe
where $R_{N}(s_{ij})$ is a single function of the Mandelstam invariants. We have labeled the amplitude by ${\cal R}$ instead of ${\cal A}$ to distinguish it from more general R-charge assignments. Note that the $\eta_1^8 \eta_3^8$ component of ${\cal R}_N$, which corresponds to $N-2$ $\delta\tau$ particles and 2 $\delta\bar\tau$ particles, can be related the same component of ${\cal R}_{N-k}$ in the soft limit where $k$ of the $\delta\tau$ particles are taken to have vanishing momentum.  Consequently, these amplitudes admit a low energy expansion which has the schematic form
\ie{}
R_{N}(s_{ij}) = r_0^{(N)}(\tau,\bar{\tau}) + r^{(N)}_4(\tau,\bar\tau)\underline{s}_{ij}^2 + r^{(N)}_6(\tau,\bar\tau)\underline{s}_{ij}^3 + \cdots ,
\fe
with $\cdots$ representing higher-order terms in the momentum expansion. Here, the coefficient $r^{(N)}_{2p}(\tau,\bar\tau)$ is a weight $(N-4,4-N)$ non-holomorphic modular form under the $SL(2,\bZ)$ duality group that multiplies the $(\delta\tau)^{N-4} D^{2p}R^4 $ vertex in the quantum effective action. To be precise, there can be several such coefficients for each value of $p$, which corresponds to the set of independent kinematic structures at each order in the momentum expansion. As mentioned above, the coefficients for a given value of $p$ but different $N$ are related via soft theorems\cite{Green:2019rhz}.\footnote{As a minor note of caution, the $f_{2p}$ are not strictly the same as $r^{(N)}_{2p}$ for $N=4$, since the former are defined as coefficients in the low energy expansion of $M(s,t;\tau,\bar\tau)$, with the supergravity piece factored out, whereas the latter appear in that of the full amplitude $R_4(s,t)$.}

\subsection{D-instanton effects}
\label{sec:cnnlo}

We now consider the contribution of a single D-instanton to the higher-point MRV amplitude. As before, we specialize to the $\eta_1^8\eta_3^8$ component of the superamplitude, which consists of $N$ axion-dilatons $\delta\tau$ and $2$ complex conjugate particles $\delta\bar\tau$, with the momenta of the latter given by by $p_1,p_3$. For a single D-instanton, the diagrammatics of the MRV $N$-point amplitude mirror those of the $4$-point amplitude in Section~\ref{sec:diagrammatics}. Concretely, the D-instanton contribution to the axion-dilaton amplitude takes the form
\ie{}
{\cal R}_{\delta\bar\tau\delta\tau\delta\bar\tau\delta\tau^{N-3}}^{(1,0)} = {\cal N}_D e^{2\pi i \tau} \int d^{10}x d^{16}\theta \sum_{L=0}^\infty \tau_2^{-L}A^{(L)}_{\delta\bar\tau\delta\tau\delta\bar\tau\delta\tau^{N-3}}(x,\theta),
\fe
where $A^{(L)}_{\delta\bar\tau\delta\tau\delta\bar\tau\delta\tau^{N-3}}(x,\theta)$ captures the contributions of worldsheet diagrams of Euler characteristic $-L$ with D-instanton boundary conditions $(x,\theta)$. The factor of ${\cal N}_D$ as well as the moduli $(x,\theta)$ are identical to the $4$-point case since they are universal to any process mediated by a single D-instanton. Due to the same arguments as before, integration over the bosonic moduli restores momentum conservation via $i (2\pi)^{10}\delta^{10}(P)$. Coincidentally, integration over the fermionic moduli also mimics the $4$-point case, transforming $\delta\bar\tau(p_1)\delta\bar\tau(p_3)$ into $\delta\tau(p_1)\delta\tau(p_3)$ while picking up an overall factor proportional to $s_{13}^2$. This is due to the fact that the additional $\delta\tau$ insertions in the $N$-point amplitude are blind to the broken supercharges $\widehat{Q}_-^\alpha$. The net result is that the amplitude takes on a familiar form
\ie{}
{\cal R}_{\delta\bar\tau\delta\tau\delta\bar\tau\delta\tau^{N-3}}^{(1,0)} = \pi^{16} s_{13}^4 {\cal N}_D e^{2\pi i\tau} \sum_{L=0}^\infty \tau_2^{-L} A^{(L)}_{\delta\tau^{N}},
\label{eq:higherptmod}
\fe
where $A^{(L)}_{\delta\tau^{N}}$ is a sum over worldsheet diagrams of Euler characteristic $-L$ with $N$ $\delta\tau$ insertions, and with the D-instanton boundary condition $x^\mu = 0$.

A key point is that the D-instanton contributions to higher-point MRV amplitudes share the same connected diagrams up to some order in the open string loop expansion. In particular, the $N$-point amplitude and $4$-point amplitude are constructed from the same connected diagrams up to order $\tau_2^{-1}$ with $L \leq 1$, and so our results from the previous section immediately apply at these orders.

\subsubsection*{Leading order contribution}

At leading order, $A^{(0)}_{\delta\tau^{N}}$ consists of $N$ copies of the disc 1-point diagram \eqref{eq:disc1pt}. Using \eqref{eq:higherptmod}, we find that the amplitude is
\ie{}
R_{N}^{(1,0)}(s_{ij})\big|_{\text{LO}} = i^N 2^{2N-{33 \over 2}} \pi^{N-{33 \over 4}}\kappa^{N-{1 \over 2}} \tau_2^{N-4} e^{2\pi i \tau}.
\label{eq:higherptLOsl2z}
\fe

\subsubsection*{Next-to-leading order contribution}

At next-to-leading order, $A^{(1)}_{\delta\tau^{N}}$ receives contributions separately from the disc 2-point diagram and the annulus 1-point diagram, as well as the 3-holed sphere and 1-holed torus. As was the case before, the latter three vanish and hence not contribute, and the momentum-independent part of $A^{(1)}_{\delta\tau^{N}}$ encounters the same type of ambiguity. For the disc 2-point, we must sum over distinct pairs $1 \leq i < j \leq N$ in \eqref{eq:disc2pt} and multiply by $N-2$ copies of the disc 1-point diagram \eqref{eq:disc1pt}. Its contribution to the momentum-dependent part of the amplitude meanwhile is unambiguous, taking the form 
\ie{}
R_{N}^{(1,0)}(s_{ij})\big|_{\text{NLO}} &=  i^N 2^{2N-{37 \over 2}} \pi^{N-{37 \over 4}}\kappa^{N-{1\over2}}  e^{2\pi i \tau}\tau_2^{N-5} \\
&~~~\times \left[-\sum_{1 \leq i < j \leq N} \alpha' s_{ij} \psi\left(1 - {\alpha' s_{ij} \over 4}\right) +64C^{(N)}_1 \right],
\label{eq:higherptNLO}
\fe
where $C^{(N)}_1$ is the undetermined constant inaccessible from the on-shell approach. As we shall soon see, this constant is intimately related to that of the 4-point amplitude. The amplitude admits a low energy expansion given by
\ie{}
R_{N}^{(1,0)}(s_{ij})\big|_{\text{NLO}} &=i^N 2^{2N-{37 \over 2}} \pi^{N-{37 \over 4}}\kappa^{N-{1\over2}}  e^{2\pi i \tau}\tau_2^{N-5}\left[\sum_{p=2}^\infty { \tau_2^{{p \over 2}} \over 2^{2p-3}} {\cal O}_{N}^{(p)} +64C^{(N)}_1 \right],
\label{eq:higherptNLOexp}
\fe
where we have introduced the kinematic structures 
\ie{}
{\cal O}_{N}^{(p)} \equiv \frac12 \sum_{1 \leq i < j \leq N} \underline{s}_{ij}^p.
\label{eq:kinstruc}
\fe
For the case of 4-point scattering, these reduce to the familiar ones ${\cal O}_{4}^{(p)} = \underline{s}^p + \underline{t}^p +\underline{u}^p$.

\subsubsection*{Consequences of soft relations}

Amplitudes in type IIB string theory have well-known soft behavior that relates amplitudes in the limit where the momenta of several of the external particles are taken to zero \cite{Chen:2014cuc, Wang:2015aua, Bianchi:2016viy}. For instance, the $(N+1)$-point MRV amplitude with a soft $\delta\tau(p_i)$ is related to the $N$-point MRV amplitude without this particle by
\ie{}
{\cal R}_{N+1}(X,\delta\tau(p_i))\big|_{p_i \to 0} = -2 i \kappa {\cal D}_{N-4} {\cal R}_N(X),
\fe
where ${\cal D}_w$ is the modular covariant derivative that takes modular forms of weight $(w,\tilde{w})$ to those of weight $(w+1,\tilde{w}-1)$, as defined in Appendix~\ref{sec:modforms}, and $X$ denotes the remaining $N$ particles with finite momenta. Given the leading-order \eqref{eq:higherptLOsl2z} and next-to-leading order \eqref{eq:higherptNLOexp} D-instanton contributions, the soft theorem implies that the unknown constant satisfies a recursion relation given by
\ie{}
C_1^{(N+1)} = C_1^{(N)} + {4-N \over 32}.
\fe
with solution
\ie\label{eq:softconst}{}
C_1^{(N)} = {3 \over 256} - {(N-4)(N-5) \over 64}.
\fe
In the above expression, we have used the fact that $C_1^{(4)}$ is precisely the same constant $C_1$ which appears in \eqref{nextleadmst}.

\subsection{Implications for higher-point effective couplings}

In the previous section, we found that D-instanton amplitudes have relatively simple dependence on the momenta, at least for low orders in the open string loop expansion. For instance, the leading-order contribution, which consists of a product of disc 1-point diagrams, has no momentum dependence whatsoever. Similarly, at next-to-leading order the disc 2-point diagram, which involves the kinematic structures ${\cal O}_{N}^{(p)}$, entirely captures the momentum dependence at this order. The fact that only a few kinematic structures enter at low orders can be seen as a consequence of the diagrammatics of D-instanton perturbation theory, i.e. that disconnected diagrams contribute to the connected amplitude. The structure of the diagrammatics in turn has implications for the higher-point effective couplings, ruling out D-instanton contributions to certain kinematic structures.

As a nice example, we shall explore consequences of D-instanton diagrammatics for the 6-point MRV amplitude. The tree-level contribution to the low-energy expansion of the amplitude can be written as\cite{Green:2019rhz}\footnote{Our conventions for Newton's constant are related to the ones in \cite{Green:2019rhz} by $\kappa_{\text{ours}} = 16 \kappa_{\text{theirs}}.$} 
\ie{}
R_6^{\text{tree}}(s_{ij}) \propto {15 \zeta(3) \over 2}\tau_3^{\frac32}  + {35\zeta(5)  \over 64}\tau_2^{\frac52}{\cal O}_6^{(2)} + { \zeta(3)^2 \over 512}  \tau_2^3  {\cal O}_{6,1}^{(3)} + \cdots,
\fe
where the overall normalization of the amplitude is irrelevant for our analysis. In the above expression, the kinematic structure ${\cal O}^{(3)}_{6,1}$ is given by
\ie
{\cal O}^{(3)}_{6,1}&\equiv {1 \over 32} \left(10 \sum_{1 \leq i < j \leq 6}\underline{s}_{ij}^3 + 3 \sum_{1 \leq i<j<k \leq 6} \underline{s}_{ijk}^3\right),
\label{eq:kinstruct1}
\fe
where $s_{ijk} = -(p_i+p_j+p_k)^2$, with the underlined quantity written in Planck units. In analogy with the 4-point case, the first few terms in the momentum expansion of the 6-point amplitude are protected by supersymmetry and can be determined by $SL(2,\bZ)$ covariance \cite{Green:2019rhz}. That is, the momentum expansion of the 6-point amplitude takes the form
\ie{}
R_6(s_{ij})\big|_{\text{analytic}} &\propto r_{0}^{(6)}(\tau,\bar{\tau}) + r_{4}^{(6)}(\tau,\bar{\tau}){\cal O}^{(2)}_6 + r_{6,1}^{(6)}(\tau,\bar{\tau}) {\cal O}^{(3)}_{6,1} +  r_{6,2}^{(6)}(\tau,\bar{\tau}) {\cal O}^{(3)}_{6,2} + \cdots,
\label{eq:6ptmom}
\fe
where $r_0^{(6)}, r_4^{(6)},r_{6,1}^{(6)},r_{6,2}^{(6)},\cdots$ are weight $(2,-2)$ modular forms which multiply the $\delta\tau^2 R^4$, $\delta\tau^2 D^4 R^4$, $\delta \tau^2 D^6 R^4$,$\cdots$ terms in the quantum effective action (see Appendix~\ref{sec:modforms} for more details). To be precise, there are two kinematic structures ${\cal O}_{6,i}^{(3)}$ with $i=1,2$ that appear in the six-point amplitude, which we refer to as $(\delta \tau)^2 D_i^6 R^4$.  The structure with $i=1$ is given by \eqref{eq:kinstruct1}, while the structure with $i=2$ takes the form
\ie
{\cal O}^{(3)}_{6,2}&\equiv 2 \sum_{1 \leq i < j \leq 6}\underline{s}_{ij}^3 - \sum_{1 \leq i<j<k \leq 6} \underline{s}_{ijk}^3 .
\label{eq:kinstruct2}
\fe
Note that the functional form of $r_{6,2}^{(6)}(\tau,\bar{\tau})$ is only determined up to an overall multiplicative constant $c_1$ which cannot be fixed by the tree-level contribution and $SL(2,\bZ)$-covariance alone.

Using our results for the D-instanton diagrammatics from the previous section, we can proceed to determine this unknown constant. According to \eqref{eq:6ptmom} and \eqref{eq:d6r4MRV}, the single D-instanton is expected to contribute up to eighth order in the momentum expansion, and up to next-to-leading order in $\tau_2^{-1}$ as
\ie{}
R_6^{(1,0)}(s_{ij}) &\propto e^{2\pi i \tau} \bigg[\pi^3 \tau_2^2 -{5 \pi^2 \over 16} \tau_2 + {\cal O}(1) \\
&~~~~~~~~~~~+ \left({\pi^2 \zeta(2) \over 8}  \tau_2^2+ {\cal O}(\tau_2) \right) {\cal O}^{(2)}_{6} \\
&~~~~~~~~~~~+ {\pi^2 \zeta(3) \over 32} \tau_2^{\frac52}{\cal O}^{(3)}_{6,1} + {\pi^2 \zeta(3)c_1 \over 96} \tau_2^{\frac52}{\cal O}^{(3)}_{6,2}  + {\cal O}(\tau_2^{\frac32})  \\
&~~~~~~~~~~~+{\cal O}(\underline{s}_{ij}^4)\bigg].
\label{eq:predictedDLO}
\fe
As mentioned, the disc 2-point diagram is solely responsible for the nontrivial momentum dependence at these orders, and so ${\cal O}^{(p)}_6 = \frac12 \sum_{i<j}\underline{s}_{ij}^p$ is the only  kinematic structure which can appear in the expression above. The two kinematic structures $\sum_{i<j}\underline{s}_{ij}^3$ and $\sum_{i<j<k}\underline{s}_{ijk}^3$ that enter into ${\cal O}^{(3)}_{6,i}$ are independent, and thus the coefficient of the latter in \eqref{eq:predictedDLO} is necessarily zero, which immediately implies that the constant is
\ie{}
c_1 = {9 \over 32}.
\label{eq:c1result}
\fe
Note that this result does not rely on the specific value of the disc 2-point diagram, nor any other diagrams which enter at this order. Nevertheless, as a consistency check we can substitute in this value for $c_1$ in \eqref{eq:predictedDLO}, which agrees with our D-instanton calculations in \eqref{eq:higherptLOsl2z} and \eqref{eq:higherptNLO} for $N=4$. Furthermore, it was mentioned in \cite{Green:2019rhz} that $c_1$ is independently determined by the one-loop amplitude, which would serve as a nontrivial test of the D-instanton diagrammatics.

\section{Effects of a D-instanton/anti-D-instanton pair}
\label{sec:dantid}

In this section we analyze the contribution of a D-instanton and an anti-D-instanton pair, or D-$\overline{\text{D}}$ pair for short, to the 4-point supergraviton scattering amplitude, at leading order in both the open string loop and low energy expansions. The final result is presented in \eqref{eq:F11WS} and matches precisely with the coefficient of $D^6R^4$, as expected from supersymmetry and S-duality.

\subsection{Diagrammatic expansion}

We shall once again focus on the scalar components of the multiplet with amplitude $\delta\bar\tau(p_1)+\delta\tau(p_2)\to\delta\bar\tau(p_3)+\delta\tau(p_4)$. The contribution of a D-$\overline{\text{D}}$ pair to the $2\to2$ scattering amplitude is given by the formal expression
\ie
{\cal A}^{(1,1)}_{\delta\bar\tau\delta\tau\delta\bar\tau\delta\tau} = {\cal N}_{D}{\cal N}_{\bar{D}} e^{-4\pi  \tau_2} \int d^{10}x_1 d^{16}\theta_1 d^{10}x_2 d^{16}\theta_2 \sum_{L=0}^\infty \tau_2^{-L} A^{(L)}_{\delta\bar\tau\delta\tau\delta\bar\tau\delta\tau}(x_1,x_2,\theta_1,\theta_2) ,
\label{eq:(1,1)amplitude}
\fe
where the supermoduli space $\widetilde{\cal{M}}_{1,1}={\mathbb{R}}^{20|32}$ is parametrized by 10+16 collective coordinates $(x_1^\mu,\theta_{1\alpha})$ for the D-instanton and 10+16 collective coordinates $(x_2^\mu,\theta_{2\alpha})$ for the anti-D-instanton. The notation here follows that of Section~\ref{sec:1D4pt},  with the sum over $L$ denoting the open string loop expansion. The normalization of the supermoduli space measure factorizes into a product ${\cal N}_{D}{\cal N}_{\bar{D}} $, where ${\cal N}_{\bar{D}}={\cal N}_{{D}}$ is the normalization for that of the anti-D-instanton.

The integrand on the RHS takes the form of a sum over (disconnected) worldsheet diagrams with boundary ending on either the D-instanton or anti-D-instanton. The empty disc diagrams, which come in pairs with net zero $\tau_1$ charge, exponentiate to give an overall factor of $e^{-4\pi\tau_2}$. The contributions from empty annuli with both ends on the same (anti-)D-instanton have been absorbed into the overall normalization (${\cal N}_{\bar{D}}$) ${\cal N}_{D}$.  

A new feature of scattering amplitudes mediated by the D-$\overline{\text{D}}$ pair is the contribution from empty annuli whose boundaries lie on different D-instantons. In particular, we denote the annulus diagram with one boundary on the D-instanton and the other on the anti-D-instanton by $C_{D_1 \bar{D}_2}$. In the sum over Riemann surfaces, such diagrams exponentiate to give an overall factor of $e^{2C_{D_1{\bar D}_2}}$, thereby providing a nontrivial measure on $\widetilde{\cal{M}}_{1,1}$\footnote{The factor of $2$ accounts for the two opposite orientations of open strings stretched between the D-$\overline{\text{D}}$ pair.}.  While in principle there are other empty diagrams which can correct the super-moduli space measure, these appear at subleading orders in $\tau_2^{-1}$ and hence will not be considered in our analysis.

As in \eqref{(1,0)def}, the fermionic moduli contribute through an insertion of the form
\ie
e^{-\Delta S_{\mathrm{WS,R}}(\theta_1)-\Delta S_{\mathrm{WS,R}}(\theta_2)},
\label{eq:DDbarferm}
\fe
where the R-sector vertex operators in $S_{\mathrm{WS,R}}(\theta_1)$ correspond to open strings with endpoints on the D-instanton, and analogously for $S_{\mathrm{WS,R}}(\theta_2)$ and the anti-D-instanton. 

The leading order contribution $A^{(0)}_{\delta\bar\tau\delta\tau\delta\bar\tau\delta\tau}(x_1,x_2,\theta_1,\theta_2)$ comes from the diagram consisting of four disconnected discs, each with one bulk puncture, together with the exponentiated annulus diagram for the D-$\overline{\text{D}}$ pair. This is given by
\ie
&A^{(0)}_{\delta\bar\tau\delta\tau\delta\bar\tau\delta\tau}(x_1,x_2,\theta_1,\theta_2) \\
&~~~=  e^{2C_{D_1,{\bar D}_2}}\left\langle \! \left\langle e^{-\Delta S^{(1)}_{\text{WS,R}}(\theta_1)-\Delta S^{(2)}_{\text{WS,R}}(\theta_2)}V_{\delta\bar\tau(p_1)}V_{\delta\tau(p_2)}V_{\delta\bar\tau(p_3)}V_{\delta\tau(p_4)}\right\rangle \! \right\rangle^{D^2_1 \sqcup D^2_1 \sqcup D^2_1 \sqcup D^2_1}_{x_1\oplus x_2}.
\label{eq:leadingampDDbar}
\fe
where the notation here follows closely with that of Section~\ref{sec:1D4pt} (for instance, see \eqref{eq:leadingamp}). Recall that the double-angled brackets denote a CFT correlation function evaluated over worldsheet with disconnected components, which at leading order corresponds to four disconnected discs each with a bulk puncture, and with boundary lying on the D-instanton (as parameterized by $x_1$) or the anti-D-instanton (as parameterized by $x_2$).  Finally, the dependence on the fermionic moduli is given by \eqref{eq:DDbarferm}.

\subsection{Integration over the fermionic moduli}

We shall first discuss how to perform the integral over the $\theta_i$ coordinates.  The Berezin integral on the RHS of \eqref{eq:leadingampDDbar} gives
\ie{}
&\int d^{16}\theta_1 d^{16}\theta_2 e^{-\Delta S_{\mathrm{WS,R}}(\theta_1)-\Delta S_{\mathrm{WS,R}}(\theta_2)}  = \pi^{32} \widehat{Q}_+^{16} \widehat{Q}_-^{16},
\label{eq:DDbarsusycharges}
\fe
where the supercharge operators are given in \eqref{eq:deltaQ}. We identify $\widehat{Q}^\alpha_{s}$ as the spacetime supercharges broken by the D-instanton ($s=-1$) and anti-D-instanton ($s=+1$), respectively, which as before are defined modulo additive contributions from the preserved supercharges. They are topological in the sense that they can be deformed in a worldsheet diagram so long as they do not cross any bulk closed string insertions.

We proceed to evaluate the amplitude using the approach of Section~\ref{sec:1D4pt}, where the supercharge contours are taken to surround the closed string vertex operators, acting as supersymmetry transformations on the 1-particle states. As was the case for the single D-instanton, $\widehat{Q}_-^{16}$ acts only on the $V_{\delta\bar\tau(p_i)}$, converting the two vertex operators to $V_{\delta\tau(p_i)}$ with an overall factor of $t^4$.  What remains are four $V_{\delta\tau(p_i)}$ insertions surrounded by the sixteen supercharges broken by the anti-D-instanton, namely $\widehat{Q}_+^{16}$.  Here we need to appeal to the details of the D-instanton/anti-D-instanton boundary conditions. In particular, the disc 1-point diagram vanishes for all massless insertions except for $\delta\tau$ and $\delta\bar\tau$. Consequently, the only nontrivial configurations are given by two $\delta\tau$ insertions and two $\delta\bar\tau$ insertions, where the latter correspond to eight supercharges acting on $V_{\delta\tau(p_i)}$. Using the same type of argument in the COM frame, it is straightforward to show that the remaining supercharges convert $V_{\delta\tau(p_i)},V_{\delta\tau(p_j)}$ into $V_{\delta\bar\tau(p_i)},V_{\delta\bar\tau(p_j)}$ with an overall factor of $(p_i+p_j)^8$. There are a total of $6 = {4 \choose 2}$ such configurations, which naturally decompose into $s,t,u$ channels corresponding to which particles are paired. Overall, we find that the leading-order D-$\overline{\text{D}}$ contribution takes the form
\ie{}
\left.{\cal A}^{(1,1)}_{\delta\bar\tau\delta\tau\delta\bar\tau\delta\tau}\right|_{\text{LO}} = 2\pi^{32}t^4 {\cal N}_D {\cal N}_{\bar{D}}e^{-4\pi\tau_2} \left[s^4 A_{\delta\bar\tau\delta\bar\tau\delta\tau\delta\tau}^{(0),s} + t^4 A_{\delta\bar\tau\delta\tau\delta\bar\tau\delta\tau}^{(0),t} + u^4 A_{\delta\bar\tau\delta\tau\delta\tau\delta\bar\tau}^{(0),u}\right],
\label{eq:2to2DDball}
\fe
where the contribution from each of the three channels is given by
\ie
A_{\delta\bar\tau\delta\bar\tau\delta\tau\delta\tau}^{(0),s}&=\int d^{10} x_1d^{10} x_2 e^{2C_{D_1,{\bar D}_2}}\langle c_0V_{\delta\bar\tau(p_1)}\rangle^{D^2}_{x_2}\langle c_0V_{\delta\bar\tau(p_2)}\rangle^{D^2}_{x_2} \langle c_0V_{\delta\tau(p_3)}\rangle^{D^2}_{x_1}\langle c_0V_{\delta\tau(p_4)}\rangle^{D^2}_{x_1},
\\
A_{\delta\bar\tau\delta\tau\delta\bar\tau\delta\tau}^{(0),t}&=\int d^{10} x_1d^{10} x_2 e^{2C_{D_1,{\bar D}_2}}\langle c_0V_{\delta\bar\tau(p_1)}\rangle^{D^2}_{x_2}\langle c_0V_{\delta\tau(p_2)}\rangle^{D^2}_{x_1}\langle c_0V_{\delta\bar\tau(p_3)}\rangle^{D^2}_{x_2}\langle c_0V_{\delta\tau(p_4)}\rangle^{D^2}_{x_1},
\\
A_{\delta\bar\tau\delta\tau\delta\tau\delta\bar\tau}^{(0),u}&=\int d^{10} x_1d^{10} x_2 e^{2C_{D_1,{\bar D}_2}}\langle c_0V_{\delta\bar\tau(p_1)}\rangle^{D^2}_{x_2}\langle c_0V_{\delta\tau(p_2)}\rangle^{D^2}_{x_1}\langle c_0V_{\delta\tau(p_3)}\rangle^{D^2}_{x_1}\langle c_0V_{\delta\bar\tau(p_4)}\rangle^{D^2}_{x_2}.
\label{eq:2to2DDbfull}
\fe
which consists of the exponentiated annulus diagram together with four 1-punctured discs, with the vertex operators distributed such that the discs with D-instanton boundary conditions contain $V_{\delta\tau(p_i)}$, while those with anti-D-instanton boundary conditions contain $V_{\delta\bar\tau(p_j)}$. Furthermore, for the disc topology the anti-D-instanton boundary conditions are identical to those of the D-instanton other than a sign flip relating the spin fields, and so the disc 1-point diagrams are given by
\ie
\langle c_0V_{\delta\tau(p)}\rangle^{D^2}_{x_1}&= A_{\delta\tau(p)}^{D^2} e^{i p \cdot x_1},~~~~\langle c_0V_{\delta\tau(p)}\rangle^{D^2}_{x_2}= 0,
\\
\langle c_0V_{\delta\bar{\tau}(p)}\rangle^{D^2}_{x_1}&=0,~~~~~~~~~~~~~~~ \langle c_0V_{\delta\bar{\tau}(p)}\rangle^{D^2}_{x_2}=A_{\delta\tau(p)}^{D^2} e^{i p \cdot x_2},
\label{eq:disc1tautaub}
\fe
where $A_{\delta\tau(p)}^{D^2}$ is given the axion-dilaton disc 1-point diagram \eqref{eq:disc1pt} with boundary condition $x^\mu=0$, and $x_1$ denotes the boundary condition for the D-instanton at $x_1$,  and similarly for the anti-D-instanton at  $x_2$.

\subsection{The measure on moduli space}

We now discuss how to compute the annulus diagram $C_{D_1{\bar D}_2}$, which contributes nontrivially to the measure on the moduli space for the D-$\overline{\text{D}}$ pair.

The boundary conditions for the D-instanton with bosonic modulus $x$ can be embedded in a GSO-even boundary state of the form
\ie
\left|D_{x}\right\rangle=\frac{1}{\sqrt{2}}\left(\left|\mathrm{NSNS};x\right\rangle+\left|\mathrm{RR};x\right\rangle\right),
\label{eq:Dbdrystate}
\fe
where $\left|\mathrm{NSNS};x\right\rangle$ and $\left|\mathrm{RR};x\right\rangle$ denote the contribution to the boundary state coming from closed strings in the NSNS and RR sectors, respectively.  For simplicity,  we have set the fermionic modulus $\theta = 0$. The anti-D-instanton has opposite charge with respect to the RR axion field, and therefore its boundary state is given by
\ie
\left|\bar{D}_{x}\right\rangle=\frac{1}{\sqrt{2}}\left(\left|\mathrm{NSNS};x\right\rangle-\left|\mathrm{RR};x\right\rangle\right).
\fe

Let us first consider the cylinder diagram between two D-instantons with bosonic collective coordinates $x_1,x_2$.  In the closed string channel, this is given by the overlap of two D-instanton boundary states, which takes the form\footnote{We take the RR sector component of the D-instanton boundary state $\ket{D_i}$ to have picture number $\left(-\frac{1}{2},-\frac{3}{2}\right)$, while its bra $\bra{D_i}$ has picture number $\left(-\frac{3}{2},-\frac{1}{2}\right)$. This ensures that the overlap \eqref{eq:cylclosedchannel} has total picture number $(-2,-2)$,  as required for a non-zero result \cite{Sen:2021tpp}.} 
\ie
Z_{D_1D_2}(t)=\left\langle D_1\right|e^{-\frac{\pi}{t}\left(L_0+\widetilde{L}_0\right)} b_0 c_0\left|D_2\right\rangle,
\label{eq:cylclosedchannel}
\fe
where $D_i$ labels the D-instanton with collective coordinate $x_i$. In \eqref{eq:cylclosedchannel}, we are working with a cylinder of length $1/(2t)$ and circumference $2\pi$. The ghost insertion $b_0 c_0$ is required for a nonzero result,  acting as the projector onto the ghost ground state annihilated by $b_0$. Here, $L_0$ ($\widetilde{L}_0$) denotes the zero Fourier mode of the stress tensor $T$ ($\widetilde{T}$). Under a modular transformation of the cylinder, \eqref{eq:cylclosedchannel} is related to the open string partition function for a cylinder of length $\pi$ and circumference $2\pi t$, i.e. 
\ie
Z_{D_1D_2}(t)=\left(\mathrm{Tr}_{\mathcal{H}_0^{\mathrm{NS}}}-\mathrm{Tr}_{\mathcal{H}_0^{\mathrm{R}}}\right)\frac{1+(-1)^F}{2}(-)^{N_{bc}+N_{\beta\gamma}}b_0c_0e^{-2\pi t L_0}.
\label{eq:cylopenchannel}
\fe
In the above expression, the trace is taken with respect to the NS and R sectors of the Hilbert space of open strings stretched between two D-instantons at positions $x_1,x_2$, and the factor $(1+(-)^F)/2$ implements the type IIB GSO projection.  The minus sign in front of the Ramond sector contribution has the usual interpretation of spacetime fermions running in the loop. The insertion $(-)^{N_{bc}+N_{\beta\gamma}}b_0c_0$ projects onto the ghost ground state,  and is needed to obtain a non-zero result.

In \eqref{eq:cylopenchannel}, the terms with the $(-)^F$ insertion map under the inverse modular transformation to closed string states with periodic boundary conditions along the circle, and therefore correspond to the RR sector contributions to the boundary state \eqref{eq:Dbdrystate}. On the other hand, the terms without the $(-)^F$ insertion map to closed string states in the NSNS component of the boundary state. 

For now, we focus on the contribution coming from the exchange of closed strings in the NSNS sector, which corresponds to open strings with anti-periodic boundary condition in the time direction. The ghosts give a contribution that cancel one pair of oscillators as usual. It follows that \cite{Polchinski:1995mt}
\ie
Z^{NS}_{D_1D_2}(t)&=\frac{1}{2}\left(\mathrm{Tr}_{\mathcal{H}_0^{\mathrm{NS}}}-\mathrm{Tr}_{\mathcal{H}_0^{\mathrm{R}}}\right)(-)^{N_{bc}+N_{\beta\gamma}}b_0c_0e^{-2\pi \tau L_0}\\
&=\frac{1}{2}e^{-t\frac{(x_{12})^2}{2\pi}}\left(\eta(it)\right)^{-8}\left(\left[\frac{\vartheta_3(it)}{\eta(it)}\right]^4-\left[\frac{\vartheta_2(it)}{\eta(it)}\right]^4\right),
\label{eq:cylopenchannel2}
\fe
where $x_{12}\equiv {x}_1-{x}_2$ is the relative position between the D-instantons. Note that there is no overall factor of $i$ since each D-instanton is localized in Euclidean target space. Overall, we find the NSNS contribution to the amplitude between two D-instantons is
\ie
C^{NS}_{D_1D_2}&=\int_0^{\infty}\frac{dt}{2t}Z^{NS}_{D_1,D_2}(t)
\\
&=\int_0^{\infty}\frac{dt}{4t}e^{-t\frac{(x_{12})^2}{2\pi}}\left(\eta(it)\right)^{-8}\left(\left[\frac{\vartheta_3(it)}{\eta(it)}\right]^4-\left[\frac{\vartheta_2(it)}{\eta(it)}\right]^4\right).
\\
&=\frac{1}{4}\int_0^\infty dt ~t^3 e^{-t\frac{(x_{12})^2}{2\pi}}\left[16+\mathcal{O}\left(e^{-\frac{2\pi}{t}}\right)\right]
\\
&=\frac{24 (2\pi)^4}{\left(x_{12}\right)^8}+\mathcal{O}\left(\frac{e^{-2|x_{12}|}}{\left(x_{12}\right)^4}\right).
\label{eq:DDcylmassless}
\fe
In the second line we performed a modular transformation of the cylinder on \eqref{eq:cylopenchannel}. The first term in the last line of \eqref{eq:DDcylmassless} gives the contribution to the amplitude due to massless closed strings in the NSNS sector. The other terms are exponentially suppressed at large $x_{12}$ and can be interpreted as the contribution of massive closed strings. As we will see, it is only the massless exchange which is relevant for our analysis.

The contribution to \eqref{eq:cylopenchannel} coming from the RR closed strings must cancel that of the NSNS closed strings,  since the potential between the D-instantons vanishes by supersymmetry, and so
\ie
C^{\text{NS}}_{D_1D_2}+C^{\text{R}}_{D_1D_2}=0.
\fe
The anti-D-instanton boundary state differs from the D-instanton boundary state by a minus sign multiplying the RR component, which implies that the annulus amplitude for the D-$\overline{\text{D}}$ pair is given by twice the NSNS contribution \eqref{eq:DDcylmassless}, i.e.
\ie
C_{D_1\bar{D}_2}&=2\int_0^{\infty}\frac{dt}{4t}e^{-t\frac{(x_{12})^2}{2\pi}}\left(\eta(it)\right)^{-8}\left(\left[\frac{\vartheta_3(it)}{\eta(it)}\right]^4-\left[\frac{\vartheta_2(it)}{\eta(it)}\right]^4\right)
\\
&=\frac{48 (2\pi)^4}{\left(x_{12}\right)^8}+\mathcal{O}\left(\frac{e^{-2|x_{12}|}}{\left(x_{12}\right)^4}\right).
\label{eq:annulusDDb}
\fe
The measure on moduli space is given by the exponential of this diagram, which thus takes the form
\ie
\exp\left[\frac{96 (2\pi)^4}{\left(x_{12}\right)^8}+\mathcal{O}\left(\frac{e^{-2|x_{12}|}}{\left(x_{12}\right)^4}\right)\right]=\sum_{n=1}^{\infty}\frac{1}{n!}\left[\frac{96 (2\pi)^4}{\left(x_{12}\right)^8}\right]^n+\mathcal{O}\left(\frac{e^{-2|x_{12}|}}{\left(x_{12}\right)^4}\right),
\label{eq:DDbarmeasure}
\fe
where we continue to ignore the contribution of massive closed strings. 

\subsection{Integration over the bosonic moduli}

At leading order in $\tau_2^{-1}$, the moduli space measure is given by the exponentiated annulus diagram $e^{2C_{D_1\bar{D}_2}}$, where $C_{D_1\bar{D}_2}$ is given by \eqref{eq:annulusDDb}. From \eqref{eq:annulusDDb}, we observe that the integral over $t$ develops a logarithmic divergence when the D-$\overline{\text{D}}$ pair is separated by a distance
\ie
\left(x_{12}\right)^2=2\pi^2.
\fe
This has the interpretation of an open string stretched between the D-$\overline{\text{D}}$ pair going on-shell. Furthermore, this divergent behavior becomes tachyonic for $\left(x_{12}\right)^2<2\pi^2$.  Thus, the integration over the full range of $x_1,x_2$ in \eqref{eq:2to2DDbfull} is only well-defined for a choice of contour avoiding these singularities.

Instead of worrying about the choice of contour, we shall focus only on the part of \eqref{eq:annulusDDb} corresponding to asymptotic separation of the D-$\overline{\text{D}}$ pair. As we shall see, such contributions are unambiguous. Using \eqref{eq:DDbarmeasure}, we find that the contribution of massless closed string states to the $s$-channel amplitude is given by
\ie
A_{\delta\bar\tau\delta\bar\tau\delta\tau\delta\tau}^{(0),s}&=\left(A_{\delta\tau(p)}^{D^2}\right)^4 \int_{|x_{12}|>a} d^{10} x_1d^{10} x_2 \sum_{n=0}^{\infty}\frac{1}{n!}\left[\frac{96 (2\pi)^4}{\left(x_{12}\right)^8}\right]^ne^{i \left(p_1+p_2\right)\cdot x_2+i \left(p_3+p_4\right)\cdot x_1},
\label{eq:2to2DDb}
\fe
where we have cut out a finite domain of the moduli space. In the above expression, higher-order terms in the sum over $n$ contribute only to the higher-order terms in the low energy expansion of the scattering amplitude. The lowest order term, $n=0$, contributes only to the disconnected part of the amplitude, and can be neglected. 

The next term in the sum, $n=1$, is given by
\ie
&i(2\pi)^{10}\delta^{10}(P)\int_{|x_{12}|>a} d^{10} x_{12}\frac{96 (2\pi)^4}{\left(x_{12}\right)^8}e^{-i x_{12}\cdot \left(p_1+p_2\right)} \\&~~~~~~~~~~~~~~~~~~~~~~= -i(2\pi)^{10}\delta^{10}(P)\left[{2^{10} \pi^{9} \over s}+ {\rm analytic} \right],
\label{eq:2to2DDbn1}
\fe
where on the LHS we have integrated over the center-of-mass collective coordinates $x_1+x_2$, which restores momentum conservation. On the RHS, we omitted terms that are analytic in momenta at $s=0$, which is part of the ``analytic ambiguity" of the amplitude that involves the integration over a finite domain of the moduli space with a yet unspecified contour prescription. Meanwhile, the $s^{-1}$ term captures the contribution of massless closed string exchange at large relative separation $x_{12}$ of the D-$\overline{\text{D}}$ pair. Crucially, it is independent of $a$, and so this contribution \cite{Green:2014yxa}, at leading order in both the open string loop expansion and low energy expansion, is unambiguously defined.

A similar analysis in the $t$- and $u$-channels yields analgous results, with $s$ replaced by $t$ and $u$, respectively. Using \eqref{eq:2to2DDb} and \eqref{eq:2to2DDbn1}, we find that the leading contribution (both in the open string loop expansion, and in momentum) from the D-$\overline{\text{D}}$ pair to the axion-dilaton amplitude is 
\ie
\left.\widetilde{{\cal A}}^{(1,1)}_{\delta\bar\tau\delta\tau\delta\bar\tau\delta\tau}\right|_{\text{LO}}=- 2^{-20} \pi^{-{21 \over 2}} \kappa^5 e^{-4\pi\tau_2} \tau_2^{-2}t^4 \bigg[ (s^3 + t^3+u^3) + {\cal O}(s^4,t^4,u^4) \bigg], 
\label{eq:A11ws}
\fe
where we have used our results for $A_{\delta\tau(p)}^{D^2}$ in \eqref{eq:disc1pt} and ${\cal N}_D$ in \eqref{Dinstnorm}, and have restored the value of $\alpha'$. From this, it follows that the $2 \to 2$ amplitude reads
\ie{}
F^{(1,1)}(s,t)\big|_{\text{LO}} = - 2^{-{31 \over 2}} \pi^{-{21 \over 4}} \kappa^{\frac72} e^{-4\pi \tau_2} \tau_2^{-2} (\underline{s}^3 + \underline{t}^3+ \underline{u}^3) + {\cal O}(\underline{s}^4,\underline{t}^4,\underline{u}^4),
\label{eq:F11WS}
\fe
where we have written the Mandelstam invariants as measured in Planck units.

\subsection{Comparison against S-duality}

We now test our results for the D-$\overline{\text{D}}$ amplitude in \eqref{eq:F11WS} against the $D^6 R^4$ coefficient $f_6(\tau,\bar\tau)$ expected from supersymmetry and S-duality. The D-$\overline{\text{D}}$ contributions to the latter can be identified as the as the terms multiplying $e^{-4\pi \tau_2}$ in \eqref{eq:D4D6}, which at leading-order reads
\ie{}
2^{12}f^{(1,1)}_6(\tau,\bar{\tau}) &=- 2e^{-4\pi \tau_2}\tau_2^{-2} + {\cal O}(e^{-4\pi \tau_2}\tau_2^{-3}) .
\fe
Upon including the supergravity factor $2^{-\frac92} \pi^{-{21 \over 4}} \kappa^{\frac72}$, we find that this matches precisely with our results in \eqref{eq:F11WS}. Before moving on, we note that in retrospect it seems somewhat surprising that the $D^6 R^4$ vertex, which is protected by supersymmetry by virtue of being $\frac18$-BPS, receives contributions from the non-supersymmetric D-$\overline{\text{D}}$ instanton configuration.

\section{A test of non-perturbative unitarity}
\label{sec:DDbarunitarity}

As discussed in Section~\ref{sec:dantid}, the measure on the supermoduli space is singular due to open strings going on-shell, which in turn implies that the integration over the bosonic moduli suffers from ambiguities. So far, we have investigated D-$\overline{\text{D}}$ contributions to the $2\to2$ scattering amplitude which are insensitive to these ambiguities, which were found to be consistent with supersymmetry and $SL(2,\mathbb{Z})$ duality.

This leads us to ask whether there are other D-$\overline{\text{D}}$ contributions which are unambiguous in the on-shell approach. From the expression in \eqref{eq:DDbarmeasure} for the moduli space measure,  and following the steps leading to \eqref{eq:A11ws}, it is clear that arbitrarily massive closed strings will contribute to the scattering amplitude at higher orders in the low energy expansion. Although this would seem to suggest that all such higher order terms are ambiguous, it turns out that certain contributions are unambiguous owing to the fact that they have non-polynomial dependence on the external momenta. This is related to the idea that such amplitudes can be obtained from lower-point amplitudes by unitarity cuts. In this section, we analyze the simplest example of this phenomenon from the worldsheet perspective,  thereby providing a nontrivial check of non-perturbative unitarity in type IIB scattering amplitudes.

\subsection{A discontinuity in the D-$\overline{\text{D}}$ amplitude}

We begin by returning to the  D-$\overline{\text{D}}$ contribution to the $2 \to 2$ amplitude coming from the next term in the expansion of  \eqref{eq:2to2DDb}, i.e. the term with $n=2$. This corresponds to two copies of the annulus diagram (or more precisely, of the contribution from the massless closed strings to the annulus diagram). In the $s$-channel, this can be written as
\ie
A_{\delta\tau\delta\tau\delta\tau\delta\tau}^{(0),s} &\supset \frac12 \left(A_{\delta\tau(p)}^{D^2}\right)^4\alpha'^8 \int_{|x_{12}|>a} d^{10} x_{12}\frac{\left[96 (2\pi)^4\right]^2}{\left(x_{12}\right)^{16}}e^{-i x_{12} \cdot \left(p_1+p_2\right)} \\
&=  9 \cdot 2^{17} \pi^8 S_{8} \left(A_{\delta\tau(p)}^{D^2}\right)^4 \alpha'^8 \int_a^{\infty}dr~r^9\int_0^\pi d\theta\sin^8\theta\frac{1}{r^{16}}e^{i p r\cos\theta},
\label{eq:2to2DDbn2}
\fe
where $r = -|x_{12}|$, $p =|p_1+p_2|$, and $S_8=\frac{32\pi^4}{105}$ is the area of the 8-sphere.  We can directly evaluate the regularized expression in \eqref{eq:2to2DDbn2}, with the understanding that the terms analytic in $s$ are ambiguous, as they are sensitive to the cutoff $a$. The non-analytic piece receives contributions from
\ie{}
\int_a^{\infty}dr~r^9\int_0^\pi d\theta\sin^8\theta\frac{1}{r^{16}}e^{i p r\cos\theta} = \frac{\pi  s^3 \ln(-s) }{9 \cdot 2^{16}} + \cdots,
\fe
where $\cdots$ indicates terms analytic in $s=-(p_1+p_2)^2$. It follows that the leading-order contribution from the D-$\overline{\text{D}}$ pair to the non-analytic part of the amplitude goes like
\ie{}
\left.\widetilde{{\cal A}}^{(1,1)}_{\delta\bar\tau\delta\tau\delta\bar\tau\delta\tau}\right|_{\text{LO}}^{\text{non-analytic}} &= -4\pi^{41}S_8\left(A_{\delta\tau(p)}^{D^2}\right)^4 \alpha'^8 {\cal N}_{D}{\cal N}_{\bar{D}}e^{-4\pi\tau_2} \\ &~~~\times t^4 (s^7\ln(-s) + t^7 \ln(-t) + u^7 \ln(-u))\\
&=-2^{-35} \pi^{-{35 \over 2}}S_8  \kappa^7 e^{-4\pi\tau_2} \\
&~~~\times t^4 (s^7\ln(-s) + t^7 \ln(-t) + u^7 \ln(-u)),
\label{eq:A11log}
\fe
where the extra terms come from a similar analysis in the $t$- and $u$-channels.

In each of the $s,t,u$ channels, the logarithmic dependence gives rise to a branch cut in the corresponding complex plane. For instance, the discontinuity across the $s$-cut is given by
\ie
2\mathrm{Im}\left(\left.\widetilde{{\cal A}}^{(1,1)}_{\delta\bar\tau\delta\tau\delta\bar\tau\delta\tau}\right|^{s}_{\text{LO}}\right)= 2^{-34} \pi^{-{33 \over 2}}S_8\kappa^7e^{-4\pi\tau_2} t^{4}s^7 ,
\label{eq:ImA11}
\fe
where the superscript $s$ indicates that we keep the terms multiplying the $s$-channel contribution $A_{\delta\tau\delta\tau\delta\tau\delta\tau}^{(0),s}$ in \eqref{eq:2to2DDball}.  Furthermore, we are ignoring contributions from higher particle cuts to $A_{\delta\tau\delta\tau\delta\tau\delta\tau}^{(0),s}$ (e.g.  3-particle cuts or higher),  which come from higher order terms in the expansion of \eqref{eq:2to2DDb} (i.e.  $n\geq3$).  The discontinuity across the branch cut in \eqref{eq:ImA11} has the interpretation of massless closed strings exchanged between the D-instanton and anti D-instanton being on-shell.  In the next subsection, we shall verify this explicitly through a worldsheet calculation that relies only on scattering amplitudes mediated by a single D-instanton. This provides a nontrivial check of 
\eqref{eq:ImA11} and verifies that \eqref{eq:A11log} is insensitive to the analytic ambiguities present in the bosonic moduli integration.

\subsection{Verification of unitarity}

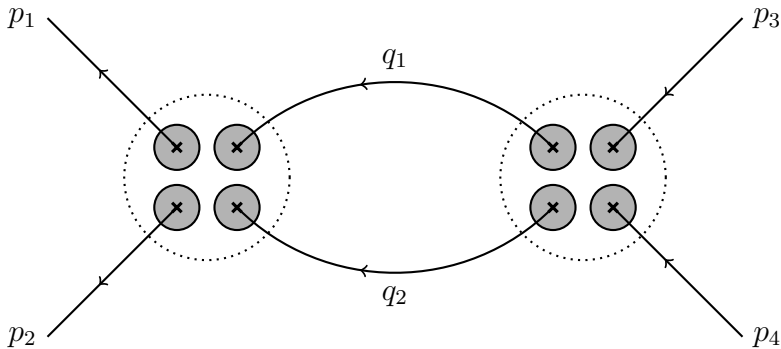
\begin{figure}[t!]
\centering
\begin{tikzpicture}
[decoration={markings,mark=at position 0.6 with {\arrow{>}}}]

\draw[color=black, thick, dotted] (0,0) circle (1.1);
\filldraw[color=black, fill=black!30, thick] (0.4,0.4) circle (0.3);
\draw node[cross=3pt, very thick] at (0.4,0.4) {};
\filldraw[color=black, fill=black!30, thick] (0.4,-0.4) circle (0.3);
\draw node[cross=3pt, very thick] at (0.4,-0.4) {};
\filldraw[color=black, fill=black!30, thick] (-0.4,0.4) circle (0.3);
\draw node[cross=3pt, very thick] at (-0.4,0.4) {};
\filldraw[color=black, fill=black!30, thick] (-0.4,-0.4) circle (0.3);
\draw node[cross=3pt, very thick] at (-0.4,-0.4) {};

\draw[color=black, thick, dotted] (5,0) circle (1.1);
\filldraw[color=black, fill=black!30, thick] (5.4,0.4) circle (0.3);
\draw node[cross=3pt, very thick] at (5.4,0.4) {};
\filldraw[color=black, fill=black!30, thick] (5.4,-0.4) circle (0.3);
\draw node[cross=3pt, very thick] at (5.4,-0.4) {};
\filldraw[color=black, fill=black!30, thick] (4.6,0.4) circle (0.3);
\draw node[cross=3pt, very thick] at (4.6,0.4) {};
\filldraw[color=black, fill=black!30, thick] (4.6,-0.4) circle (0.3);
\draw node[cross=3pt, very thick] at (4.6,-0.4) {};

\draw [color=black, thick, postaction={decorate}] (-0.4,0.4) -- (-2.12,2.12); 
\draw [color=black, thick, postaction={decorate}] (-0.4,-0.4) -- (-2.12,-2.12); 

\draw [color=black, thick, postaction={decorate}] (7.12,2.12) -- (5.4,0.4); 
\draw [color=black, thick, postaction={decorate}] (7.12,-2.12) -- (5.4,-0.4); 

\draw [color=black, thick, postaction={decorate}] (4.6,0.4) to[out=135, in=45] (0.4,0.4); 
\draw [color=black, thick, postaction={decorate}] (4.6,-0.4) to[out=-135, in=-45] (0.4,-0.4); 

\node[left] at (-2.12,2.12) {$p_1$};
\node[left] at (-2.12,-2.12) {$p_2$};
\node[right] at (7.12,2.12) {$p_3$};
\node[right] at (7.12,-2.12) {$p_4$};
\node[above] at (2.5,1.3) {$q_1$};
\node[below] at (02.5,-1.3) {$q_2$};

\end{tikzpicture}
\caption{Spacetime Feynman diagram that contributes to the unitarity cut.  After cutting the internal propagators to put the intermediate particles on-shell, the vertices are given by either the D-instanton or anti-D-instanton mediated contributions to $R^4$. Ingoing arrows denote $\delta\tau$ and outgoing arrows denote $\delta\bar\tau$.}
\label{fig:unitaritycut}
\end{figure}

In search of unitarity, we will focus on the $s$-channel cut contribution to the scattering amplitude $\delta\bar\tau(p_1)+\delta\bar\tau(p_2)\to\delta\tau(p_3)+\delta\tau(p_4)$ as represented by the diagram in Figure~\ref{fig:unitaritycut}. As a shorthand, we define
\ie{}
{\cal A}_{\delta\bar\tau\delta\bar\tau\delta\tau\delta\tau}(p_1,p_2,p_3,p_4) \equiv {\cal A}_{2\to2}\big|_{\eta_1^8 \eta_2^8},
\fe
to distinguish it from the $\eta_1^8 \eta_3^8$ component ${\cal A}_{\delta\bar\tau\delta\tau\delta\bar\tau\delta\tau}$. (This scattering amplitude is more convenient to use than the previous one since the phase space integrals turn out to be simpler to evaluate, as will be seen below.) Note that to extract the original axion-dilaton amplitude from the new one, we must make the replacement $t\to s$ in \eqref{eq:ImA11}. In principle, there are also $t$- and $u$-channel cuts, which can be obtained in an analogous fashion.  

We now proceed to evaluate the contribution from the diagram in Figure~\ref{fig:unitaritycut} with the internal lines cut, as given by\footnote{The $\delta\tau\delta\bar\tau$ propagator at momentum $p^\mu$ is given by $\frac{i}{-p^2+i\epsilon}$.}
\ie
&-2\mathrm{Re}\left(\left.{\cal A}^{(1,1)}_{\delta\bar\tau\delta\bar\tau\delta\tau\delta\tau}\right|^{s}_{\text{LO}}\right)=\frac{1}{2} \int \frac{d^{10}q_1}{(2\pi)^{10}} \frac{d^{10}q_1}{(2\pi)^{10}}(-2\pi i)^2\delta(q_1^2)\delta(q_2^2) \\ &
~~~~~~~~~~~~~~~~~~~~\times \left[\left.{\cal A}^{(1,0)}_{\delta\bar\tau\delta\bar\tau\delta\tau\delta\tau}\right|_{\text{LO}}(p_1,p_2,q_1,q_2)\left.{\cal A}^{(0,1)}_{\delta\bar\tau\delta\bar\tau\delta\tau\delta\tau}\right|_{\text{LO}}(q_1,q_2,p_3,p_4)+(p_i\leftrightarrow q_i)\right],
\label{eq:cutdiag1}
\fe
where the subscript $\text{LO}$ reminds us that the D-instanton and anti-D-instanton contributions to the $R^4$ vertex appear at leading order in the open string loop expansion. The quantity ${\cal A}_{\delta\bar\tau\delta\bar\tau\delta\tau\delta\tau}^{(0,1)}$ in the above expression captures the anti-D-instanton contributions to the scattering amplitude, which at leading order agrees with that of the D-instanton, i.e.
\ie{}
e^{-2\pi i \tau}\left.{\cal A}_{\delta\bar\tau\delta\bar\tau\delta\tau\delta\tau}^{(0,1)}\right|_{\text{LO}}=e^{2\pi i \bar\tau}\left.{\cal A}_{\delta\bar\tau\delta\bar\tau\delta\tau\delta\tau}^{(1,0)}\right|_{\text{LO}}.
\fe
Using our results for these amplitudes as presented in \eqref{eq:LOamp}, we find
\ie
&2\mathrm{Im}\left(\left.\widetilde{{\cal A}}^{(1,1)}_{\delta\bar\tau\delta\bar\tau\delta\tau\delta\tau}\right|^{s}_{\text{LO}}\right)=
\\
~&~~~~~2^{-9}\pi^{{3 \over 2}} \kappa^7 e^{-4\pi\tau_2} s^8\int \frac{d^{9}\vec q_1}{(2\pi)^{9}} \frac{d^{9}\vec q_2}{(2\pi)^{9}}\frac{1}{|\vec q_1| |\vec q_2|}  \delta^{10}(p_1+p_2+q_1+q_2),
\label{eq:cutdiag2}
\fe
where we have stripped off an overall factor of $i(2\pi)^{10}\delta(P)$ from the RHS of \eqref{eq:cutdiag1}.To evaluate this integral, we find it helpful to work in the COM frame \eqref{eq:momframe}, where we find
\ie
&\int \frac{d^{9}\vec q_1}{(2\pi)^{9}} \frac{d^{9}\vec q_2}{(2\pi)^{9}}\frac{1}{|\vec q_1| |\vec q_2|}  \delta^{(10)}(p_1+p_2+q_1+q_2) \\
~~~~&=\frac{1}{(2\pi)^{18}}\int \frac{d^{9}\vec q_1}{|q_1|^2}\delta(2 E-2 |\vec q_1|)
\\
~~~~&=\frac{S_8}{2^{25}\pi^{18}} s^3,
\label{eq:phaseint}
\fe
In the first line of the above expression, we used the $\delta$-functions to set $\vec{q}_2 = -\vec{q}_1$, and in the second we substituted out $E$ using $s=4 E^2$. Plugging \eqref{eq:phaseint} into \eqref{eq:cutdiag2}, we immediately find
\ie
2\mathrm{Im}\left(\left.\widetilde{{\cal A}}^{(1,1)}_{\delta\bar\tau\delta\bar\tau\delta\tau\delta\tau}\right|^{s}_{\text{LO}}\right)=2^{-34} \pi^{-{33 \over 2}}S_8\kappa^7e^{-4\pi\tau_2} s^{11}\label{eq:cutdiag3},
\fe
which exactly reproduces the discontinuity found in the worldsheet calculation, after replacing $t\to s$ in \eqref{eq:ImA11}.  This is a non-trivial test of the interpretation of the discontinuity in the D-instanton/anti-D-instanton mediated scattering amplitude, and of unitarity of non-perturbative scattering amplitudes in type IIB string theory.

\section{Resolution of the ambiguity via string field theory}
\label{sftsec}

It has been emphasized recently that closed SFT is a rigorous framework for string perturbation theory. In particular, it provides a fully consistent regularization of possible divergences near the boundary of the moduli space in the on-shell worldsheet formulation of scattering amplitudes \cite{Sen:2019jpm}. The situation is more dramatic in D-instanton perturbation theory, where the open+closed SFT is necessary to fix ambiguities of the naive on-shell formalism \cite{Sen:2019jpm, Sen:2019qqg, Sen:2020cef, Sen:2020eck, Sen:2021qdk, Sen:2021tpp, Sen:2021jbr}. In this section, we describe the strategy for a string field theoretic computation of (\ref{nextleadmst}). The main purpose here is to explain why such a computation is free of divergences, and that it agrees with the naive on-shell computation apart from the constant term appearing on the RHS of (\ref{nextleadmst}). We also demonstrate how SFT unambiguously computes $A_0$ in (\ref{conerest}), which agrees with the soft relations discussed in Section~\ref{sec:cnnlo}. A similar problem in the context of $c=1$ string theory was examined in detail in \cite{Sen:2019qqg, Sen:2020cef, Sen:2020eck}. The extension of this analysis to type IIB string theory requires taking into account the additional ingredients of PCOs and vertical integration \cite{Sen:2015hia}.

\subsection{General strategy of the string field theoretic computation}

The amplitudes of interest are extracted from the path integral (\ref{ocpathint}) over the open string fields, while the closed string fields are taken to be on-shell. The open+closed SFT action $S_{oc}[\Psi_o,\Psi_c]$ consists of the kinetic terms for open string fields and the string vertices for open+closed string fields. Here we briefly recap the logic of how this action is constructed.

In SFT, the worldsheet moduli space is divided into domains that correspond to distinct Feynman diagrams, each of which is formed by gluing string vertices with propagators. To specify the string vertices further requires choosing local coordinates around each of the punctures on the worldsheet surface where the string fields are inserted, as well as the loci of PCO insertions. As the closed string field insertions are on-shell in D-instanton perturbation theory, one only needs to keep track of the local coordinates around the boundary points of the worldsheet where the open string fields are inserted.

A propagator amounts to the plumbing construction that identifies the neighborhoods of a pair of punctures, on either one or two connected surfaces in the vertex region. The moduli domain corresponding to a Feynman diagram with a single propagator, referred to as the `propagator region,' meets the vertex region along a codimension 1 wall in the moduli space where the propagator shrinks to zero length. It is important that at the wall separating the propagator and vertex regions, the choices of coordinate charts around the punctures on the worldsheet, as well as the PCO locations (possibly with vertical integration), agree with one another. This requirement amounts to the so-called geometric master equation, which ensures that the SFT action constructed from the string vertices is gauge invariant. The explicit construction of the relevant string vertices in the bosonic string case is described for example in section 4 of \cite{Sen:2020eck}. For the case of the superstring D-instanton, an example of this procedure will be illustrated in Section~\ref{FeynmanSec} for the disc with two closed string insertions.

The vertex region, by design, resides away from the boundary of the moduli space where the worldsheet surface degenerates, and thus the moduli integration over the vertex region is finite, modulo potential spurious singularities in the PCO locations which can be circumvented through the vertical integration prescription of \cite{Sen:2015hia}. A string vertex ${\cal V}[\Psi_o,\Psi_c]$ is a term in the SFT action obtained by integrating a worldsheet correlator with $\Psi_o,\Psi_c$ insertions over the corresponding vertex region of the moduli space. A propagator region, on the other hand, corresponds to a Feynman diagram in which the string vertices are connected through the string field propagator.

\subsection{BV gauge condition and massless open string modes}
\label{sec:bvgauge}

The open SFT path integral (\ref{ocpathint}) is defined subject to a choice of the BV gauge condition $L$.  It is important to make sure that this gauge condition is non-singular.  As was pointed out in \cite{Sen:2020cef}, the commonly adopted Siegel gauge, in which the open string fields are annihilated by $b_0$, is singular for D-instantons and must be modified. 

To see this, we begin by inspecting the kinetic terms in the action and their corresponding propagators.
The propagator for an open string field $\Psi_o$ in NS sector is obtained by inverting its kinetic term ${1\over2}\langle\Psi_o|Q_B|\Psi_o\rangle$. In the Siegel gauge, this propagator is ${b_0\over L_0}$, where $L_0$ is proportional to the ``mass squared" of the open string field. A similar propagator that involves picture changing can be derived in the R sector. The massive open string fields have well-defined propagators; they can be integrated out perturbatively, and their propagators appear in Feynman diagrams. The massless open string fields do not have well-defined propagators in the Siegel gauge, and require special treatment.

One class of massless open string fields correspond to the collective coordinates of the D-instanton. Namely, there are ten bosonic modes $\phi^\mu$ associated with the vertex operators $ce^{-\phi}\psi_\mu$, and sixteen fermionic modes $\theta_\A$ associated with the vertex operators $ce^{-{\phi\over2}}S^\alpha$. The integration over these modes amounts to the integration over the D-instanton (super) moduli space. However, there is a subtle but important difference between these open string fields and the deformation parameters for D-instanton boundary conditions in the worldsheet CFT, which we discuss in Sections~\ref{collectiveSec} and \ref{sec:thetaintl}.

There is another class of massless open string fields that cannot be interpreted as collective coordinates, of the form
\ie\label{ghostzeromodes}
\varkappa^1\beta_{-1/2}c_0c_1|-1\rangle+\zeta^2\beta_{-1/2}c_1|-1\rangle+\zeta_1\gamma_{-1/2}c_1|-1\rangle+\varkappa_2\gamma_{-1/2}c_0c_1|-1\rangle,
\fe
where $|-1\rangle$ stands for the state correspoding to the vertex operator $e^{-\phi}$. The coefficients $\varkappa^1$ and $\varkappa_2$ are Grassmann even, whereas $\zeta_1$ and $\zeta^2$ are Grassmann odd. They will be referred to as ``ghost zero modes." In Siegel gauge, $\varkappa^1$ and $\varkappa_2$ are set to zero, while the propagating modes $\zeta_1$ and $\zeta^2$ have vanishing kinetic term. The latter leads to vanishing path integral, which seems problematic. However, it was pointed out in \cite{Sen:2020cef} that this indicates not the breakdown of D-instanton perturbation theory, but rather that the Siegel gauge condition is singular.

Instead, \cite{Sen:2020cef} adopts a different BV gauge condition for the ghost zero mode sector, which we refer to as Sen gauge, defined by setting $\zeta_1$ and $\varkappa_2$ to zero. In this gauge, the propagator for $\varkappa^1$ is finite. The propagator for $\zeta^2$ is still ill-defined. Naively, consideration of ghost number symmetry indicates that $\zeta^2$ decouples from the effective action of massless open string fields, and if one simply omits the integration over $\zeta^2$, the open string path integral would appear to be well-defined. However, $\zeta^2$ in fact has the interpretation of the Faddeev-Popov ghost associated with fixing the $U(1)$ gauge symmetry on the D-instanton, and would be coupled to non-Abelian open string modes in the presence of other D-instantons. Integrating out $\zeta^2$ would then lead to a correction to the measure in the open string fields, which is present even in the absence of other D-instantons. This extremely subtle analysis is discussed in Section~ \ref{gzmSec}.

Following \cite{Sen:2020cef}, we will work in Siegel gauge for all sectors except for the ghost zero modes (\ref{ghostzeromodes}), where we impose Sen gauge instead. Among the open string fields, we denote by $\Psi_o^f$ the modes with finite propagators, namely $\varkappa^1$ in addition to all the massive modes. The remaining open string field components that require special treatment are the collective modes $\phi^\mu,\theta_\A$, and $\zeta^2$. This leads us to consider the path integral (\ref{ocpathint}) in the form
\ie\label{ocpathint2}
\left. e^{-\Gamma[\Psi_c]} \right|_{\rm D-inst} = e^{-{C\over g_s}} \int d^{10}\phi d^{16}\theta d\zeta^2 \, \exp\left( W_{f}[\phi^\mu,\theta_\A,\zeta^2,\Psi_c] \right),
\fe
where the effective action $W_f$, defined by
\ie\label{ocpathint3}
\exp\left( W_{f}[\phi^\mu,\theta_\A,\zeta^2,\Psi_c]\right) =\int D\Psi_o^f \exp\left( - S_{oc}[\phi^\mu,\theta_\A,\zeta^2,\Psi_o^f, \Psi_c]\right),
\fe
is calculated perturbatively by Feynman diagrams with well-defined propagators.

\subsection{Feynman diagrams for $W_f$}\label{FeynmanSec}

Of concern to the amplitudes considered in this paper are order $g_s$ terms in $W_f[\phi^\mu,\theta_\A,\psi^2,\Psi_c]$. The latter involve worldsheets of the following topologies: a disc with two closed string insertions, an annulus with one closed string insertion, a sphere with three holes, and a torus with one hole, as already described in (\ref{eq:wsdiagramsNLO}). Additional open string insertions on the boundary will be handled in Appendix~\ref{sec:directint}. Note that each worldsheet topology corresponds to several SFT Feynman diagrams. We emphasize that these SFT Feynman diagrams are manifestly finite.

Except for the disc with two closed string insertions, the remaining worldsheet topologies mentioned above involve either one closed string insertion or none. As the closed string field is taken to be that of an on-shell axion-dilaton state, the corresponding SFT Feynman diagrams evaluate to constants, independent of the closed string momentum, as a simple consequence of Lorentz invariance. In other words, nontrivial momentum dependence arises only from the disc with two closed string insertions, which we now discuss.

We begin by specifying the string vertex ${\cal V}_{D^2}^{1_{NSNS},1_{NS}}$ that corresponds to a disc with one NSNS dilaton and one NS open string field insertion. Representing the disc as the upper-half plane (UHP), we place the closed string at $z=i$ and open string at $z=0$. We must further specify a local coordinate $w$ on the chart that contains the open string insertion, with the transition map $w=w(z)$. Moreover, we need to place one PCO, say at some point $z=p$ on the UHP. These data completely determine ${\cal V}_{D^2}^{1_{NSNS},1_{NS}}$.

\begin{figure}[t!]
\centering
\begin{tikzpicture}
\draw[->, line width=1pt] (0,0) -- (10,0);
\node[right] at (10,0) {$y$};
\draw[line width=1pt] (0,0.15) -- (0,-0.15);
\node[below] at (0,-0.15) {$0$};
\draw[line width=1pt] (4,0.15) -- (4,-0.15);
\node[below] at (4,-0.15) {$y(q=1)$};
\draw[line width=1pt] (9,0.15) -- (9,-0.15);
\node[below] at (9,-0.15) {$1$};
\draw[line width=1pt]  (1,1) circle (0.6);
\draw (1,1) node[cross=4pt, very thick] {};
\draw[line width=1pt, dotted]  (1,1) circle (0.2);
\draw[line width=1pt]  (3,1) circle (0.6);
\draw (3,1) node[cross=4pt, very thick] {};
\draw[line width=1pt, dotted]  (3,1) circle (0.2);
\draw[line width=1pt, snake it] (1.6,1) to (2.4,1);
\draw (1.6,1) node[cross=4pt, very thick] {};
\draw (2.4,1) node[cross=4pt, very thick] {};
\draw[line width=1pt]  (6.5,1) circle (0.6);
\draw (6.25,1) node[cross=4pt, very thick] {};
\draw (6.75,1) node[cross=4pt, very thick] {};
\draw[line width=1pt, dotted]  (6.25,1) circle (0.2);
\draw[line width=1pt, dotted]  (6.75,1) circle (0.2);
\end{tikzpicture}
\caption{Feynman diagrams and the corresponding moduli domains for the disc with two dilaton insertions. Dotted circles around the vertex operators represent holomorphic PCOs. On the worldsheet represented as UHP, one of dilatons is inserted $i$ and the other at $iy$. In the moduli space parameterized by $y$, the domain $(0,y(q=1))$ is the propagator region, while $(y(q=1),1)$ is the vertex region. }
\label{fig:disk2pt}
\end{figure}
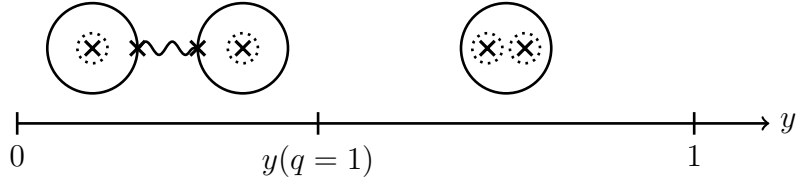

Next, we contract a pair of the vertices ${\cal V}_{D^2}^{1_{NSNS},1_{NS}}$ with an NS open string propagator to form a Feynman diagram. The corresponding family of worldsheet configurations is constructed by gluing together a pair of discs, parameterized by $z$ and $z'$ respectively, and identifying a pair of annuli around each of the open string insertions with the plumbing map $w(z)w(z')=-q$, for $0<q<1$. This results in a single disc with two dilaton insertions and two PCOs. Representing the latter as the UHP, up to a ${\rm PSL}(2,{\mathbb R})$ transformation, we can place one closed string at $z=i$ and another at $z=iy$ with $0<y<1$. The plumbing construction determines $y=y(q)$ as a function of $q$, as well as the locations of the two PCOs, $p_1(q)$ and $p_2(q)$. The family of configurations obtained via plumbing covers a domain $(0,y(q=1))$ of the moduli space parameterized by $y\in (0,1)$, as displayed in Figure \ref{fig:disk2pt}. 

There is a second Feynman diagram that is the string vertex ${\cal V}_{D^2}^{2_{NSNS}}$ itself. The corresponding worldsheet configurations are that of a disc with two NSNS closed string insertions as above, but with the modulus $y$ restricted to the domain $(y(q=1),1)$. In defining this string vertex, we must also choose the locations of the two PCOs such that they agree with $p_1(q=1)$ and $p_2(q=1)$ at the boundary of the vertex region, namely $y=y(q=1)$. If we simply place one holomorphic PCO on top of each of the closed string insertions both for ${\cal V}_{D^2}^{1_{NSNS},1_{NS}}$ and ${\cal V}_{D^2}^{2_{NSNS}}$, then ${\cal V}_{D^2}^{2_{NSNS}}$ is essentially what we computed in (\ref{NSNS2ptint}) where the lower cutoff for the integration over $y$ is taken to be $y(q=1)$, and the locations of the two PCOs are continuous at $y=y(q=1)$ so that there is no need for the vertical integration.

The first Feynman diagram of Figure \ref{fig:disk2pt} is defined not by moduli integration over the domain $(0,y(q=1))$, but by Feynman rules that involve summing over the exchange of all open string modes that have well-defined propagators, namely all massive modes together with the ghost zero mode $\varkappa^1$. It is convenient to consider a limit such that $y(q=1)\to 0$, say by rescaling $w(z)$. In this limit, the contribution from the exchange of massive open string modes vanishes, but the $\varkappa^1$-exchange diagram remains finite. The latter, however, is independent of the momenta of the dilatons.

A similar analysis applies to the diagrams involving RR axion insertions as well. If we choose to place one holomorphic PCO on top of each of NSNS closed string insertions, the final result for the disc with two $\delta\tau$'s is given by (\ref{eq:disc2ptSLZ}) with $\E=y(q=1)\rightarrow0$, up to a constant contribution from the $\varkappa^1$-exchange diagram.

In addition to the Feynman diagrams for $W_f$, we should also integrate over $\phi^\mu, \theta_\A,$ and $\zeta^2$ in (\ref{ocpathint2}). As shown in Appendix ~\ref{sec:directint}, they only contribute to the momentum-independent constant in (\ref{eq:Dinstamp}). In conclusion, the string field theoretic calculation of (\ref{eq:Dinstamp}) gives a result that can in principle differ from the naive on-shell computation only through the constant $C_1$.

\subsection{Derivation of the constant $A_0$ in MRV amplitudes}
\label{sec:sftmrv}

As discussed in Section~\ref{sec:higherptmrv}, the diagrammatics of the leading and next-to-leading order single D-instanton contributions to the $N$-point MRV amplitudes are essentially the same as those of the 4-point amplitudes. From the SFT perspective, this means that terms in the massless open string effective action $W_f$ that are relevant for the $N$-point MRV amplitudes are the same as those relevant for the 4-point amplitudes at the orders of interest. That is, the effect of integration in $(\phi^\mu, \theta_\A, \zeta^2)$ is similar to that analyzed in Appendix~\ref{sec:directint}. In particular, the momentum-dependent terms in (\ref{eq:higherptNLOexp}) are justified in the SFT formalism.

A first-principles derivation of the constant term $C_1^{(N)}$ in (\ref{eq:higherptNLOexp}), on the other hand, requires all of the SFT ingredients described in this section and Appendix~\ref{sec:directint}. The terms in $W_f$ that contribute to $C_1^{(N)}$ come from worldsheet topologies of either a disc with 1 or 2 punctures, a 1-punctured annulus, a 3-holed sphere, or a 1-holed torus. The diagrams that involve at most 1 puncture on each connected component, such as a 1-punctured annulus multiplied by $(N-1)$ copies of 1-punctured disc, contribute to $C_1^{(N)}$ at most linearly in $N$. The same conclusion holds for the corrections that arise from the integration in $(\phi^\mu, \theta_\A, \zeta^2)$: as discussed in Appendix~\ref{sec:directint}, these corrections effectively generate extra 1-point vertices for $\delta\tau$.

In contrast, the contribution to $C_1^{(N)}$ from the diagram of a 2-punctured disc multiplied by $(N-2)$ copies of 1-punctured disc is of the form ${N(N-1)\over2}A_0$, for some constant coefficient $A_0$, as in (\ref{conerest}). We now demonstrate that SFT produces a nonzero value of $A_0$ that is in precise agreement with the result (\ref{eq:softconst}), which we had determined indirectly in Section~\ref{sec:cnnlo} from the consideration of supersymmetry and soft limits.

As described in Section~\ref{FeynmanSec}, by placing one holomorphic PCO at each of the NSNS closed string insertions, the disc bulk 2-point string vertex ${\cal V}_{D^2}^{\delta\tau(p)\delta\tau(k)}$ evaluates to the same result as (\ref{eq:disc2ptSLZ}), where the moduli cutoff $\E$ is now replaced by a constant that depends on the SFT parameters. Importantly, in Sen gauge, SFT prescribes an additional contribution, namely the propagation of the ghost zero mode $\varkappa^1$ in the first Feynman diagram of Figure \ref{fig:disk2pt}. To evaluate this Feynman diagram, we shall compute the open-closed string vertex ${\cal V}_{D^2}^{\varkappa^1\delta\tau(p)}$, associated with the disc with a $\delta\tau(p)$ insertion in the bulk and a $\varkappa^1$ insertion on the boundary, as well as the propagator for $\varkappa^1$, as follows. 

The vertex operator associated with $\varkappa^1$ is 
\ie\label{kappavop}
e^{-2\phi}\partial\xi c\partial c. 
\fe
As both (\ref{kappavop}) and the vertex operator of $\delta\tau$ are (boundary and bulk, respectively) conformal primaries of zero weight, ${\cal V}_{D^2}^{\varkappa^1\delta\tau(p)}$ is independent of the choice of the local charts. Due to the unusual ghost structure of (\ref{kappavop}), only the dilaton $\delta\tau_2(p)$ contributes to ${\cal V}_{D^2}^{\varkappa^1\delta\tau(p)}$. The corresponding disc diagram has one holomorphic PCO insertion, which we place at the same point as the $\delta\tau_2$ insertion. The relevant part of the $\delta\tau_2$ vertex operator is
\ie
\lim_{y \to z} {\cal X}(y)V_{\delta\tau_2}(z,\bar{z})=-g_ce_{\mu\nu}\eta e^{\phi}\psi^\mu{\widetilde c}e^{-{\widetilde\phi}}{\widetilde\psi}^\nu e^{ip\cdot X}(z,\bar{z}) + \cdots,
\fe
where the omitted terms do not contribute to ${\cal V}_{D^2}^{\varkappa^1\delta\tau(p)}$. The resulting open-closed string vertex is
\ie
{\cal V}_{D^2}^{\varkappa^1\delta\tau(p)}&=-{i\over\sqrt{2}}g_c\varkappa^1\delta\tau(p)e_{\mu\nu}(p)\left\langle \eta e^{\phi}\psi^\mu{\widetilde c}e^{-{\widetilde\phi}}{\widetilde\psi}^\nu e^{ip\cdot X}(i,-i) ~e^{-2\phi}\partial\xi c\partial c(0)\right\rangle^{D^2}_{x^\mu=0}\\
&={i\over\sqrt{2}}g_c C_{D^2} \varkappa^1\delta\tau(p)\eta^{\mu\nu}e_{\mu\nu}(p).
\fe
Recall that in Sen gauge, the kinetic term for open string fields takes the form
\ie
{1\over2}\langle\Psi_o|Q_B|\Psi_o\rangle=-C_{D^2}\left(\varkappa^1\right)^2+ ({\rm massive~modes}),
\fe
from which we read off the propagator for $\varkappa^1$,\footnote{A highly nontrivial consistency check of the propagation of $\varkappa^1$ is seen in the computation of the effective potential for the bosonic open string collective modes in Appendix~\ref{sec:discopen4pt}.}
\ie
P_{\varkappa^1}={1\over2C_{D^2}}.
\fe
The first Feynman diagram of Figure \ref{fig:disk2pt} with a $\varkappa^1$-propagator then evaluates to
\ie
A^{D^2,~\varkappa^1}_{\delta\tau(p)\delta\tau(k)}={{\cal V}_{D^2}^{\varkappa^1\delta\tau(p)}\over\varkappa^1\delta\tau(p)}P_{\varkappa^1}{{\cal V}_{D^2}^{\varkappa^1\delta\tau(k)}\over\varkappa^1\delta\tau(k)}=-2\pi\kappa^2\tau_2.
\fe

In total, the SFT Feynman diagrams associated with the 2-punctured disc topology produces the amplitude
\ie\label{adtwosft}
A^{D^2,~\text{SFT}}_{\delta\tau(p)\delta\tau(k)}&=A^{D^2}_{\delta\tau(p)\delta\tau(k)}+A^{D^2,~\varkappa^1}_{\delta\tau(p)\delta\tau(k)}
\\
&={2\pi \kappa^2\tau_2} \bigg[4 - \sqrt{2}{ e_{\sigma\rho}(p) k^\sigma k^\rho + e_{\sigma\rho}(k) p^\sigma p^\rho \over p \cdot k}
\\
&~~~~~~~~~~~~~~ - 4\alpha'(p \cdot k) \left(\psi\left(1+{\alpha' p \cdot k \over 2}\right)+ \gamma + \log(4\epsilon) \right) \bigg].
\fe
Importantly, the first term in the square bracket on the RHS, namely the number ``4", differs from the number ``5" appearing in the naive on-shell formula (\ref{eq:disc2ptSLZ}). This is due to the contribution from the ghost zero mode via $A^{D^2,~\varkappa^1}_{\delta\tau(p)\delta\tau(k)}$. 

Multiplying (\ref{adtwosft}) by $(N-2)$ copies of disc 1-point amplitude (\ref{eq:disc1ptSLZ}), and summing over the permutations of $N$ insertions of $\delta\tau$, we obtain the following contribution to the $N$-point MRV ampitude,
\ie
R_{N,D^2}^{(1,0)}(s_{ij})\big|_{\text{NLO}} &=i^N 2^{2N-{37 \over 2}} \pi^{N-{37 \over 4}}\kappa^{N-{1\over2}}  e^{2\pi i \tau}\tau_2^{N-5}\left[\sum_{p=2}^\infty { \tau_2^{{p \over 2}} \over 2^{2p-3}} {\cal O}_{N}^{(p)} +64C^{(N)}_{1,D^2} \right],
\fe
with
\ie\label{cndtwo}
C^{(N)}_{1,D^2}=-{1\over128}\sum_{1\leq i<j\leq N}\left(4+{1\over2}\left(\ell_i\cdot p_j+\ell_j\cdot p_i\right)\right)=-{1\over128}\left(4\cdot{N(N-1)\over2}-{N\over2}\right),
\fe
where $\ell_i$ appears in the polarization tensor $e_{\mu\nu}(p_i)$ as in (\ref{dilatonPolarization}). In the above expressions, the subscript $D^2$ refers specifically to the contribution of the disc topology.

As already mentioned, while $C_1^{(N)}$ also receives contributions from other Feynman diagrams, as well as corrections that arise in the integration in $(\phi^\mu,\theta_\A,\zeta^2)$,  (\ref{cndtwo}) is solely responsible for the $N^2$ term in $C_1^{(N)}$. This unambiguously determines
\ie
A_0=-{1\over32}
\fe
in (\ref{conerest}), in perfect agreement with the expected result (\ref{eq:softconst})! 
We emphasize that the inclusion of the Feynman diagram where $\varkappa^1$ propagates was essential in obtaining the correct result, which would not have been possible in the naive on-shell approach.

\section{Discussion}

Let us recap the logic of the determination of D-instanton effects in this paper. Our working assumption has been that D-instanton contributions to closed string scattering amplitudes should be computed by the SFT of bulk closed strings and open strings on the D-instanton, to the extent in which the D-instanton effects are unambiguously defined. Certain aspects of the SFT formulation of D-instanton perturbation theory are analogous to that of the naive on-shell formulation. For instance, the integration over D-instanton moduli space is taken into account in SFT as part of the functional integration over open string fields. A priori, the on-shell approach to D-instanton amplitudes is subject to open string divergences and regularization ambiguities, which are resolved in SFT. On the other hand, the naive on-shell computation often captures the SFT result up to ambiguities of a simple form, which may be fixed by either indirect arguments or genuine SFT computations.

Indeed, most of our explicit computations, particularly concerning MRV amplitudes, are carried out in the naive on-shell formalism. We have argued that the naive on-shell results necessarily agree with the string field theoretic computation up to terms that can be fixed indirectly by considerations of spacetime supersymmetry and soft theorems concerning moduli of type IIB string vacua. This has allowed us to obtain unambiguous D-instanton contributions to MRV amplitudes.

Nonetheless, there is value in carrying out a first-principles SFT computation of the amplitudes considered in this paper, specifically the constant coefficients $A_0, A_1, A_2$ (\ref{conerest}) appearing in the NLO one-D-instanton contribution to the $N$-point MRV amplitude. This would serve to verify that the closed string vacuum preserves Poincar\'e supersymmetry, which is certainly expected for the Minkowskian vacuum of type IIB string theory at the non-perturbative level, but is not at all manifest in the string field theoretic formulaton of D-instanton perturbation theory. 

The SFT computation of $A_0$, performed in Section~\ref{sec:sftmrv}, agrees with the expectation of supersymmetry and soft limits, which is perhaps the most striking result of this paper. In particular, it includes a contribution from an SFT Feynman diagram that cannot be interpreted simply as a worldsheet correlator with vertex operator insertions, and serves as a highly nontrivial test of the SFT formalism. We hope to report on the SFT computation of $A_1$ and $A_2$, which involve more subtle aspects of D-instanton perturbation theory, in the near future.\footnote{Previously, a NLO SFT D-instanton computation was performed in the context of $c=1$ string theory in \cite{Sen:2020eck}. While there was highly non-trivial agreement with predictions of the duality with the matrix model, a mismatch in a single constant coefficient was found and remains unresolved at the moment. This constant of concern in \cite{Sen:2020eck} is analogous to $A_1$ in (\ref{conerest}).}

A few other comments are in order. From the on-shell perspective, the moduli space of multiple D-instantons typically admits singularities where new massless open string modes appear. This difficulty was evaded in the analyses of \cite{Balthazar:2019ypi, Balthazar:2022apu} in $c=1$ and type 0B string theory due to the simplicity of the moduli space of ZZ-instantons. In the setting of the D-instanton/anti-D-instanton pair analyzed in Section~\ref{sec:dantid}, it so happens that the contribution to the $D^6 R^4$ effective coupling comes from only the integration over the asymptotic region of the moduli space, which is well-defined.  Generally, the integral over multi-D-instanton moduli spaces is expected to be singular, and should be replaced by an integral over non-Abelian open string fields in the SFT framework. This was carried out to leading order in D-instanton perturbation theory in \cite{Sen:2021jbr}, and it would be very interesting to extend the analysis to subleading orders. 

A more intriguing question is to what extent the SFT framework captures D-instanton/anti-D-instanton contributions. In this case, there are regions of the moduli space where tachyonic open string modes appear, and the integration over open string fields may be ill-defined. In simple situations such as ZZ-instantons in $c=1$ string theory, this problem appears to be remedied by a Wick rotation prescription on the open string field integration contour \cite{Balthazar:2019rnh, Sen:2019qqg, Sen:2020cef}. In type IIB string theory, on the other hand, it is unclear whether the contribution from a D-instanton/anti-D-instanton pair to certain observables, such as the $D^8 R^4$ effective coupling, is even well-defined, as we do not understand the asymptotic properties of the perturbative contribution to the $D^8 R^4$ coupling nor how to separate it from the D-instanton effects. Furthermore, the understanding of open+closed SFT in the presence of both D-instantons and anti-D-instantons seems to be crucial in connecting D-instanton perturbation theory in different instanton charge sectors, which are thus far treated separately, via open string tachyon condensation \cite{Sen:1998sm}.

Finally, let us remark that the D-instanton amplitudes of the sort computed in this paper may provide useful input for the program of bootstrapping the non-perturbative string S-matrix \cite{Guerrieri:2021ivu, Gopakumar:2022kof}.

\section*{Acknowledgements}

We would like to thank Ashoke Sen for discussions. 
XY thanks Cargese Summer Institute, Aspen Center for Physics, Massachusetts Institute of Technology, VR and XY thank Kavli Institute for Theoretical Physics, for their hospitality during the course of this work.
This work is supported in part by a Simons Investigator Award from the Simons Foundation, by the Simons Collaboration Grant on the Non-Perturbative Bootstrap, and by DOE grants DE-SC0007870 and DE-SC0009924.

\appendix

\section{Worldsheet theory}
\label{app:wstheory}

\subsection{Spinor conventions}
\label{spinorsec}

In this appendix, we briefly summarize the spinor conventions used throughout the main text. The $SO(1,9)$ gamma matrices $\Gamma^\mu$ obey the Clifford algebra
\ie
\{\Gamma^\mu,\Gamma^\nu\} = 2\eta^{\mu\nu}.
\fe
In Lorentzian signature, we can work with purely real matrices with the off-diagonal form
\ie
(\Gamma^\mu)_A^{\:\:\:B} = \begin{pmatrix}
	0 & (\gamma^\mu)_{\alpha\beta} \\
	(\gamma^\mu)^{\alpha\beta} & 0
\end{pmatrix}, \\
\fe
where $A,B=1,\ldots,32$, and $\alpha,\beta=1,\ldots,16$. In this basis, the chirality matrix $\Gamma \equiv \Gamma^0 \cdots \Gamma^9$  reads
\ie{}
(\Gamma)_A^{\:\:\:B} = \begin{pmatrix}
	\delta^\alpha_\beta & 0 \\
	0 & - \delta^\alpha_\beta
\end{pmatrix}.
\fe 
The chiral matrices $(\gamma^\mu)_{\alpha\beta}$ and $(\gamma^\mu)^{\alpha\beta}$ satisfy the Clifford algebra
\ie{}
\{\gamma^\mu,\gamma^\nu\} = 2\eta^{\mu\nu},
\fe
and have the following properties:
\ie{}
(\gamma^\mu)^{\alpha\beta} = (\gamma^\mu)^{\beta\alpha}, ~~~
(\gamma^i)^{\alpha\beta} = (\gamma^i)_{\alpha\beta}, ~~~ (\gamma^0)_{\alpha\beta} = -(\gamma^0)^{\alpha\beta} = \delta_{\alpha\beta}.
\fe
Products of such matrices have a natural index structure, i.e. $(\gamma^\mu\gamma^\nu)_\alpha^{\:\:\:\beta} = (\gamma^\mu)_{\alpha\delta}(\gamma^\nu)^{\delta\beta}$ and  $(\gamma^\mu\gamma^\nu)^\alpha_{\:\:\:\beta} = (\gamma^\mu)^{\alpha\delta}(\gamma^\nu)_{\delta\beta}$.

In certain instances we will work with an explicit representation of the gamma matrices, where
\ie{}
(\gamma^0)^{\alpha\beta} = \begin{pmatrix} 1_{8 \times 8} & 0_{8 \times 8} \\ 0_{8 \times 8} & 1_{8 \times 8} \end{pmatrix}, \quad 
(\gamma^9)^{\alpha\beta} = \begin{pmatrix} 1_{8 \times 8} & 0_{8 \times 8} \\ 0_{8 \times 8} & -1_{8 \times 8} \end{pmatrix}.
\fe
In Euclidean signature, we perform a Wick rotation and replace $\gamma^0$ with $i\gamma^0$ in all of the pertinent formulas. Note that in this case the spinors are now in general complex-valued.

\subsection{NSR formalism}

In the NSR formalism, the worldsheet of the type II string is described by a 2d ${\cal N}=(1,1)$ unitary superconformal field theory with central charge $+15$ coupled to the $b,c,\beta,\gamma$ ghost system. The theory enjoys a global fermionic BRST symmetry generated by $Q_B+\widetilde{Q}_B$ with
\ie
Q_B = \oint_C {dz \over 2\pi i} &\bigg[c \left( T_m(z) - \frac12 \partial \phi(z)^2 - \partial^2 \phi(z) - \eta \partial \xi(z) \right) \\
&~+ \eta e^{\phi}G_m(z) + bc \partial c (z)  - \eta \partial \eta b e^{2\phi}(z) \bigg],
\fe
where $T_m$ and $G_m$ are the stress tensor and supercurrent of the matter CFT, respectively, and $C$ is a contour encircling the origin counterclockwise. The fields $\eta,\xi,\phi$ appear in the re-bosonized $\beta\gamma$ ghost system with dictionary
\ie{}
\gamma = \eta e^{\phi}, \quad \beta = e^{-\phi}\partial \xi,
\fe
where $e^{(2q-1)\phi}$ with $q \in \mathbb{Z}$ is taken to be Grassmann odd.
We also work with the picture changing operators (PCO) ${\cal X}$ and $\widetilde{{\cal X}}$ given by
\ie
{\cal X}(z) &\equiv \{Q_B,\xi(z)\} = c\partial \xi + e^{\phi}G_m - \partial \eta b e^{2\phi} - \partial(\eta b e^{2\phi}).
\fe
The type IIB theory is specified by the chiral GSO projection
\ie{}
(-1)^F = (-1)^{\widetilde{F}} = +1,
\fe
where $(-1)^F$ and $(-1)^{\widetilde{F}}$ are the holomorphic and antiholomorphic worldsheet fermion numbers.

\subsubsection*{Matter sector}

The matter CFT of the flat Minkowski background consists of ten free noncompact bosons $X^\mu$ and ten free Majorana fermions $\psi^\mu,\widetilde{\psi}^\mu$ with $\mu=0,\ldots,9$.  Its stress tensor and supercurrent are given by
\ie
T_m &= -\partial X_\mu \partial X^\mu - \frac12 \psi_\mu \partial \psi^\mu, \\
G_m &= i\sqrt{2}\psi_\mu \partial X^\mu,
\fe
where we are working in units of $\alpha'=1$ throughout the appendix. The elementary fields obey
\ie
X^\mu(z,\bar{z}) X^\nu(w,\bar{w}) &\sim -{1 \over 2} \eta^{\mu\nu}\log|z-w|^2, \quad
\psi^\mu(z)\psi^\nu(w) \sim {\eta^{\mu\nu} \over z-w},
\fe
where $\sim$ denotes the singular part of the operator product expansion (OPE), and $\eta^{\mu\nu} = \text{diag}(-1,+1,\cdots,+1)$ is the Minkowski metric in mostly plus signature. The free fermion theory also admits a pair of conjugate Weyl spinors $S^\alpha$ and $S_\alpha$. By pairing these with the R-sector operators in the $\phi$ CFT, we can form operators with well-defined GSO parity and worldsheet statistics. In particular, we shall use $e^{-\frac12\phi}S^\alpha$ and $e^{+\frac12\phi}S_\alpha$, which are both Grassmann odd and GSO even.  

\subsubsection*{Spacetime supersymmetry}

The type IIB string has an ${\cal N}=(2,0)$ target space supersymmetry generated by 32 chiral supercharges $\widehat{Q}^\alpha_{(-{1 \over 2})}$ and $\widehat{\widetilde{Q}}{}^\alpha_{(-{1 \over 2})}$ in the $(-{1 \over 2})$ picture given by 
\ie
\widehat{Q}^\alpha_{(-\frac12)} \equiv \oint_C {dz \over 2\pi i} e^{-\frac12\phi}S^\alpha(z),
\label{eq:susychargescontour1}
\fe
and similarly for $\widehat{\widetilde{Q}}{}^\alpha_{(-{1 \over 2})}$. We also work with the supercharges in the $(+{1 \over 2})$ picture as given by
\ie
\widehat{Q}^\alpha_{(+{1 \over 2})} \equiv -i (\gamma_\mu)^{\alpha\beta} \oint_C {dz \over 2\pi i} \partial X^\mu  e^{+\frac12\phi}S_\beta(z).
\label{eq:susychargescontour2}
\fe
Together they satisfy the super-Poincar\'e algebra
\ie{}
\{ \widehat{Q}^\alpha_{(-\frac12)}, \widehat{Q}^\beta_{(+{1 \over 2})} \} &= - \frac12 (\gamma_\mu)^{\alpha\beta} P_{(0)}^\mu, \\
\{ \widehat{\widetilde{Q}}{}^\alpha_{(-{1 \over 2})}, \widehat{\widetilde{Q}}{}^\beta_{(+\frac12)} \}&= - \frac12 (\gamma_\mu)^{\alpha\beta} \widetilde{P}_{(0)}^\mu,\\
\{\widehat{Q}^\alpha_{(\pm {1 \over 2})}, \widehat{\widetilde{Q}}{}_{(\mp \frac12)}^\beta\} &= 0,
\label{eq:SUSYalg}
\fe
where $P_{(0)}^\mu$ and $\widetilde{P}_{(0)}^\mu$ generate target space translations as
\ie{}
P_{(0)}^\mu \equiv {2i}\oint_C {dz \over 2\pi i} \partial X^\mu(z).
\fe
Note that in a noncompact target space $P_{(0)}^\mu$ and $\widetilde{P}_{(0)}^\mu$ act identically on momentum eigenstates as $P^\mu e^{ip \cdot X} = p^\mu e^{ip \cdot X}$.

\subsection{D-instanton boundary conditions}

In this part of the appendix, we outline our worldsheet conventions for the D-instanton. The single D-instanton solution is characterized by a family of BRST-invariant boundary conditions parametrized by ten bosonic moduli $x^\mu$ and sixteen fermionic moduli $\theta_\alpha$. For simplicity, we focus only on the boundary conditions with $\theta_\alpha = 0$,  since those with nonzero $\theta_\alpha$ can be described in terms of massless R-sector boundary deformations.  To be concrete, we take the worldsheet to be the disc $D^2$ represented by the upper half plane $\text{Im}(z) > 0$ with boundary parametrized by $u = \text{Re}(z)$. Compatibility with BRST invariance implies that the ghosts obey boundary conditions given by
\ie{}
b(u) &= \widetilde{b}(u), \quad c(u) = \widetilde{c}(u), \\ \eta(u) &= \widetilde{\eta}(u), \quad \xi(u) = \widetilde{\xi}(u), \quad \phi(u) = \widetilde{\phi}(u).
\label{eq:BCghosts}
\fe
In the matter sector, the D-instanton imposes Dirichlet boundary conditions in all ten (Euclidean) spacetime directions such that
\ie{}
\partial X^\mu(u) = -\bar{\partial}X^\mu(u), \quad \psi^\mu(u) = -\widetilde{\psi}^\mu(u).
\label{eq:BCmatter}
\fe
This implies that there is a family of boundary conditions with $\theta_\alpha = 0$ parametrized by $x^\mu \in \mathbb{R}^{10}$ given by
\ie{}
X^\mu(u) = x^\mu.
\label{eq:Dinstxmu}
\fe
Furthermore, consistency of the boundary conditions for $\psi^\mu$ in \eqref{eq:BCmatter} with the $\psi^\mu S^\alpha$ OPE \eqref{eq:psiSS} imply that the spin fields necessarily obey
\ie{}
e^{-\frac12\phi}S^\alpha(u) = is e^{-\frac12 \widetilde{\phi}}\widetilde{S}^\alpha(u), \quad\quad s = \pm 1,
\label{eq:BCspinfields}
\fe
where the factor of $i$ arises from Wick rotation to Euclidean signature. We shall take $s = +1$ to correspond with the D-instanton, and $s = -1$ with the anti-D-instanton. Note that this choice is merely a matter of convention in any background where the RR zero-form potential vanishes.

In general, the D-instanton boundary conditions $(x^\mu,\theta_\alpha)$ preserve half of the target space supersymmetries. The associated preserved supercharges are given by
\ie{}
\widehat{Q}^\alpha_{(\pm \frac12),+} =  \widehat{Q}^\alpha_{(\pm \frac12)} + i \widehat{\widetilde{Q}}{}^\alpha_{(\pm \frac12)}.
\fe
It is convenient to organize the remaining supercharges into 
\ie{}
\widehat{Q}^\alpha_{(\pm \frac12),-} =  \widehat{Q}^\alpha_{(\pm \frac12)} - i \widehat{\widetilde{Q}}{}^\alpha_{(\pm \frac12)}.
\fe
With respect to this basis, the super-Poincar\'e algebra takes the form
\ie{}
&\left\{\widehat{Q}^\alpha_{(\pm \frac12),+}, \widehat{Q}^\alpha_{(\mp \frac12),-} \right\} = -(\gamma_\mu)^{\alpha\beta}(P_{(0)}^\mu + \widetilde{P}_{(0)}^\mu) , \\
&\left\{\widehat{Q}^\alpha_{(\pm \frac12),+}, \widehat{Q}^\alpha_{(\pm \frac12),+} \right\} = 0, \\
&\left\{\widehat{Q}^\alpha_{(\pm \frac12),-}, \widehat{Q}^\alpha_{(\pm \frac12),-} \right\} = 0.
\fe

\subsection{Supergraviton vertex operators}
\label{app:sugramultiplet}

In this appendix, we review the vertex operators appearing at mass level zero in the BRST cohomology, which correspond to the on-shell states of the supergraviton multiplet. 

\subsubsection*{NSNS sector}

In the NSNS sector, we work with the BRST representative 
\ie{}\label{NSNS(-1,-1)}
V^{(-1,-1)}_{\text{NSNS}} = g_c  c \widetilde{c} \: \varepsilon_{\mu\nu}(p)e^{-\phi}\psi^\mu e^{-\widetilde{\phi}} \widetilde{\psi}^\nu e^{ip \cdot X}, \quad p^2 = 0,
\fe
where $g_c$ is the closed string coupling. The polarization tensor $\varepsilon_{\mu\nu}$ obeys
\ie{}
\varepsilon_{\mu\nu}(p) p^\mu = \varepsilon_{\mu\nu}(p) p^\nu = 0, \quad \varepsilon_{\mu\nu}(p)\varepsilon^{\mu\nu}(p) = 1,
\fe
where the first condition ensures that the vertex operator is BRST-closed, and the second reflects our choice of normalization for the one-particle states. Decomposing $\varepsilon_{\mu\nu}$ into irreducible representations of the $SO(8)$ little group yields a symmetric, traceless tensor $h_{\mu\nu}$ (the graviton), an antisymmetric tensor $b_{\mu\nu}$ (the Kalb-Ramond B-field), and a scalar $e_{\mu\nu}$ (the dilaton $\delta \tau_2$); together, these contribute $35+28+1 = 64$ bosonic states to the supergraviton multiplet. 

Of particular interest to this work is the dilaton, whose vertex operator we denote by $V_{\delta\tau_2(p)}$ with $\varepsilon_{\mu\nu} =e_{\mu\nu}$ in \eqref{NSNS(-1,-1)}. By introducing a lightlike vector $\ell$,  its polarization tensor can be written explicitly as\footnote{Being a scalar, the dilaton should have a polarization tensor with only a single degree of freedom. The freedom to choose $\ell^\mu$, which superficially contradicts this statement, is merely an artifact from our choice of BRST representative. In particular, $\ell$ is expected to drop out of any on-shell amplitudes calculated using \eqref{dilatonPolarization}. It can also be removed once and for all by adding to \eqref{NSNS(-1,-1)} a BRST-exact term
\ie{}
g_c c \tilde{c} \left( (\ell_\mu p_\nu + \ell_\nu p_\mu)e^{-\phi}\psi^\mu e^{-\widetilde{\phi}} \widetilde{\psi}^\nu  +   (e^{-2\phi} \partial \xi \tilde{\eta} +  e^{-2\tilde{\phi}} \bar{\partial} \tilde{\xi}\eta) \right) e^{ip\cdot X} ,
\fe
in which case the scalar nature of the dilaton is manifest in the vertex operator
\ie{}
g_c  c \widetilde{c} \left(e^{-\phi}\psi^\mu e^{-\widetilde{\phi}} \widetilde{\psi}_\mu   + e^{-2\phi} \partial \xi \widetilde{\eta} +  e^{-2\widetilde{\phi}} \bar{\partial} \widetilde{\xi} \eta \right) e^{ip \cdot X} .
\fe
Both this vertex operator and \eqref{NSNS(-1,-1)} should produce identical on-shell amplitudes, although intermediate diagrams will in general differ from one another. In practice, we shall stick with the latter since it is easier to use in our computations. 
}
\ie{}
e_{\mu\nu}(p) = {1 \over \sqrt{8}}\left( \eta_{\mu\nu} - \ell_\mu p_\nu - \ell_\nu p_\mu \right) , \quad \ell \cdot p = 1.
\label{dilatonPolarization}
\fe

For the amplitudes under consideration we will also need several picture-raised variants of the NSNS vertex operator. The picture $(0,-1)$ vertex operator is given by taking a holomorphic PCO coincident with \eqref{NSNS(-1,-1)}
\ie{} \label{NSNS(0,-1)}
V_{\text{NSNS}}^{(0,-1)}(z,\bar{z}) &\equiv \lim_{y \to z} {\cal X}(y)V^{(-1,-1)}_{\text{NSNS}}(z,\bar{z}) \\
&=-\sqrt{2}g_c  \varepsilon_{\mu\nu}c \widetilde{c} \left( i \partial X^\mu + \frac12 p \cdot \psi \psi^\mu \right) e^{-{\widetilde{\phi}}}\widetilde{\psi}^\nu e^{ip \cdot X} + \cdots ,
\fe
Similarly, the picture $(0,0)$ vertex operator is given by taking a pair of holomorphic and antiholomorphic PCOs coincident
\ie{} \label{NSNS(0,0)}
V^{(0,0)}_{\text{NSNS}} &\equiv \lim_{\bar{y} \to \bar{z}} \widetilde{\cal X}(\bar{y})V_{\text{NS}}^{(0,-1)}(z,\bar{z})\\
&= 2g_c  \varepsilon_{\mu\nu}c \widetilde{c}   \bigg( i \partial X^\mu + \frac12 p \cdot \psi \psi^\mu \bigg) \bigg( i \bar{\partial} X^\nu + \frac12 p \cdot \widetilde{\psi} \widetilde{\psi}^\nu \bigg) e^{ip \cdot X} + \cdots .
\fe
The operators appearing in $\cdots$ for both \eqref{NSNS(0,-1)} and \eqref{NSNS(0,0)} involve $\eta$ and $\widetilde{\eta}$ and do not contribute to the amplitudes in the main text.

\subsubsection*{RR sector}

In the RR sector, there is a unique choice of BRST representative given by
\ie\label{RR(-1/2,-1/2)}
V^{(-\frac12,-\frac12)}_{\text{RR}} = g_c  c \tilde{c} \: f_{\alpha\beta}(p) e^{-\frac12\phi} S^\alpha e^{-\frac12\widetilde{\phi}}\widetilde{S}^\beta e^{ip \cdot X}, \quad p^2 = 0,
\fe
where the polarization tensor $f_{\alpha\beta}$ obeys
\ie{}
f_{\alpha\beta}(p) \slashed{p}^{\beta\gamma} = f_{\alpha\beta} (p)\slashed{p}^{\alpha\gamma} = 0, \quad\quad \slashed{p} \equiv p_\mu \gamma^\mu .
\label{eq:RRtransversality}
\fe
As a bispinor, $f_{\alpha\beta}$ decomposes into odd-rank forms according to
\ie{}
f_{\alpha\beta}^{(r)}(p) = {i \over r!} F^{(r)}_{\mu_1 \cdots \mu_r}(p) (\gamma^{\mu_1 \cdots \mu_r})_{\alpha\beta},
\fe
where $F^{(r)}$ is related to $F^{(10-r)}$ by Hodge duality, with the 5-form $F^{(5)}$ naturally being self-dual.  Working with vector indices, the transversality constraints \eqref{eq:RRtransversality} reduce to the Bianci identities 
\ie{}
p_{[\nu} F^{(r)}_{\mu_1 \cdots \mu_r]}(p) = 0.
\fe
The coefficents $F^{(r)}_{\mu_1 \cdots \mu_r}$ thus serve as field strengths associated to the RR gauge potentials, which consist of a 4-form $c_{\mu\nu\sigma\rho}$, a 2-form $c_{\mu\nu}$, and a scalar $\delta\tau_1$ (the axion), which together comprise the other $35+28+1=64$ bosonic states of the supergraviton multiplet. 

We are primarily interested in the axion, whose vertex operator we denote by $V_{\delta\tau_1(p)}$. Its polarization tensor $f^{(0)}$ is fixed by Lorentz invariance to be
\ie{}
f^{(0)}_{\alpha\beta}(p) = -{\slashed{p}_{\alpha\beta} \over 4\sqrt{2}},
\label{eq:axionpolarization}
\fe
where the overall factor has been chosen such that the axion and dilaton 1-particle states share the same normalization conventions.

\subsubsection*{NSR and RNS sectors}

Although the NSR and RNS sectors do not play a direct role in our computations, we shall include them here for the sake of completeness. We can write the level zero BRST representatives as
\ie{}\label{NSR(-1,-1)}
V^{(-1,-\frac12)\oplus(-\frac12,-1)}_{\text{NSR}}= g_c  c \widetilde{c} \left[u_{1;\mu\alpha}(p)e^{-\phi}\psi^\mu e^{-\frac12 \widetilde{\phi}} \widetilde{S}^\alpha + u_{2;\mu\alpha}(p)e^{-\frac12 \phi}S^\alpha e^{-\widetilde{\phi}} \widetilde{\psi}^\mu \right] e^{ip \cdot X},
\fe
where $p^2 = 0$, and where the polarization tensors $u_i$ satisfy
\ie{}
p^\mu u_{i;\mu}(p) = u_{i;\mu}(p)\slashed{p} = 0.
\label{eq:RNStransverse}
\fe
They admit a decomposition into $SO(8)$ representations given by
\ie{}
u_{i;\mu\alpha}(p) = (\gamma_\mu)_{\alpha\beta} \lambda_i^\beta(p) + \psi_{i;\mu\alpha}(p) , \quad \slashed{\psi}_i(p) = 0,
\fe
which obey \eqref{eq:RNStransverse}
\ie{}
\slashed{p}\lambda_i(p) =0, \\
\gamma^{\mu\nu\rho}p_\nu \psi_{i\rho}(p) = 0.
\fe
These are the momentum-space Dirac and Rarita-Schwinger equations, respectively. In other words, the RNS and NSR states comprise two massless spin $1/2$ fermions $\lambda_i$ (the dilatinos), and two massless spin $3/2$ fermions $\psi_i$ (the gravitinos), which together constitute the $2 \times (8 + 56) = 128$ fermions of the supergraviton multiplet.

\subsection{OPEs and correlation functions}
\label{opesec}

In this part of the appendix, we collect various OPEs relevant for the disc amplitudes considered in this work. For instance, we have 
\ie{}
e^{-\phi}\psi^\mu(z) e^{-\frac12\phi}S^\alpha(w) &\sim -{(\gamma^\mu)^{\alpha\beta} \over \sqrt{2}(z-w) } e^{-\frac12\phi} S_{\beta}(w), \\
e^{-\phi}\psi^\mu(z) e^{-\frac12\phi} S_{\alpha}(w) &\sim -{(\gamma^\mu)_{\alpha\beta} \over \sqrt{2}(z-w) } e^{-\frac12\phi}S^\beta(w),  \\
e^{-\frac12\phi}S^\alpha(z)e^{-\frac12\phi}S^\beta(w) &\sim {(\gamma_\mu)^{\alpha\beta} \over \sqrt{2}(z-w) } e^{-\phi}\psi^\mu(w), \\
e^{-\frac32\phi}S_{\alpha}(z)e^{-\frac12\phi}S^\beta(w) &\sim {\delta^\beta_\alpha \over (z-w)^2} e^{-2\phi}(w) - {3 \over 2(z-w)}e^{-2\phi}\partial \phi(w) \\&-{(\gamma_{\mu\nu})_\alpha^{\:\:\:\beta}  \over 2(z-w)}e^{-2\phi}\psi^\mu\psi^\nu(w), \\
\psi^\mu\psi^\nu(z)e^{-\phi}\psi^\rho(w) &\sim {\eta^{\nu\rho}\delta^\mu_\sigma - \eta^{\mu\rho}\delta^\nu_\sigma  \over z-w} e^{-\phi}\psi^\sigma(w),\\
\psi^\mu\psi^\nu(z) e^{-\frac12\phi} S^\alpha(w) &\sim -\frac12{(\gamma^{\mu\nu})^\alpha_{\:\:\:\beta} \over z-w}e^{-\frac12\phi}S^\beta(w).
\fe
These OPEs subsequently determine various correlators, such as the 3-point function
\ie{}
\label{eq:psiSS}
\left\langle e^{-\phi}\psi^\mu(z_1) e^{-\frac12\phi}S^\alpha(z_2) e^{-\frac12\phi}S^\beta(z_3) \right\rangle^{S^2}_{\mathbf{chiral}} = -{(\gamma^\mu)^{\alpha\beta} \over \sqrt{2}z_{12}z_{13}z_{23}},
\fe
as well as the 4-point function
\ie{} 
&\left\langle e^{-\frac12\phi}S^{\alpha_1}(z_1) e^{-\frac12\phi}S^{\alpha_2}(z_2)e^{-\frac12\phi}S^{\alpha_3}(z_3) e^{-\frac12\phi}S^{\alpha_4}(z_4)\right\rangle^{S^2}_{\mathbf{chiral}} \\
&\qquad=  -{(\gamma_\mu)^{\alpha_1\alpha_2}(\gamma^\mu)^{\alpha_3\alpha_4} \over 2z_{12} z_{23}z_{24}z_{34}}+{(\gamma_\mu)^{\alpha_1\alpha_3}(\gamma^\mu)^{\alpha_2\alpha_4} \over 2z_{13} z_{32}z_{34}z_{24}}-{(\gamma_\mu)^{\alpha_1\alpha_4}(\gamma^\mu)^{\alpha_2\alpha_3} \over 2z_{14} z_{42}z_{43}z_{23}}.
\label{eq:SSSS}
\fe

\section{Modular forms and $SL(2,\bZ)$ covariance}
\label{sec:modforms}

In this appendix, we collect various results for the vertices appearing in the low-energy expansion of the quantum effective action of type IIB string theory. Their coefficients transform under the $SL(2,\bZ)$ duality group as non-holomorphic forms of weight $(w,\tilde{w})$, i.e.
\ie{}
f^{(w,\tilde{w})}\left( {a\tau+b \over c\tau+d} \right) = (c\tau+d)^w(c\bar{\tau}+d)^{\tilde{w}} f^{(w,\tilde{w})}(\tau), \quad a,b,c,d \in \mathbb{Z}, \quad ad-bc =1.
\fe
We shall also need the (holomorphic) modular covariant derivative on the upper-half $\tau$-plane, 
\ie{}
{\cal D}_w = i \tau_2 \partial_\tau + {w \over 2},
\fe
which takes non-holomorphic forms of weight $(w,\tilde{w})$ to those of weight $(w+1,\tilde{w}-1)$.

We first consider the $\frac12$-BPS $R^4$ and $\frac14$-BPS $D^4 R^4$ vertices. Their coefficients take the form of modular functions, i.e. non-holomorphic modular forms of weight $(0,0)$, which satisfy a homogenous Laplace equation on the upper half plane
\ie{}
\left(\tau_2^2 \partial_\tau \partial_{\bar\tau} - \frac14 s(s-1) \right) f(\tau,\bar{\tau}) = 0, \quad s \in \mathbb{C},
\fe
subject to the boundary condition $f(\tau,\bar{\tau}) = {\cal O}(\tau_2^p)$ for $p \in \mathbb{R}$. Its solution is given by the non-holomorphic Eisenstein series
\ie
E_s(\tau,\bar\tau) &= \sum_{(m,n) \neq (0,0)} {\tau_2^{s} \over |m+n\tau|^{2s}} \\
&= 2\zeta(2s)\tau_2^s + 2\sqrt{\pi}{\Gamma(s-1/2)\zeta(2s-1) \over \Gamma(s)}\tau_2^{1-s} \\&~~~+ {4\pi^s \over \Gamma(s)} \tau_2^{1/2} \sum_{n \neq 0} |n|^{s-\frac12} \sigma_{1-2s}(|n|) e^{2\pi i n\tau_1}K_{s-\frac12}(2\pi|n|\tau_2),
\fe
where $K_\alpha(z)$ is the K-Bessel function, and $\sigma_z(n)$ is the divisor function
\ie{}
\sigma_z(n) = \sum_{d|n}d^{2z}, \quad \sigma_{-z}(n) = n^{-z} \sigma_z(n).
\fe
In the weak-coupling limit, $E_s$ admits an expansion in $\tau_2^{-1}$ given by
\ie{}
E_s(\tau,\bar{\tau}) &=  2\zeta(2s)\tau_2^s + 2\sqrt{\pi}{\Gamma(s-1/2)\zeta(2s-1) \over \Gamma(s)}\tau_2^{1-s} \\
&+ (e^{2\pi i \tau}+e^{-2\pi i \tau}) \left( {2\pi^s \over \Gamma(s)} + {s(s-1) \over 2\Gamma(s)}\tau_2^{-1} + {\cal O}(\tau_2^{-2}) \right) + {\cal O}(e^{-4\pi \tau_2}).
\label{eq:eisensteinexp}
\fe
The $R^4$ and $D^4 R^4$ coefficients $f_0$ and $f_4$ have $s=3/2$ and $s=5/2$, and under our conventions are given by 
\ie{}
f_0 = {1 \over 2^6}E_{\frac32}(\tau,\bar{\tau}), \quad 
f_4 =  {1 \over 2^{11}}E_{\frac52}(\tau,\bar{\tau}).
\fe

The coefficient $f_6$ of the $\frac18$-BPS $D^6 R^4$ interaction meanwhile satisfies an inhomogeneous Laplace equation of the form
\ie{}
\left(\tau_2^2 \partial_\tau \partial_{\bar\tau} - 12 \right) {\cal E}_3(\tau,\bar{\tau}) = -E_{\frac32}(\tau,\bar{\tau})^2,
\fe
together with the weak-coupling boundary condition $f_6(\tau,\bar\tau) = {\cal O}(\tau_2^3)$ as $\tau_2 \to \infty$. Its solution takes the form of a modular function that can be written in the weak-coupling limit ($\tau_2 \to \infty$) as \cite{Green:2014yxa}
\ie{}
{\cal E}_3(\tau,\bar{\tau}) &= {2 \zeta(3)^2 \over 3} \tau_2^3 + {4\zeta(2) \zeta(3) \over 3}\tau_2 + {8\zeta(2)^2 \over 5} \tau_2^{-1} + {4 \zeta(6) \over 28}\tau_2^{-3} \\
&+ (e^{2\pi i \tau}+e^{-2\pi i \tau_2})\left(8\zeta(3)\tau_2^{\frac12} + {\cal O}(1) \right) - e^{-4\pi \tau_2} \left(2 \tau_2^{-2} + {\cal O}(\tau_2^{-3}) \right) \\
&+ {\cal O}(e^{-6\pi\tau_2}) .
\label{eq:f6weak}
\fe
Under our conventions, the $D^6 R^4$ coefficient appearing in the main text is given by
\ie{}
f_6 = {1 \over 2^{12}} {\cal E}_3(\tau,\bar{\tau}).
\fe

In our discussion on higher-point amplitudes, we will also need the coefficients of the $N$-point MRV vertices, which transform as weight $(N-4,4-N)$ non-holomorphic modular forms. To describe such forms, we introduce the generalized Eisenstein series $E^{(w)}_s$ as given by
\ie{}
E^{(w)}_{s}(\tau,\bar{\tau})
&= \sum_{(m,n) \neq (0,0)} \left( {m+n \bar{\tau} \over m+n \tau} \right)^w {\tau_2^s \over |m+n\tau|^{2s}} \\
&= {2^w \Gamma(s) \over \Gamma(s+w)} {\cal D}_{w-1} \cdots {\cal D}_0 E_s(\tau,\bar{\tau}) .
\fe
Following \cite{Green:2019rhz}, the $\delta\tau^2 R^4$  and $\delta \tau^2 D^4 R^4$ coefficients are proportional to
\ie{}
r_0^{(6)} = {15 \over 2^8} E_{\frac32}^{(2)}(\tau,\bar{\tau}), \quad
r_4^{(6)} ={35 \over 2^{13}}E_{\frac52}^{(2)}(\tau,\bar{\tau}),
\label{eq:d2r4MRV}
\fe
where the overall choice of normalization does not factor into our analysis in the main text.

It was shown in \cite{Green:2019rhz} that the coefficients multiplying $\delta\tau^2 D_i^6 R^4$, with the associated kinematic structures ${\cal O}^{(3)}_{6,i}$ in \eqref{eq:kinstruct1} and \eqref{eq:kinstruct2}, satisfy inhomogeneous Laplace equations of the form
\ie{}
\left(\tau_2^2 \partial_\tau \partial_{\bar\tau} - 10 \right){\cal E}_{2,1} &= -{15 \over 2} \left(E^{(0)}_{\frac32}E^{(2)}_{\frac32} + \frac35 \left(E^{(1)}_{\frac32}\right)^2 \right),  \\
\left(\tau_2^2 \partial_\tau \partial_{\bar\tau} - 10 \right){\cal E}_{2,2} &= -{5c_1 \over 2} \left(E^{(0)}_{\frac32}E^{(2)}_{\frac32} -  \left(E^{(1)}_{\frac32}\right)^2 \right) ,
\fe
where $c_1$ is not determined by supersymmetry and $SL(2,\bZ)$-covariance alone. Their solutions can be written in terms of the $D^6 R^4$ coefficient as
\ie{}
{\cal E}_{2,1}  &= 4 {\cal D}_1 {\cal D}_0 {\cal E}_{3}, \\
{\cal E}_{2,2}  &= {c_1 \over 5}\left({\cal E}_{2,1} - {1 \over 2} \left(E^{(1)}_{\frac32} \right)^2\right).
\fe
In our conventions, the  $\delta\tau^2 D_i^6 R^4$ coefficients $r^{(6)}_{6,i}$ are proportional to
\ie{}
r^{(6)}_{6,1} = {1 \over 2^{12}} {\cal E}_{2,1}(\tau,\bar{\tau}), \quad r^{(6)}_{6,2} = {1 \over 2^{12}} {\cal E}_{2,2}(\tau,\bar{\tau}),
\label{eq:d6r4MRV}
\fe
where only the relative factors between these and the lower-order coefficients is of relevance.

\section{Integration over $\zeta^2,\phi^\mu$, and $\theta_\A$}\label{sec:directint}

In this appendix, we discuss how to directly integrate over $\zeta^2,\phi^\mu$, and $\theta_\A$ in (\ref{ocpathint2}).

\subsection{Integration over $\zeta^2$}\label{gzmSec}

The Grassmann-odd ghost zero mode $\zeta^2$, as already mentioned, has the interpretation as the Faddeev-Popov ghost associated with gauge fixing the $U(1)$ symmetry on the D-instanton. However, there are no charged open strings on a single D-instanton, and so $\zeta^2$ is absent in the effective action $W_f$ (\ref{ocpathint3}). Formally,
the integration over $\zeta^2$ in (\ref{ocpathint2}) gives zero, and one may be tempted to simply drop the $\zeta^2$-integral, but this leaves an ambiguous (possibly background field dependent) normalization. 

A more careful treatment that fixes the normalization requires introducing a spectator D-instanton \cite{Sen:2020cef} (which we refer to as the D$^s$-instanton), so that there are charged open string modes with respect to the $U(1)$ gauge symmetry on the original D-instanton. One can then calculate the $\zeta^2$-integral, which is now nonzero, and move the spectator D$^s$-instanton to infinity in the end.\footnote{In \cite{Sen:2020cef}, the integration over $\zeta^2$ is interpreted as the division by the volume of the $U(1)$ gauge group.}

Indeed, at leading order in $g_s$, $W_f$ now contains couplings between $\zeta^2$, D-D$^s$ open string fields $\chi$, and D$^s$-D open string fields $\chi^*$, of the schematic form
\ie
\zeta^2 \chi\chi^*,
\fe
which is computed by a disc diagram with $\chi, \chi^*, \zeta^2$ insertions. The integration over $\zeta^2$ is now no longer singular and gives the factor $\chi\chi^*$.

At the first subleading order in $g_s$, there is a contribution to $W_f$ from the analogous disc diagram with an extra closed string field insertion. Such a coupling that is relevant to the amplitude considered in this paper is of the form
\ie\label{gsydisc}
g_s Y \zeta^2 \chi\chi^* \Big[ \delta\tau(x) + \cdots \Big],
\fe
where $Y$ is a constant that depends on the string field theoretic parameters that enter into the definition of the string vertices, $\delta\tau(x)$ is the axion-dilaton field at the D-instanton location $x^\mu$, and $\cdots$ represents terms involving $\theta_\A$ and other components of the supergraviton multiplet. As we will discuss in Section~\ref{sec:thetaintl}, to leading order in $g_s$, the insertion of a fermionic open string field $\theta_\A$ on the boundary can be replaced by that of a supercharge ${\widehat Q}^\alpha_{(\pm \frac12),-}$. In the axion-dilaton background, the linear combination of $\delta\tau$ and $\delta\bar\tau$ that appear in (\ref{gsydisc}) is determined by the nonlinearly realized super-Poincar\'e symmetry to be $\left(e^{i\theta_\A {\cQ}_-^\A}\delta\tau\right)(x)$, which contains a term of order $\theta^8$ that involves $\delta\bar\tau$. Here, $\cQ_-^\A$ is the supercharge acting as a raising operator on the spacetime fields, which can be regarded as a dual of the supercharge ${\widehat Q}_-^\A$ acting as a lowering operator on the one-particle states.
We emphasize that the computation of $Y$ is nontrivial and requires considering contributions from multiple Feynman diagrams.

Upon integration over $\zeta^2$, this corrects (\ref{ocpathint2}) by the factor
\ie\label{gsytheta}
1 + g_s Y \left( e^{i\theta_\A Q_-^\A}\delta\tau \right) (x).
\fe
In the end, it leads to a correction to the D-instanton amplitude ${\cal A}_{2\to 2}^{(1,0)}$ by the factor
\ie
1 + 4 g_s Y ({\cal A}_{\delta\tau}^{D^2})^{-1},
\fe
which only contributes to the momentum-independent constant in (\ref{eq:Dinstamp}).

\subsection{Integration over $\phi^\mu$}\label{collectiveSec}

\begin{figure}[h!]
\centering
\begin{tikzpicture}
\draw (0,0) node { $\displaystyle\int d\Phi_o^m  \left[\left( \vphantom{\frac{1}{1}} \right.\right.$ };
\filldraw[fill=black!30, thick]  (1.4,0) circle (0.4);
\draw (1.4,0) node[crossbl=2pt, thick] {};
\draw (2.2,0) node {$+$};
\filldraw[fill=black!30, thick]  (3,0) circle (0.4);
\draw (3,0) node[crossbl=2pt, thick] {};
\draw (3.4,0) node[bcross=2pt, thick] {};
\draw (3.8,0) node {$+$};
\filldraw[fill=black!30, thick]  (4.6,0) circle (0.4);
\draw (4.6,0) node[crossbl=2pt, thick] {};
\draw (4.2,0) node[bcross=2pt, thick] {};
\draw (5,0) node[bcross=2pt, thick] {};
\draw (5.4,0) node {$+$};
\filldraw[fill=black!30, thick]  (6.2,0) circle (0.4);
\draw (6.2,0) node[crossbl=2pt, thick] {};
\draw (5.8,0) node[bcross=2pt, thick] {};
\draw (6.2,0.4) node[bcross=2pt, thick] {};
\draw (6.6,0) node[bcross=2pt, thick] {};
\draw (7,0) node {$+$};
\filldraw[fill=black!30, thick]  (7.8,0) circle (0.4);
\draw (7.8,0) node[crossbl=2pt, thick] {};
\draw (7.4,0) node[bcross=2pt, thick] {};
\draw (7.8,0.4) node[bcross=2pt, thick] {};
\draw (7.8,-0.4) node[bcross=2pt, thick] {};
\draw (8.2,0) node[bcross=2pt, thick] {};
\draw (8.6,0) node {$+$};
\draw (9,0) node {$...$};
\draw (9.4,0) node {$\displaystyle \left. \vphantom{\frac{1}{1}} \right)$};
\end{tikzpicture}

\vspace{0.3cm}

\begin{tikzpicture}
\draw (0,0) node {$\times$};
\draw (0.6,0) node {$\displaystyle \left( \vphantom{\frac{1}{1}} \right.$};
\filldraw[fill=black!30, thick]  (1.4,0) circle (0.4);
\draw (1.4,0) node[crossbl=2pt, thick] {};
\draw (2.2,0) node {$+$};
\filldraw[fill=black!30, thick]  (3,0) circle (0.4);
\draw (3,0) node[crossbl=2pt, thick] {};
\draw (3.4,0) node[bcross=2pt, thick] {};
\draw (3.8,0) node {$+$};
\filldraw[fill=black!30, thick]  (4.6,0) circle (0.4);
\draw (4.6,0) node[crossbl=2pt, thick] {};
\draw (4.2,0) node[bcross=2pt, thick] {};
\draw (5,0) node[bcross=2pt, thick] {};
\draw (5.4,0) node {$+$};
\filldraw[fill=black!30, thick]  (6.2,0) circle (0.4);
\draw (6.2,0) node[crossbl=2pt, thick] {};
\draw (5.8,0) node[bcross=2pt, thick] {};
\draw (6.2,0.4) node[bcross=2pt, thick] {};
\draw (6.6,0) node[bcross=2pt, thick] {};
\draw (7,0) node {$+$};
\filldraw[fill=black!30, thick]  (7.8,0) circle (0.4);
\draw (7.8,0) node[crossbl=2pt, thick] {};
\draw (7.4,0) node[bcross=2pt, thick] {};
\draw (7.8,0.4) node[bcross=2pt, thick] {};
\draw (7.8,-0.4) node[bcross=2pt, thick] {};
\draw (8.2,0) node[bcross=2pt, thick] {};
\draw (8.6,0) node {$+$};
\draw (9,0) node {$...$};
\draw (9.4,0) node {$\displaystyle \left. \vphantom{\frac{1}{1}} \right)$};
\end{tikzpicture}

\vspace{0.3cm}

\begin{tikzpicture}
\draw (0,0) node {$\times$};
\draw (0.6,0) node {$\displaystyle \left( \vphantom{\frac{1}{1}} \right.$};
\filldraw[fill=black!30, thick]  (1.4,0) circle (0.4);
\draw (1.2,0) node[crossbl=2pt, thick] {};
\draw (1.6,0) node[crossbl=2pt, thick] {};
\draw (2.2,0) node {$+$};
\filldraw[fill=black!30, thick]  (3,0) circle (0.4);
\draw (2.8,0) node[crossbl=2pt, thick] {};
\draw (3.2,0) node[crossbl=2pt, thick] {};
\draw (3.4,0) node[bcross=2pt, thick] {};
\draw (3.8,0) node {$+$};
\filldraw[fill=black!30, thick]  (4.6,0) circle (0.4);
\draw (4.4,0) node[crossbl=2pt, thick] {};
\draw (4.8,0) node[crossbl=2pt, thick] {};
\draw (4.2,0) node[bcross=2pt, thick] {};
\draw (5,0) node[bcross=2pt, thick] {};
\draw (5.4,0) node {$+$};
\filldraw[fill=black!30, thick]  (6.2,0) circle (0.4);
\draw (6,0) node[crossbl=2pt, thick] {};
\draw (6.4,0) node[crossbl=2pt, thick] {};
\draw (5.8,0) node[bcross=2pt, thick] {};
\draw (6.2,0.4) node[bcross=2pt, thick] {};
\draw (6.6,0) node[bcross=2pt, thick] {};
\draw (7,0) node {$+$};
\filldraw[fill=black!30, thick]  (7.8,0) circle (0.4);
\draw (7.6,0) node[crossbl=2pt, thick] {};
\draw (8,0) node[crossbl=2pt, thick] {};
\draw (7.4,0) node[bcross=2pt, thick] {};
\draw (7.8,0.4) node[bcross=2pt, thick] {};
\draw (7.8,-0.4) node[bcross=2pt, thick] {};
\draw (8.2,0) node[bcross=2pt, thick] {};
\draw (8.6,0) node {$+$};
\draw (9,0) node {$...$};
\draw (9.4,0) node {$\displaystyle \left.\left. \vphantom{\frac{1}{1}} \right)\right]$ };

\end{tikzpicture}

\vspace{0.3cm}

\begin{tikzpicture}
\draw (0,0) node {$ \displaystyle = \int d^{10}x \left[ \vphantom{\frac{1}{1}} \right. ~$ };
\filldraw[fill=black!30, thick]  (1.4,0) circle (0.4);
\draw (1.2,0) node[gcross=2pt, thick] {};
\draw (1.6,0) node[gcross=2pt, thick] {};
\filldraw[fill=black!30, thick]  (2.4,0) circle (0.4);
\draw (2.4,0) node[gcross=2pt, thick] {};
\filldraw[fill=black!30, thick]  (3.4,0) circle (0.4);
\draw (3.4,0) node[gcross=2pt, thick] {};
\draw (4.2,0) node {$+$};
\draw (5,0) node {$g_s\Delta B$};
\filldraw[fill=black!30, thick]  (6.2,0) circle (0.4);
\draw (6.2,0) node[gcross=2pt, thick] {};
\filldraw[fill=black!30, thick]  (7.2,0) circle (0.4);
\draw (7.2,0) node[gcross=2pt, thick] {};
\filldraw[fill=black!30, thick]  (8.2,0) circle (0.4);
\draw (8.2,0) node[gcross=2pt, thick] {};
\filldraw[fill=black!30, thick]  (9.2,0) circle (0.4);
\draw (9.2,0) node[gcross=2pt, thick] {};
\draw (9.8,0) node { $\displaystyle \left. \vphantom{\frac{1}{1}} \right]$ };

\end{tikzpicture}

\caption{Diagrams contributing to the momenta dependent part of order $g_s$ D-instanton amplitude for four closed strings. They come with arbitrary number of insertions of open string field collective modes $\Phi_o^m$ whose corresponding vertex operators are represented by blue crosses. Black crosses are linear combination of closed string vertex operators for $\delta\tau$ and $\delta{\bar\tau}$. Red crosses are vertex operators for $\delta\tau$ and $\Delta B$ is a string field theory parameter dependent constant. All the diagrams are at the D-instanton location $x^\mu$.}
\label{fig:4ptAmp}
\end{figure}

After integrating out $\zeta^2$ in (\ref{ocpathint2}) according to the prescription in the previous subsection, we are left with
\ie\label{oscintv}
e^{-{C\over g_s}} \int d^{10}\phi d^{16}\theta \, e^{W[\phi,\theta,x;\Psi_c]},
\fe
where we have indicated the explicit dependence of the integrand on the D-instanton moduli $x^\mu$. $W[\phi,\theta,x;\Psi_c]$ includes diagrams with an arbitrary number of $\phi^\mu$ and $\theta_\A$ insertions even at a given order in $g_s$ (Figure \ref{fig:4ptAmp}). Even though the integration over $\theta_\A$ picks out only the terms proportional to $\theta^{16}$ in $W[\phi,\theta,x;\Psi_c]$, they still include arbitrarily many powers of $\phi^\mu$. In this section, we discuss how to carry out the sum over infinitely many such terms by an appropriate change of variables for $\phi^\mu$. We then discuss the effects of $\theta_\A$ insertions in the next section.

Heuristically, the integration in $\phi$ should be equivalent to an integration over the D-instanton moduli space. One way to understand their relation is through the background independence of SFT \cite{Sen:1990hh, Sen:1993kb, Sen:2017szq}: a deformation of the boundary moduli $\delta x^\mu$ can be absorbed by an open string field redefinition $(\phi^\mu,\theta_\A) \to (\phi^\mu + \delta\phi^\mu, \theta_\A + \delta \theta_\A)$, where
\ie\label{varxphi}
\delta\phi^\mu = \delta x^\nu f^\mu{}_\nu[\phi, \theta, x; \Psi_c],~~~~ \delta\theta_\A = \delta x^\nu g_{\A\nu}[\phi,\theta, x; \Psi_c].
\fe
In other words, different points on a hypersurface obtained by integrating the equation (\ref{varxphi}) represent equivalent string field configurations. Transporting along this hypersurface from $x$ to $x'$, $(\phi,\theta)$ turn into $(\phi',\theta')$, while the integration measure (\ref{oscintv}) is invariant, namely
\ie
d^{10}\phi d^{16}\theta \,e^{W[\phi,\theta, x;\Psi_c]} = d^{10}\phi' d^{16}\theta' \, e^{W[\phi',\theta',x';\Psi_c]}.
\fe
Now we can transport along the hypersurface to $\phi=0$, and write the integral in (\ref{oscintv}) equivalently as 
\ie\label{rewrjac}
\int d^{10}\phi d^{16}\theta\, e^{W[\phi,\theta,x;\Psi_c]} = \int d^{10}x d^{16}\theta\, \det\Big( f^\mu{}_\nu[0,\theta,x;\Psi_c] \Big) \,e^{W[0,\theta,x;\Psi_c]}.
\fe
where the dependence on the bosonic open string field has now been eliminated from the integrand on the RHS.\footnote{This in particular implies that $W[\phi,\theta,x;\Psi_c]$ does not include terms consisting of only $\phi^\mu$'s. In Appendix~\ref{sec:discopen4pt}, we perform an explicit computation of the $\phi^4$ term in $W[\phi,\theta,x;\Psi_c]$ and show that it is indeed zero.}

For the purpose of extracting the Jacobian factor $\det f$ to first order in $g_s$, we can replace $\theta_\A$ insertions with that of the supercharges, which amounts to viewing $\theta$ as fermionic moduli rather than open string fields. This eliminates the need for considering the shift $\delta\theta$ in (\ref{varxphi}). Expanding
\ie
f[0,\theta,x;\Psi_c] = 1 + g_s f_1[\theta,x;\Psi_c] ,
\fe
we expect the $\theta$-dependence of $f_1$ to be dictated by the nonlinearly realized super-Poincar\'e symmetry, similar to (\ref{gsytheta}). We can determine $f_1[0,x;\Psi_c]$ from the equation\footnote{This is a special case of the analog of (4.12) of \cite{Sen:1993kb} for open string field theory, restricted to $\phi=0$. Here we also assumed $\partial_\phi f^\mu{}_\nu\big|_{\phi=0}=0$, which follows from the general construction of  \cite{Sen:1993kb}.}
\ie{}
\left(f^\mu{}_\nu[0,0, x;\Psi_c] {\partial\over \partial\phi^\mu} 
+ {\partial\over \partial x^\nu} \right) W[\phi, 0, x; \Psi_c]\bigg|_{\phi=0} = 0.
\fe
Expanding $W=\sum_{n=0}^\infty g_s^n W^{(n)}$, we have at order $g_s$ the relation
\ie\label{fwweq}
(f_1)^\mu{}_\nu[0,x;\Psi_c] {\partial\over \partial\phi^\mu} W^{(0}[\phi,0,x;\Psi_c]\bigg|_{\phi=0}  =   {\partial\over \partial\phi^\nu} W^{(1)}[\phi,0,x;\Psi_c]\bigg|_{\phi=0} - {\partial\over \partial x^\nu} W^{(1)}[0,0,x;\Psi_c]  .
\fe
In a closed string background where only supergraviton modes are turned on, the order $g_s^0$ term in the effective action gives
\ie
W^{(0)}[\phi,0,x;\Psi_c] = {\cal A}^{D^2}_{\delta\tau} \, \delta\tau(x+\phi).
\fe
At first order in $g_s$, a calculation of SFT Feynman diagrams analogous to those of Section~\ref{FeynmanSec} gives
\ie\label{wwdiff}
 {\partial\over \partial\phi^\nu} W^{(1)}[\phi,0,x;\Psi_c]\bigg|_{\phi=0} - {\partial\over \partial x^\nu} W^{(1)}[0,0,x;\Psi_c] = \Delta U  \delta\tau(x) \partial_\nu \delta\tau(x) + \Delta U' \partial^\mu \delta\tau(x) \partial_\mu\partial_\nu \delta\tau(x) ,
\fe
where $\Delta U$ and $\Delta U'$ are constants that depend on SFT parameters. From (\ref{fwweq}) we solve
\ie
(f_1)^\mu{}_\nu[0,x;\Psi_c] =  ({\cal A}^{D^2}_{\delta\tau})^{-1} \left[ \Delta U \delta^\mu_\nu \delta\tau(x) + \Delta U' \partial^\mu\partial_\nu\delta\tau(x) \right] .
\fe
After restoring the $\theta$-dependence by super-Poincar\'e symmetry, we obtain the Jacobian factor 
\ie\label{defgs}
\det f[0,\theta,x,\Psi_c] = 1 +10 g_s ({\cal A}^{D^2}_{\delta\tau})^{-1} \Delta U \left( e^{i\theta_\A {\cQ}_-^\A} \delta\tau \right) (x)
\fe
that appears on the RHS of (\ref{rewrjac}). Note that $\Delta U'$ drops out due to the on-shell condition of the background field $\delta\tau(x)$.
Similar to the correction factor (\ref{gsytheta}), the Jacobian factor (\ref{defgs}) contributes only to the constant term in (\ref{eq:Dinstamp}).

\subsection{Integration over $\theta_\A$}
\label{sec:thetaintl}

As already mentioned, the integration over the fermionic open string collective modes $\theta_\A$ is similar to inserting the spacetime supercharge ${\widehat Q}^\alpha_{(\pm \frac12),-}$ represented as a contour integral of the spin field along the boundary of the worldsheet, but they are not the same beyond leading order in $g_s$. In particular, while the former is unambiguously defined in SFT, the latter is subject to an ambiguity in the location of the PCOs that accompany the supercharge insertion, as already encountered in the on-shell computation of Section~\ref{sec:1D4pt}.

For instance, at leading order in $g_s$, $W$ contains an effective string vertex that couples $\theta$ to the dilatino $\lambda$, of the form
\ie
{\cal A}^{D^2}_{\delta\tau} \theta_\A\lambda^\A(x) ,
\fe
where the coefficient, which is computed by the disc diagram with a fermion open string collective mode and a dilatino insertion, is equal to ${\cal A}^{D^2}_{\delta\tau}$, the disc amplitude with a single $\delta\tau$ insertion. In this case, integrating out $\theta_\A$ has the same effect as acting on the dilatino field with the supercharge ${\cQ}_+^\A$ acting as a lowering operator.

At the next order in $g_s$, there is an effective string vertex of the form
\ie
g_s \int d^{10} p d^{10} p' \,e^{i(p+p')\cdot x} V(p,p')  \theta_\A \widetilde{\lambda^\A}(p) \widetilde{\delta\tau}(p').
\fe
where $V(p,p')$ is computed by summing over certain SFT diagrams, which consist either of an open-closed-closed vertex by itself, or of Feynman diagrams corresponding to open-closed vertices connected by massive open string or $\varkappa^1$ propagators, whose corresponding worldsheet diagram is that of a disc with a fermionic open string collective mode, a dilatino, and a $\delta\tau$ insertion. The resulting quantity is different from that of a disc with the two bulk insertions, $\delta\tau(p)$ and $\delta\tau(p')$. However, by a calculation similar to that of (\ref{wwdiff}), one finds that the difference takes the factorized form
\ie
g_s \left(\Delta V \theta_\A \lambda^\A(x) \delta\tau(x)+ \Delta V' \theta_\A \partial_\mu\lambda^\A(x) \partial^\mu\delta\tau(x)\right),
\fe
where $\Delta V$ and $\Delta V'$ are constants that depend on the SFT parameters. When summed over permutations of the four closed string insertions, the $\Delta V'$ term does not contribute due to momentum conservation. Therefore, the difference between integration over $\theta$ and inserting spacetime supercharges on the boundary in the on-shell computation only enters through the constant term in (\ref{eq:Dinstamp}).

\section{Open string background independence and Sen gauge}
\label{sec:discopen4pt}

As discussed in Section~\ref{collectiveSec}, in order to integrate over the massless bosonic open string fields $\phi^\mu$, we perform a field redefinition trading the $\phi^\mu$ for the bosonic moduli $x^\mu$ \eqref{rewrjac}. The existence of such a field redefinition relies on the fact that $W_f[\phi^\mu,\theta_\alpha,\zeta^2,\Psi_c]$ vanishes upon setting the other fields to zero, i.e. it does not contain a potential for $\phi^\mu$. Although this result is anticipated from open string background independence, it is by no means obvious in the SFT framework.\footnote{We stress that $\phi^\mu$ is not the same as a moduli deformation of $x^\mu$, and so the fact that the potential vanishes is not simply a consequence of conformal perturbation theory.}

In this appendix, we shall explicitly demonstrate the vanishing of such terms in the massless open string effective action. In particular, we set our sights on the tree-level quartic coupling appearing in $W_f$, whose contribution is fixed by Lorentz invariance to take the form
\ie{}
{1 \over 4}g_{\text{tree}} (\phi^2)^2 \subset W_f[\phi^\mu,\theta_\alpha,\zeta^2,\Psi_c].
\fe
In the standard perturbative framework, we know that $g_{\text{tree}}$ enters into the 4-point amplitude as
\ie{}
A_{\phi^\mu \phi^\nu\phi^\sigma\phi^\rho}^{D^2} = 2g_{\text{tree}}S^{\mu\nu\sigma\rho} , \quad
S^{\mu\nu\sigma\rho} = \eta^{\mu\nu}\eta^{\sigma\rho} + \eta^{\mu\sigma}\eta^{\nu\rho} + \eta^{\mu\rho}\eta^{\sigma\nu}.
\fe
In the EFT analysis, this amplitude receives contributions from the elementary 4-point Feynman vertex together with a set of Feynman diagrams consisting of 3-point Feynman vertices $A_{\phi^\mu\phi^\nu \psi}^{D^2}$ for two $\phi^\mu$ fields and one massive field $\psi \in \Psi_o^f$, stitched together by an open string propagator $P_{\psi}$, i.e.
\ie{}
A_{\phi^\mu\phi^\nu\phi^\sigma\phi^\rho}^{D^2} &= A_{\phi^\mu\phi^\nu\phi^\sigma\phi^\rho}^{D^2}\bigg|_{\text{vertex}} \\
&~~~+ \sum_{\psi \in \Psi_o^f} \left(A_{\phi^\mu\phi^\nu \psi}^{D^2}P_{\psi} A_{\psi\phi^\sigma\phi^\rho}^{D^2} + \text{5 permutations of $\mu,\nu,\sigma,\rho$}\right)\bigg|_{\text{propagator}}.
\fe
From the perspective of the moduli space integration, the first term on the RHS of \eqref{eq:4ptamp} contributes to the ``vertex region'' of the amplitude, and the terms in the sum to the ``propagator region.'' The 6 permutations of the Lorentz indices reflects the decomposition of the propagator region into three disconnected components, each with two boundaries. By a judicious choice of 3-point vertex, we can completely integrate out all of the massive open strings with nonzero weight, leaving only the ghost zero mode $\varkappa^1$, for which the amplitude becomes
\ie{}
A_{\phi^\mu\phi^\nu\phi^\sigma\phi^\rho}^{D^2} &= A_{\phi^\mu\phi^\nu\phi^\sigma\phi^\rho}^{D^2}\bigg|_{\text{vertex}}  \\
&~~~+ \left(V_{\phi^\mu\phi^\nu \varkappa^1}^{D^2}P_{\varkappa^1} V_{\varkappa^1\phi^\sigma\phi^\rho}^{D^2} + \text{5 permutations of $\mu,\nu,\sigma,\rho$}\right)\bigg|_{\text{propagator}}.
\label{eq:4ptamp}
\fe
Our goal is thus to show that this expression vanishes.\footnote{The 4-point amplitude has been previously confirmed to vanish, albeit for a choice of vertices where the ghost zero mode does not contribute. \cite{Maccaferri:2018vwo, Maccaferri:2019ogq}}

\subsubsection*{3-point vertex}

In order to define the elementary 3-point Feynman vertex for NS-sector string fields, we need to specify a set of local coordinate charts around the punctures as well as the location of the PCO. As before, we take the disc to be parametrized by global coordinate $z$ in the UHP. We shall employ the same set of coordinate maps as in \cite{Sen:2020cef}, with
\ie{}
f_0(w_0) = {2w_0 \over 2\alpha + w_0}, \quad 
f_1(w_1) = {2\alpha+w_1 \over 2\alpha -w_1}, \quad 
f_\infty(w_\infty) = {w_\infty - 2\alpha \over 2w_\infty},
\fe
where $w_i$ labels the coordinates on three half-discs, and $\alpha \in \mathbb{R}$ is an SFT parameter. These functions map $w_i = 0$ to the points $z=0,1,\infty$ on the UHP, and are cyclically permuted under $z \to (1-z)^{-1}$. Their inverses are given by
\ie{}
w_0(z) = \frac{2 \alpha  z}{2-z} \,, \quad w_1(z) = \frac{2 \alpha  (z-1)}{z+1}  \,, \quad w_\infty(z) = \frac{2 \alpha }{1-2z} \,.
\label{eq:localcoords}
\fe
With this choice of coordinates, we take the PCO to be located at the permutation-invariant point $z = p_0$ with
\ie{}
p_0 = e^{+{i \pi \over 3}}.
\fe

In practice, we take the SFT parameter $\alpha$ to be arbitrarily large so that the massive open string modes do not contribute to the effective vertex.\footnote{We have also repeated the calculation of this appendix at finite $\alpha$, in which case all of the massive open string fields contribute, with the same conclusion (\ref{adphphphres}), but will not present its lengthy details here.} To be precise, a mode of weight $h>0$ will have a propagator proportional to $\alpha^{-h}$ which vanishes in the limit $\alpha \to \infty$. On the other hand, the $\varkappa^1$ propagator survives since its associated vertex operator $c \partial c e^{-2\phi}\partial \xi$ is a weight zero conformal primary. Using our choice of local coordinates and PCO location, its 3-point Feynman vertex is given by the amplitude
\ie{}
A_{\phi^\mu\phi^\nu \varkappa^1}^{D^2} = \left\langle {\cal X}(p_0) ce^{-\phi}\psi^\mu(0) ce^{-\phi}\psi^\nu(1) c\partial c e^{-2\phi}\partial \xi (\infty) \right\rangle^{D^2}.
\fe
By direct computation, we find
\ie{}
A_{\phi^\mu\phi^\nu\varkappa^1}^{D^2} = C_{D^2} \eta^{\mu\nu} {1+p_0 \over 1-p_0},
\fe 
and so the contribution of $\varkappa^1$ to the propagator region of the 4-point amplitude in \eqref{eq:4ptamp} is
\ie{}
C_{D^2}S^{\mu\nu\sigma\rho} \left( {1+p_0 \over 1-p_0}\right)^2,
\label{eq:k1cont}
\fe
where we have used the ghost propagator $P_{\varkappa^1} = {1 \over 2 C_{D^2}}$.

\subsubsection*{4-point vertex}

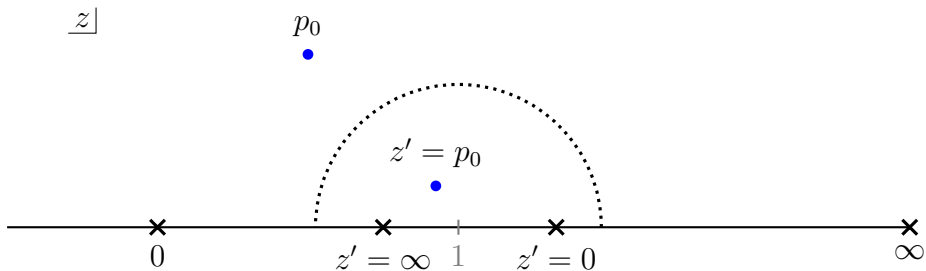
\begin{figure}
\centering
\begin{tikzpicture}
\draw[color=black, very thick, dotted] (4,0) circle (1.9);
\filldraw [color=white] (-2,0) -- (8,0) -- (8,-2) -- (-2,-2); 

\draw [color=black, thick] (-2,0) -- (10,0); 

\draw node[cross=4pt, very thick] at (0,0) {};
\node[below] at (0,-0.1) {$0$};
\draw [color=gray, thick] (4,0.1) -- (4,-0.1);
\node[below] at (4,-0.1) {\color{gray}$1$};
\draw node[cross=4pt, very thick] at (10,0) {};
\node[below] at (10,-0.1) {$\infty$};

\draw node[cross=4pt, very thick] at (3,0) {};
\node[below] at (3,-0.1) {$z'=\infty$};
\draw node[cross=4pt, very thick] at (5.3,0) {};
\node[below] at (5.3,-0.1) {$z'=0$};

\filldraw[very thick, color=blue] (2,2.3) circle (0.05);
\node[above] at (2,2.4) {$p_0$};
\filldraw[very thick, color=blue] (3.7,0.55) circle (0.05);
\node[above] at (3.7,0.65) {$z'=p_0$};

\node(n)[inner sep=2pt] at (-1,2.75) {$z$};
\draw[line cap=round](n.south west)--(n.south east)--(n.north east);

\end{tikzpicture}
\caption{Plumbing configuration of the 4-point propagator region for a finite choice of the SFT parameter $\alpha$. Vertex operator insertions are marked with black crosses, while PCO insertions are marked with blue dots.}
\label{fig:plumb}
\end{figure}

In order to define the 4-point vertex, we must first introduce the family of worldsheet configurations corresponding to two 3-point vertices joined together by an open string propagator. That is to say that the range of integration for the vertex region is given by the complement of the propagator region. The aforementioned configurations consist of two discs, parametrized by global coordinates $z,z'$ in the UHP, sewn together by plumbing maps involving the local coordinates in \eqref{eq:localcoords}. Since the 3-point vertex is defined by summing over permutations of identical open string fields, we need only consider a single plumbing configuration, e.g. (see Figure \ref{fig:plumb})
\ie{}
w_1(z) w_1(z') = -q, \quad 0 < q < 1.
\fe
We subsequently perform an $SL(2,\bR)$ transformation that maps three of the punctures on the $z$-disc to $0,1,\infty$. This transformation maps the fourth puncture to some function $x(q)$ of the gluing parameter $q$, which can be identified with the modulus of the 4-punctured disc. Similarly, the PCO locations are mapped to $p_1(q),p_2(q)$, which depend explicitly on the gluing parameter and hence implicitly on the modulus. Depending on the $SL(2,\bR)$ transformation, $x(q)$ is mapped to one of three disconnected regions, conventionally referred to as the $s,t,u$-channel contributions to the disc 4-point amplitude. Up to order ${\cal O}(\alpha^{-4})$, the propagator region consists of the components
\ie{}
s\text{-channel}: \quad &x\in (-\alpha^{-2},\alpha^{-2}), \\
t\text{-channel}: \quad &x\in (1-\alpha^{-2},1+\alpha^{-2}), \\
u\text{-channel}: \quad &x\in (-\infty, -\alpha^2) \cup (\alpha^2, \infty).
\fe
For instance, in the $s$-channel the fourth puncture is arranged to be located at 
\ie{}
x = x_s(q), \quad x_s(q) = \frac{q}{\alpha ^2} + {\cal O}(\alpha^{-4}),
\label{eq:scoord}
\fe
while the two PCOs are located at 
\ie{}
p_1 &= p_{1s}(q), \quad p_{1s}(q) = p_0 - {p_0q \over 2\alpha^2} + {\cal O}(\alpha^{-4}), \\
p_2 &= p_{2s}(q), \quad p_{2s}(q) = {p_0-1 \over p_0} + {q \over 2p_0 \alpha^2} + {\cal O}(\alpha^{-4}).
\fe
Similarly, in the $t$-channel, the fourth puncture is located at
\ie{}
x = x_t(q), \quad x_t(q) = 1-\frac{q}{\alpha ^2} + {\cal O}(\alpha^{-4}),
\label{eq:tcoord}
\fe
and the PCOs reside at
\ie{}
p_1 &= p_{1t}(q), \quad p_{1t}(q) = 1 - {p_0q \over \alpha^2(p_0-1)} + {\cal O}(\alpha^{-4}), \\
p_2 &= p_{2t}(q), \quad p_{2t}(q) = p_0 - {p_0 q \over 2 \alpha^2} + {\cal O}(\alpha^{-4}).
\fe
Strictly speaking, the range $0<q<1$ only covers half of the $s,t,u$ regions, and so we must also consider the same $SL(2,\bR)$ transformations under $q \to -q$.  

We now return to the construction of the 4-point elementary vertex. Using Lorentz invariance, we can write the contribution of the vertex region as\footnote{In principle, we must specify a set of local coordinates around the open string punctures together as well as the two PCO locations which are necessarily compatible with those of the 3-point vertex. However, since the 4-point vertex involves only on-shell fields, it is insensitive to the choice of coordinate maps.}
\ie{}
A^{D^2}_{\phi^\mu\phi^\nu\phi^\sigma\phi^\rho}\bigg|_{\text{vertex}} &= 3 \left[\int_{\alpha^{-2}}^{1-\alpha^{-2}}  \Omega_x dx + \text{VI}_s + \text{VI}_t\right].
\fe
Let us briefly unpack this expression. The first term takes the form of an integrated correlator with integrand
\ie{}
\Omega_x = {1 \over 24}\sum_{\text{perm of $\mu,\nu,\sigma,\rho$}}\left\langle {\cal B}_x {\cal X}(p_1) {\cal X}(p_2) ce^{-\phi}\psi^\mu(0) ce^{-\phi}\psi^\nu(x) ce^{-\phi}\psi^\sigma(1) ce^{-\phi}\psi^\rho(\infty) \right\rangle^{D^2}.
\label{eq:4ptcorr}
\fe 
By averaging over the 24 permutations of the spacetime Lorentz indices in \eqref{eq:4ptcorr}, we can restrict the vertex region to lie between the $s$- and $t$-channels, i.e. $x_s(1) < x < x_t(1)$, where $x_s(q)$ and $x_t(q)$ are the local coordinates covering half of the $s$- and $t$-channel regions, as defined in \eqref{eq:scoord} and \eqref{eq:tcoord}. We have chosen the PCOs to reside at fixed locations
\ie{}
p_1 = p_0 - {p_0 \over 2\alpha^2}, \quad p_2 = 1 +{p_0 \over \alpha^2(1-p_0)} ,
\label{eq:pcoloc}
\fe
which are precisely $p_{1s}(q=1)$ and $p_{1t}(q=1)$, respectively. This ensures that the location of one PCO agrees on the boundary between the $s$-channel region and the vertex region, and similarly for PCO 2 with the $t$-channel.  Finally, ${\cal B}_x dx$ is the Beltrami differential associated to the modulus $x$. For two PCOs on the disc, it generically takes the form
\ie\label{beltramidiff}
{\cal B}_x = \int_x \frac{dz}{2\pi i}b(z) + \frac{1}{{\cal X}(p_1)} \frac{\partial p_1}{\partial x} \partial \xi(p_1) + \frac{1}{{\cal X}(p_2)} \frac{\partial p_2}{\partial x} \partial \xi(p_2) \,,
\fe
where $1/{\cal X}(p)$ should be understood as a formal operator which removes the corresponding PCO ${\cal X}(p)$. However, we have chosen PCO locations in \eqref{eq:pcoloc} that do not depend on the moduli, and so the  $\partial \xi$ terms in the above expression drop out.

Following a straightforward application of Wick contractions, we find
\ie{}
\Omega_x &= \frac13S^{\mu\nu\sigma\rho} \left[ {F_0(p_1,p_2) \over x^2} + {F_1(p_1,p_2) \over (1-x)^2} +F_\infty(p_1,p_2) \right], \\
F_0(p_1,p_2) &= -C_{D^2}\frac{p_1 p_2 \left(p_1+p_2-2\right)}{ \left(p_1-p_2\right){}^2} \,, \\
F_1(p_1,p_2) &= C_{D^2}\frac{\left(p_1-1\right) \left(p_2-1\right) \left(p_1+p_2\right)}{ \left(p_1-p_2\right){}^2} \,, \\
F_\infty(p_1,p_2) &=C_{D^2} \frac{p_1 \left(2 p_2-1\right)-p_2}{ \left(p_1-p_2\right){}^2} \,.
\fe
Integrating this expression gives a contribution that is subleading in $\alpha$, namely
\ie{}
\int_{\alpha^{-2}}^{1-\alpha^{-2}} \Omega_x dx = {\cal O}(\alpha^{-2}).
\fe

At each of the boundaries between the propagator and vertex regions there is a single PCO whose location assumes some value $p$ on one side and $p'$ on the other. This issue of non-agreement can be fixed following \cite{Sen:2015hia}, where one closes the gap by integrating along the PCO direction. This amounts to integrating $\partial \xi$ in \eqref{beltramidiff}, leading to $\xi(p) - \xi(p')$. For the vertex under consideration, we must perform such a vertical integration at the $s$-channel boundary, which contributes as
\ie{}
\text{VI}_s &= {1 \over 24}\sum_{\text{perm of $\mu,\nu,\sigma,\rho$}}\langle {\cal X}(p_1) \big[\xi(p_{1t}(q=1))-\xi(p_{2s}(q=1))\big] \\&~~~\times ce^{-\phi}\psi^\mu(0) ce^{-\phi}\psi^\nu(x) ce^{-\phi}\psi^\sigma(1) ce^{-\phi}\psi^\rho(\infty)\rangle^{D^2} \\
&= -\frac13 C_{D^2}S^{\mu\nu\sigma\rho}{1+p_0 \over (1-p_0)^2} + {\cal O}(\alpha^{-2}),
\fe
as well as the $t$-channel boundary, which contributes as
\ie{}
\text{VI}_t &= {1 \over 24}\sum_{\text{perm of $\mu,\nu,\sigma,\rho$}} \langle \big[\xi(p_{1s}(q=1))-\xi(p_{2t}(q=1))\big]{\cal X}(p_2)  \\&~~~\times ce^{-\phi}\psi^\mu(0) ce^{-\phi}\psi^\nu(x) ce^{-\phi}\psi^\sigma(1) ce^{-\phi}\psi^\rho(\infty)\rangle^{D^2} \\
&= -\frac13 C_{D^2}S^{\mu\nu\sigma\rho}{p_0(1+p_0) \over (1-p_0)^2} + {\cal O}(\alpha^{-2}).
\fe
From this it follows that the vertex region contributes as
\ie{}
A^{D^2}_{\phi^\mu\phi^\nu\phi^\sigma\phi^\rho}\bigg|_{\text{vertex}} = -C_{D^2} S^{\mu\nu\sigma\rho} \left({1+p_0 \over 1-p_0}\right)^2.
\fe
Comparing this with \eqref{eq:k1cont}, we find that the vertex region exactly cancels that of the propagator region to give
\ie\label{adphphphres}
A^{D^2}_{\phi^\mu\phi^\nu\phi^\sigma\phi^\rho} = 0,
\fe
and so the 4-point vertex in the massless open string effective action vanishes, as promised.

\bibliographystyle{JHEP}
\bibliography{Dinst}

\end{document}